\documentclass[twocolumn]{aastex62}
\let\pwiflocal=\iffalse \let\pwifjournal=\iffalse

\usepackage{amsmath,amssymb}
\usepackage{epsfig}    
\usepackage{graphicx}    
\usepackage{lineno}
\usepackage{natbib}
\usepackage{bigints}
\usepackage[outdir=./]{epstopdf}

\usepackage[T1]{fontenc}
\pwifjournal\else
  \usepackage{microtype}
\fi

\pwifjournal\else
  \makeatletter
  \renewcommand\plotone[1]{%
    \centering \leavevmode \setlength{\plot@width}{0.95\linewidth}
    \includegraphics[width={\eps@scaling\plot@width}]{#1}%
  }%
  \makeatother
\fi

\makeatletter

\newcommand\@simpfx{http://simbad.u-strasbg.fr/simbad/sim-id?Ident=}

\newcommand\MakeObj[4][\@empty]{
  \pwifjournal%
    \expandafter\newcommand\csname pkgwobj@c@#2\endcsname[1]{\protect\object[#4]{##1}}%
  \else%
    \expandafter\newcommand\csname pkgwobj@c@#2\endcsname[1]{\href{\@simpfx #3}{##1}}%
  \fi%
  \expandafter\newcommand\csname pkgwobj@f#2\endcsname{#4}%
  \ifx\@empty#1%
    \expandafter\newcommand\csname pkgwobj@s#2\endcsname{#4}%
  \else%
    \expandafter\newcommand\csname pkgwobj@s#2\endcsname{#1}%
  \fi}%

\newcommand\MakeTrunc[2]{
  \expandafter\newcommand\csname pkgwobj@t#1\endcsname{#2}}%

\newcommand{\obj}[1]{%
  \expandafter\ifx\csname pkgwobj@c@#1\endcsname\relax%
    \textbf{[unknown object!]}%
  \else%
    \csname pkgwobj@c@#1\endcsname{\csname pkgwobj@s#1\endcsname}%
  \fi}
\newcommand{\objf}[1]{%
  \expandafter\ifx\csname pkgwobj@c@#1\endcsname\relax%
    \textbf{[unknown object!]}%
  \else%
    \csname pkgwobj@c@#1\endcsname{\csname pkgwobj@f#1\endcsname}%
  \fi}
\newcommand{\objt}[1]{%
  \expandafter\ifx\csname pkgwobj@c@#1\endcsname\relax%
    \textbf{[unknown object!]}%
  \else%
    \csname pkgwobj@c@#1\endcsname{\csname pkgwobj@t#1\endcsname}%
  \fi}

\makeatother

\pwifjournal\else
  \usepackage{etoolbox}
  \makeatletter
  \patchcmd{\NAT@citex}
    {\@citea\NAT@hyper@{%
       \NAT@nmfmt{\NAT@nm}%
       \hyper@natlinkbreak{\NAT@aysep\NAT@spacechar}{\@citeb\@extra@b@citeb}%
       \NAT@date}}
    {\@citea\NAT@nmfmt{\NAT@nm}%
     \NAT@aysep\NAT@spacechar\NAT@hyper@{\NAT@date}}{}{}
  \patchcmd{\NAT@citex}
    {\@citea\NAT@hyper@{%
       \NAT@nmfmt{\NAT@nm}%
       \hyper@natlinkbreak{\NAT@spacechar\NAT@@open\if*#1*\else#1\NAT@spacechar\fi}%
         {\@citeb\@extra@b@citeb}%
       \NAT@date}}
    {\@citea\NAT@nmfmt{\NAT@nm}%
     \NAT@spacechar\NAT@@open\if*#1*\else#1\NAT@spacechar\fi\NAT@hyper@{\NAT@date}}
    {}{}
  \makeatother
\fi

\providecommand{\adsurl}[1]{\href{#1}{ADS}}

\newcommand{\um}{$\mu$m}

\newcommand\teff{\ensuremath{T_\text{eff}}}

\usepackage[outdir=./]{epstopdf}

\newcommand{\mks}{$M_{K_S}$}
\newcommand{\mmk}{$M_{K_S}$\textendash$M_*$}
\newcommand{\mmkc}{$M_{K}$\textendash$M_*$}
\newcommand{\order}{5}
\newcommand{\mtot}{$M_{\rm{tot}}$}
\newcommand{\mpred}{$M_{\rm{tot,pre}}$}
\newcommand{\mdyn}{$M_{\rm{tot,dyn}}$}
\newcommand{\degree}{$^{\circ}$}
\usepackage{mathrsfs}
\usepackage{amsmath}
\usepackage{bigints} 
\usepackage{nicefrac}
\usepackage[hang]{footmisc}
\shorttitle{Mass-luminosity-metallicity relation of cool stars}
\shortauthors{Mann et al.}
\accepted{November 13, 2018}

\begin{document}

\title{How to Constrain Your M dwarf II:\\ the Mass-Luminosity-Metallicity Relation from 0.075$M_\odot$ to 0.70$M_\odot$ }

\correspondingauthor{Andrew W. Mann}
\email{awmann@unc.edu}

\author[0000-0003-3654-1602]{Andrew W. Mann}
\affiliation{Department of Physics and Astronomy, University of North Carolina at Chapel Hill, Chapel Hill, NC 27599-3255, USA}

\author[0000-0001-9823-1445]{Trent Dupuy}
\affiliation{Gemini Observatory, Northern Operations Center, 670 N. Aohoku Place, Hilo, HI 96720, USA}
 
\author[0000-0001-9811-568X]{Adam L. Kraus}
\affiliation{Department of Astronomy, The University of Texas at Austin, Austin, TX 78712, USA}

\author[0000-0002-5258-6846]{Eric Gaidos}
\affiliation{Department of Geology \& Geophysics, University of Hawaii at M\={a}noa, Honolulu, Hawaii 96822 USA}

\author[0000-0003-4142-9842]{Megan Ansdell}%
\affiliation{Center for Integrative Planetary Science, Berkeley, CA 94720, USA}
\affiliation{Department of Astronomy, University of California at Berkeley, Berkeley, CA 94720, USA}

\author[0000-0002-6194-043X]{Michael Ireland}%
\affiliation{Research School of Astronomy \& Astrophysics, Australian National University, Canberra ACT 2611, Australia}

\author[0000-0001-9982-1332]{Aaron C. Rizzuto}
\altaffiliation{51 Peg b Fellow}
\affiliation{Department of Astronomy, The University of Texas at Austin, Austin, TX 78712, USA}

\author[0000-0002-6879-3639]{Chao-Ling Hung}
\affiliation{Department of Physics, Manhattan College, 4513 Manhattan College Parkway, Riverdale, NY 10471, USA}

\author[0000-0001-7730-2240]{Jason Dittmann}
\altaffiliation{51 Peg b Fellow}
\affiliation{Massachusetts Institute of Technology, 77 Massachusetts Avenue, Cambridge, Massachusetts 02138, USA}

\author[0000-0002-8332-8516]{Samuel Factor}
\affiliation{Department of Astronomy, The University of Texas at Austin, Austin, TX 78712, USA}

\author[0000-0002-2012-7215]{Gregory Feiden}
\affiliation{Department of Physics, University of North Georgia, Dahlonega, GA 30597, USA}

\author[0000-0001-6301-896X]{Raquel A. Martinez}
\affiliation{Department of Astronomy, The University of Texas at Austin, Austin, TX 78712, USA}

\author[0000-0003-3573-8163]{Dary Ru\'iz-Rodr\'iguez}
\affiliation{Chester F. Carlson Center for Imaging Science, School of Physics \& Astronomy, and Laboratory for Multiwavelength Astrophysics,
Rochester Institute of Technology, 54 Lomb Memorial Drive, Rochester NY 14623 USA}
\affiliation{Research School of Astronomy and Astrophysics, Australian National University, Canberra, ACT 2611, Australia}

\author[0000-0001-5729-6576]{Pa Chia Thao}
\affiliation{Department of Physics and Astronomy, University of North Carolina at Chapel Hill, Chapel Hill, NC 27599-3255, USA}

\begin{abstract}
The mass-luminosity relation for late-type stars has long been a critical tool for estimating stellar masses. However, there is growing need for both a higher-precision relation and a better understanding of systematic effects (e.g., metallicity). Here we present an empirical relationship between \mks\ and $M_*$ spanning $0.075M_\odot<M_*<0.70M_\odot$. The relation is derived from 62 nearby binaries, whose orbits we determine using a combination of Keck/NIRC2 imaging, archival adaptive optics data, and literature astrometry. From their orbital parameters, we determine the total mass of each system, with a precision better than 1\% in the best cases. We use these total masses, in combination with resolved $K_S$ magnitudes and system parallaxes, to calibrate the \mmk\ relation. The resulting posteriors can be used to determine masses of single stars with a precision of 2-3\%, which we confirm by testing the relation on stars with individual dynamical masses from the literature. The precision is limited by scatter around the best-fit relation beyond measured $M_*$ uncertainties, perhaps driven by intrinsic variation in the \mmk\ relation or underestimated uncertainties in the input parallaxes. We find that the effect of [Fe/H] on the \mmk\ relation is likely negligible for metallicities in the solar neighborhood (0.0$\pm$2.2\% change in mass per dex change in [Fe/H]). This weak effect is consistent with predictions from the Dartmouth Stellar Evolution Database, but inconsistent with those from MESA Isochrones and Stellar Tracks (at $5\sigma$). A sample of binaries with a wider range of abundances will be required to discern the importance of metallicity in extreme populations (e.g., in the Galactic halo or thick disk). 
\end{abstract}

\NewPageAfterKeywords

\section{Introduction}\label{sec:intro}
Over the past decade, M dwarfs have become critical for a wide range of astrophysics. On small scales, M dwarfs are attractive targets for the identification and characterization of exoplanets. The small size, low mass, and low luminosity of late-type stars facilitate the discovery of small planets \citep[e.g.][]{Muirhead2012,Martinez:2017aa,Mann:2018} in their circumstellar habitable zone \citep[e.g.,][]{Tarter2007,Shields:2016aa,Dittmann2017b}. Close-in, rocky planets are also significantly more common around M dwarfs than their Sun-like counterparts \citep{Dressing2013,2013PNAS..11019273P,Mulders2015,Gaidos2016b}

On larger scales, the properties of both the Milky Way and more distant galaxies are inexorably linked to parameters of their most numerous constituents \citep[$>70$\% of stars in the solar neighborhood are M dwarfs;][]{Henry:1994fk,Reid:2004lr}. Late-type dwarfs weigh heavily on the Galactic mass function \citep[e.g.,][]{Covey:2008lr} and are useful probes of the Milky Way's structure \citep[e.g.,][]{2008ApJ...673..864J,2017ApJ...843..141F}, kinematics \citep[e.g.,][]{2007AJ....134.2418B,2015RAA....15..860Y}, and chemical evolution \citep{Woolf:2012lr,2015AJ....149..140H}. Although K and M dwarfs are much fainter than their higher-mass counterparts, they measurably contribute to the integrated spectra of massive galaxies; thus, M dwarf fundamental properties have become an essential component to studies of the initial mass function \citep[e.g.,][]{2012ApJ...747...69C,2016ApJ...821...39M} and mass-to-light ratio \citep{2015MNRAS.452L..21S} of nearby galaxies. Additionally, M dwarf-white dwarf pairs are a plausible progenitor for Type Ia supernovae \citep{2012ApJ...758..123W}, and hence late-type stars may be important for cosmology.

For all these areas, it is essential that we have a method to estimate the fundamental parameters of late-type dwarfs. In exoplanet research, this means stellar radii for planet radii in transit surveys, stellar masses for planet masses in radial velocity surveys, and both (stellar densities) for determining planet occurrence rates \citep[e.g.,][]{2010exop.book...55W,Gaidos2013}, internal structure \citep[e.g.,][]{Rogers:2011lr}, and habitability \citep[e.g.,][]{Gaidos2013a,Kane2017}. Spectra, photometry, and distances of stars provide a relatively direct means to measure \teff\ \citep[e.g.,][]{Rojas-Ayala:2012uq,Mann2013c}, luminosity \citep[e.g.,][]{2002AJ....124.2721R}, metallicity \citep[e.g.,][]{Bonfils:2005,RojasAyala:2010}, and radius \citep[e.g., via Stefan-Boltzmann, ][]{Newton2015A,Kesseli2018b}. Masses are much more difficult to infer from observations alone, yet they are one of the most important and fundamental properties of a star.  

In the case of a binary, it is possible to directly determine the mass of a star from its orbital parameters. For systems with reasonably short orbital periods, the motions of binary components can be monitored to determine their orbits. Radial velocity variation can yield individual stellar masses but only modulo the sine of the orbital inclination \citep[e.g.,][]{Torres2002,Kraus2011,Stevens:2018aa}. In systems where binary components are spatially resolved, monitoring of their position angle and separation can yield a measurement of the total system mass, assuming that the parallax is known \citep[e.g.,][]{Soderhjelm1999,Woi2003,2009ApJ...699..168D}. Absolute orbital astrometry (measured with respect to background stars) can yield both individual masses and a direct measurement of the system's parallax \citep[e.g.,][]{Koh2012,Benedict2016}. 

Microlensing can provide mass measurements for single stars \citep[e.g.,][]{2016ApJ...825...60Z,2017ApJ...838..154C,2017AJ....154..176S}. Unfortunately, this method cannot be used to target specific M dwarfs of interest, and detected microlensing events are both rare and primarily limited to distant ($\sim$Kpc) targets in crowded fields, where follow-up is difficult. 

Stellar evolution models can provide mass estimates of targeted single stars \citep[e.g.,][]{Muirhead2012a}. However, differences between empirical and model-predicted mass-radius and luminosity-radius relations for late-type stars \citep[e.g.,][]{Boyajian2012,Feiden2012a} raise concerns about the reliability of model-based masses. Further, the masses derived depend on both the model grid used \citep{Spada2013,MIST1}, and the observed parameter over which the interpolation is done \citep[e.g., color vs. luminosity,][]{Mann:2012,Mann2015b}. Ultimately, these models need to be tested empirically; differences between the models and empirical determinations can reveal important missing physics or erroneous assumptions in the model assumptions. 

An empirical approach to estimating single-star masses is accomplished through a relation between mass and luminosity \citep[e.g.,][]{Hen1993, Delfosse2000}, calibrated with dynamical mass measurements from binary stars. Absolute magnitude can be used as a proxy for luminosity and is generally easy to measure for visual binaries from the same data used to establish the orbit (resolved images/astrometry and a parallax). Deriving such relations for Sun-like stars is difficult, as the scatter is dominated by evolution \citep[e.g.,][]{1991A&ARv...3...91A,2010A&ARv..18...67T} leading to the need for a mass-luminosity-age relation. Because main-sequence late-type stars evolve negligibly over the age of the Universe, age becomes a negligible factor and the stellar locus in mass-luminosity space is tight for a fixed metallicity. Adopting near-infrared (NIR) instead of optical magnitudes as a proxy for luminosity mitigates the effect of metallicity, as abundance variations have a weaker effect on the absolute flux levels of M dwarfs past 1.2\um\ when compared to optical regions \citep{Delfosse2000,Bonfils:2005}. Combined with the favorable Strehl ratios in adaptive optics imaging at $K$-band, this has made the $M_K-M_*$ relation the most precise and commonly used technique for estimated masses of late K and M dwarfs. 

Empirical $M_K-M_*$ relations from \citet{Hen1993} and \citet{Delfosse2000} provided mass determinations to $\simeq$10\% precision, with more recent improvements by \citet{Benedict2016}. However, as fields that rely on M dwarf parameters have pushed to higher precision, there has been an increasing need for proportionate improvements in stellar mass precision. Until recently, the lack of precise distances to M dwarfs were the dominant source of error when estimating masses using the \mmk\ relation. With the arrival of {\it Gaia} parallaxes, many late-type dwarfs beyond the solar neighborhood have $<1\%$ parallaxes; the lower precision of existing \mmk\ relations is now the dominant source of uncertainty when estimating masses this way. Existing relations also have gaps in their calibration sample, particularly below $0.1M_\odot$, where there is need for stellar masses to match new exoplanet surveys \citep[e.g.,][]{Gillon2017}. Methods to measure metallicities of M dwarfs have become increasingly precise \citep[e.g.,][]{RojasAyala:2010, 2014A&A...568A.121N}, making it possible to explore the impact of metallicity on the \mmk\ relation. Most importantly, both existing models and empirical measurements of inactive M dwarfs have found tight ($<5$\% intrinsic scatter) relations for mass-radius \citep[e.g.,][]{Bayless2006,Spada2013,2017AJ....154..100H} and luminosity-radius \citep[e.g.,][]{Boyajian2012,2015ApJ...802L..10T,Mann2015b}, suggesting that similar improvements in the \mmk\ relation are achievable.

Here we present a revised empirical relation between $M_*$, $M_{K_S}$, and [Fe/H], spanning almost an order of magnitude in mass, from 0.075$M_\odot$ to 0.70$M_\odot$ covering $-0.6<$[Fe/H]$<+0.4$. The relation is built on orbital fits to visual binaries from a combination of AO imaging and astrometric measurements in the literature with metallicities estimated from the stars' near-infrared spectra. In Section~\ref{sec:targets} we detail our selection of nearby late-type binaries with orbits amenable to mass determinations. We overview our astrometric and spectroscopic observations in Section~\ref{sec:obs}, including those from telescope archives. We explain our procedure for computing separations and position angles, and incorporating similar measurements from the literature in Section~\ref{sec:astrometry}. Our orbit-fitting procedure is explained in Section~\ref{sec:orbit}. We describe our method for determining other parameters of each system ([Fe/H], distance, and \mks) in Section~\ref{sec:params}. Our technique to fit the \mmk\ relation from these binaries is described in Section~\ref{sec:relation}, including an analysis of the errors as a function of \mks, tests of our relation on binaries with individual masses, a detailed look at the effects of [Fe/H], and a comparison to earlier similar mass-luminosity relations. We conclude in Section~\ref{sec:discussion} with a brief summary and a discussion of the important caveats and complications to consider when using our relation, as well as future directions we are taking to expand on the current work.

\textit{If you want to use the relations in this manuscript, we advise at least reading Section~\ref{sec:caveats} to understand the potential limitations of the provided program and posteriors.\footnote{\href{https://github.com/awmann/M_-M_K-}{https://github.com/awmann/M\_-M\_K-}}}

\section{Sample Selection}\label{sec:targets}
\begin{figure}[htb]
\begin{center}
\includegraphics[width=0.47\textwidth]{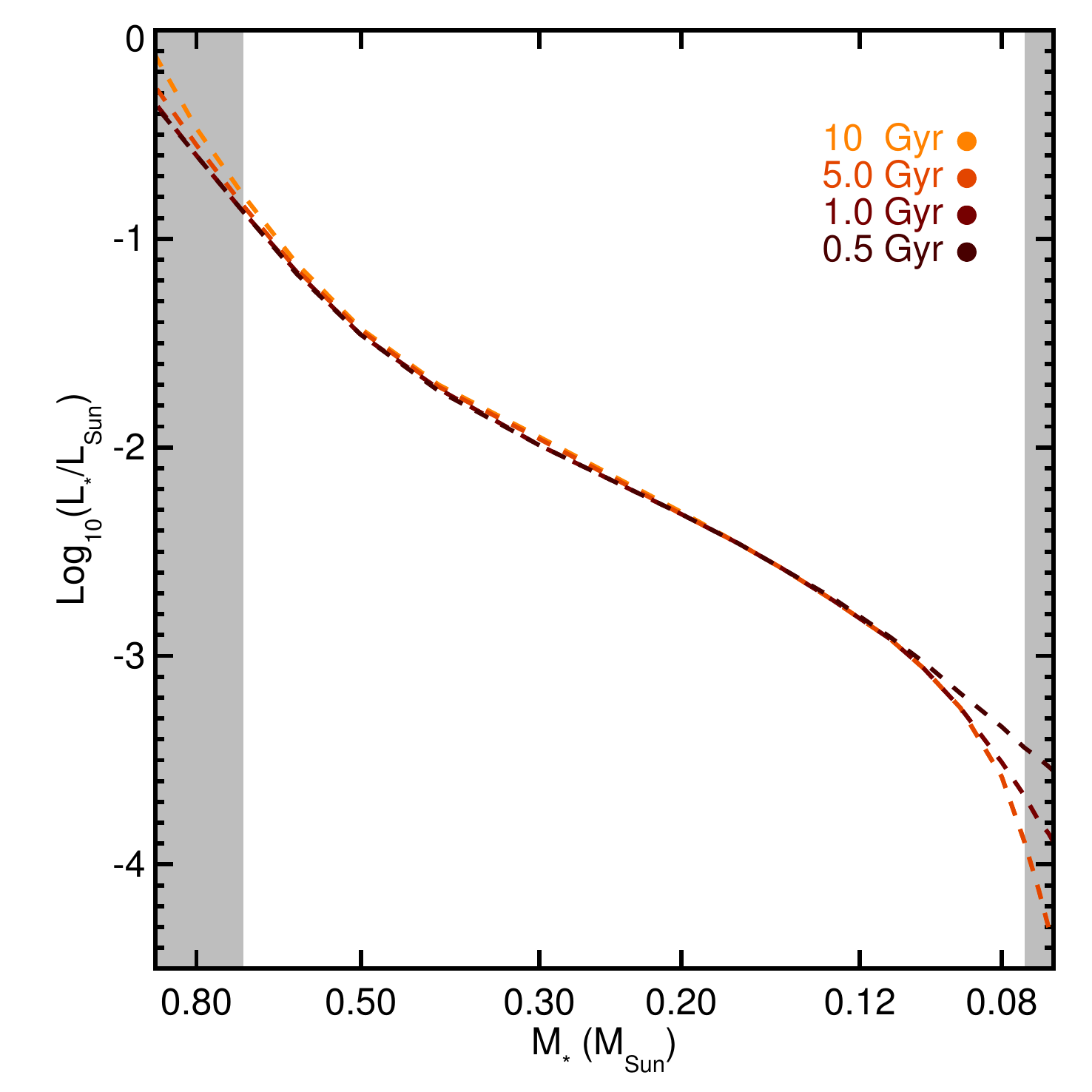}
\caption{Stellar luminosity as a function of mass predicted by the \citet{BHAC15} models, color-coded by age (metallicity fixed at Solar). The grey regions denote masses excluded by this study due to a significant age dependence on the mass-luminosity relation. We have a lower cut on the low-mass end then implied by the tracks, as some spread at this low-mass end is due to a longer pre-main-sequence lifetime and empirical studies suggest a lower-mass boundary between stars and brown dwarfs. }
\label{fig:age}
\end{center}
\end{figure}

Our selection of binaries was designed to sample the region of mass space over which the mass-luminosity relation should not evolve significantly between the zero-age main sequence and the age of the Galactic disk ($\sim$10~Gyr). We quantified this using the \citet{BHAC15} models (Figure~\ref{fig:age}). Above $0.70M_\odot$, a fixed luminosity (the observable) could correspond to a $\simeq$5\% range in masses over 1-10\, Gyr. Stars below $\simeq0.1M_\odot$ take a long time (100-1000\,Myr) to arrive on the main sequence, but obey a tight relation beyond this point. Those objects below $\simeq0.08M_\odot$ are predicted to never reach the main sequence and hence obey no mass-luminosity relation. However, this transition likely depends on metallicity, and empirical studies have found a limit closer to 0.075$M_\odot$ \citep[e.g.,][]{Dieterich2014,Dupuy2017}. Therefore, we attempted to select systems spanning $0.075M_\odot \lesssim M_* \lesssim 0.70M_\odot$.

We first selected systems by cross-matching catalogs of nearby M dwarfs \citep{Lepine2013, Gaidos2014, 2014ApJ...784..156D,Winters2015}, with the fourth catalog of interferometric measurements of binary stars \citep[INT4,][]{Hartkopf:2001}, and adaptive optics (AO) images from the Keck Observatory Archive (KOA). As part of this cross-match, we also included targets matching the M dwarf selection criteria of \citet{Gaidos2014}, but with a bluer color cut ($V-J>1.8$) to incorporate additional late-K dwarfs. We kept any binaries with separations less than 5\arcsec. We then added in other known late-type binaries from \citet{2008MNRAS.384..150L}, \citet{Jnn2012}, \citet{Jnn2014}, and \citet{Ward-Duong2015}. This provided a list of more than 300 multi-star systems.

From here we selected binaries amenable to orbital characterization on a reasonable (few year) timescale. To this end, we assumed that the average of available (literature) separation measurements approximates the semi-major axis of the system. Next, we identified systems for which the time between the first available observation and our final observation would span at least 30\% of the orbit (based on our rough semi-major axis estimate), including the two years of our orbital monitoring program with Keck. This cut accounted for existing data. As a result, long-period binaries with extensive previous observations were included, depending on the baseline available, while those with only recent epochs would generally need to have orbits of $\ll$10 yr to be targeted. These cuts left us with 129 systems. We then removed 36 systems at $\delta<-30$\degree\ that were difficult to observe from Maunakea, leaving us with 93 systems to be included in our observing program. Three systems south of this limit (Gl 54, Gl 667, and GJ 1038) were included int our final sample, as they had enough astrometry without our additional monitoring at Keck.

We removed 16 systems from our analysis because of an unresolved tertiary (or quaternary) component noted in the literature \citep[e.g.,][]{2002A&A...382..118T,2010ApJ...720.1727L,2018ApJS..235....6T}. In their current form, such systems were not useful for our analysis, as we had no $\Delta K$ magnitudes or mass ratios for the unresolved components. Since many of these are double- or triple-lined systems, it is possible to recover their parameters with multiepoch radial velocities, and some systems have the necessary data in the literature \citep[e.g.,][]{Sgr2000}. We continued to monitor these systems with high-resolution NIR spectrographs \citep{2010SPIE.7735E..1MY, 2012SPIE.8446E..2CR, Park2014}, but they were excluded from the analysis done here. High-order systems where all components are resolved (e.g., GJ 2005 ABC) were retained, although we only focus on the tighter pairs in this work. 

A total of 17 systems were flagged as young, i.e., affiliated with nearby young moving groups or clusters \citep{Shkolnik2012, Kraus2014, Gagne2014, Malo2014a, Gagne2015, 2017AJ....153...95R, 2017AJ....154...69S, Rizzuto2017, 2018MNRAS.475.2955L}, or those that are known to be pre-main-sequence \citep[e.g., LP 349-25][]{2009ApJ...705.1416R}. We monitored these targets even after flagging them as young, but they were not included in the analysis for the current work. Many of these either are pre-main-sequence stars, and hence will not follow the same mass-luminosity relation, or are atypically active compared to other stars in the solar neighborhood \citep[e.g.,][]{Malo2014a}. Because these cuts generally only remove extremely young stars, the sample may include some young field stars. 

After the completion of our observing program, we removed targets with fewer than six independent astrometric measurements and those lacking a precise parallax ($\sigma_{\pi}>7\%$). We attempted to fit orbits of the remaining 57 systems (Section~\ref{sec:orbit}). Two of the resulting orbital parameters yielded system/total masses ($M_{tot}$) for the system too imprecise ($>20\%$) to be useful for our analysis. This left us with 55 binaries (110 stars).

Our method uses $M_{tot}$ (as opposed to individual masses) to derive the \mmk\ relation (explained in Section~\ref{sec:methods}). As a result, constraints on the mass ratio through radial velocities or absolute astrometry are not required to be included in the final sample (just separations and position angles). Since most systems do not have the data required for individual masses, this decision is important to keep the sample size large and the analysis homogeneous. 

We added seven targets with orbits from \citet{Dupuy2017} to fill in the sample around the end of the M dwarf sequence. These seven were selected because they are theoretically massive enough to sustain hydrogen fusion and satisfy all our other selection criteria. Systems from \citet{Dupuy2017} also had their orbits fit using a nearly identical method to our own, often using similar or identical sources of data and analysis methods (primarily Keck/NIRC2). Two additional systems in \citet{Dupuy2017} matched our initial cut, but were still omitted from this analysis. These were LP415-20, which \citet{Dupuy2017} suggest is an anomalous system and possibly an unresolved triple, and 2M1847+55, which has a relatively imprecise orbit compared to the rest of the sample.

Parameters of the final 62 systems included in our analysis are given in Table~\ref{tab:sample}.

\begin{deluxetable*}{l l l l l l l l l l l l l l l l l l l l l l l }
\tablewidth{\linewidth}
\tablecaption{Binary Sample \label{tab:sample}}
\tablehead{
    \colhead{Name} & \colhead{Comp} & \colhead{R.A.} & \colhead{Decl.} & \colhead{System $K_S$} 
 & \colhead{$\Delta K_S$} & \colhead{$M_{\rm tot}$} & \colhead{[Fe/H]$^a$} & \colhead{Plx}  & \colhead{Plx} \\
    \colhead{} & \colhead{} & \colhead{J2000} & \colhead{J2000} & \colhead{(mag)}
 & \colhead{(mag)} & \colhead{($M_\odot$)} & \colhead{(dex)} & \colhead{(mas)} & \colhead{Ref} 
}
\startdata
\multicolumn{10}{c}{Systems analyzed in this paper}  \\
\hline
GJ 1005 & AB & 00:15:28.0 & $-$16:08:01 & \phantom{0} 6.390$\pm$0.016 &  1.145$\pm$0.016 &   0.3188\phantom{0}$\pm$\phantom{0}0.0023 & $-$0.41 & 166.60\phantom{0}$\pm$0.30 & 3\\
GJ 2005 & BC & 00:24:44.1 & $-$27:08:24 & \phantom{0} 9.371$\pm$0.050$^e$ &  0.320$\pm$0.016 &   0.1567\phantom{0}$\pm$\phantom{0}0.0055 & $-$0.08 &  128.5\phantom{00}$\pm$1.5 & 3\\
Gl 22 & AC & 00:32:29.2 & $+$67:14:08 & \phantom{0} 6.037$\pm$0.023 &  2.060$\pm$0.035 &   0.572\phantom{00}$\pm$\phantom{00}0.011 & $-$0.24 & \phantom{0}99.20\phantom{0}$\pm$0.60 & 3\\
Gl 54 & AB & 01:10:22.8 & $-$67:26:42 & \phantom{0} 5.132$\pm$0.024 &  0.697$\pm$0.036 &   0.7507\phantom{0}$\pm$\phantom{0}0.0100 & $+$0.17 & 126.90\phantom{0}$\pm$0.40 & 3\\
GJ 1038 & AB & 01:25:01.8 & $-$32:51:04 & \phantom{0} 6.207$\pm$0.021 &  0.058$\pm$0.016 &   1.23\phantom{000}$\pm$\phantom{000}0.16 & $+$0.03 & \phantom{0}39.8\phantom{00}$\pm$1.6 & 2\\
Gl 65 & AB & 01:39:01.2 & $-$17:57:02 & \phantom{0} 5.343$\pm$0.021 &  0.161$\pm$0.019 &   0.2374\phantom{0}$\pm$\phantom{0}0.0053 & $+$0.04 &  373.7\phantom{00}$\pm$2.7 & 5\\
Gl 84 & AB & 02:05:04.8 & $-$17:36:52 & \phantom{0} 5.662$\pm$0.020 &  3.262$\pm$0.016 &   0.523\phantom{00}$\pm$\phantom{00}0.028 & $-$0.14 &  109.4\phantom{00}$\pm$1.9 & 2\\
2M0213+36 & AB & 02:13:20.6 & $+$36:48:50 & \phantom{0} 8.518$\pm$0.018 &  1.493$\pm$0.018 &   0.246\phantom{00}$\pm$\phantom{00}0.035 & $-$0.07 & \phantom{0}74.6\phantom{00}$\pm$3.5 & 6\\
\hline
\multicolumn{10}{c}{Systems from \citet{Dupuy2017} }  \\
\hline
LHS1901 & AB & 07:11:11.4 & $+$43:29:58 & \phantom{0} 9.126$\pm$0.018 &  0.094$\pm$0.010 &   0.2029\phantom{0}$\pm$\phantom{0}0.0090 & $-$0.41 & \phantom{0}76.4\phantom{00}$\pm$1.1 & 4\\
2M0746+20 & AB & 07:46:42.5 & $+$20:00:32 & \phantom{0}10.468$\pm$0.022 &  0.357$\pm$0.025 &   0.1535\phantom{0}$\pm$\phantom{0}0.0017 & $-$0.18$^d$ & \phantom{0}81.24\phantom{0}$\pm$0.25 & 4\\
2M1017+13 & AB & 10:17:07.5 & $+$13:08:39 & \phantom{0}12.710$\pm$0.023 &  0.113$\pm$0.024 &   0.149\phantom{00}$\pm$\phantom{00}0.016 & $-$0.35$^d$ & \phantom{0}32.2\phantom{00}$\pm$1.2 & 4\\
\enddata
\tablecomments{Table \ref{tab:sample} is available in its entirety in the ancillary files with the arXiv submission. A portion is shown here for guidance regarding its form and content.\\
Parallax references: 1 = This work (MEarth), 2 = \citet{van-Leeuwen:2007yq}, 3 = \citet{Benedict2016}, 4 = \citet{Dupuy2017}, 5 = \citet{van-Altena1995}, 6 = \citet{2016AJ....151..160F}, 7 = \citet{2006ApJS..166..341G}, 8 = \citet{Soderhjelm1999}, 9 = \citet{2017AJ....154..151B}, 10 = \citet{Riedel2010}, 11 = \citet{gaiadr1}, 12 = companion to star in \citet{GaiaDr2}.}
\tablenotetext{$a$}{Errors on [Fe/H] are limited primarily by the calibration \citep{Mann2013a,Mann2014}, and are 0.08~dex for all targets unless otherwise noted. }
\tablenotetext{$b$}{Synthetic $K_S$ magnitudes derived from spectra. All other $K_S$ magnitudes are from 2MASS. }
\tablenotetext{$c$}{Abundance derived from lower-resolution IRTF spectrum, $\sigma$[Fe/H] estimated to be 0.12~dex. }
\tablenotetext{$d$}{L dwarfs are beyond the calibration range of \citet{Mann2014} ; [Fe/H] values should be used with caution. }
\tablenotetext{$e$}{$K_S$ magnitude from 2MASS contains a third star, listed $K_S$ magnitude has third star's flux removed. }
\end{deluxetable*}

\section{Observations and Data Reduction}\label{sec:obs} 

\subsection{Near-infrared Spectra with IRTF/SpeX}

To estimate the metallicities of our targets we obtained near-infrared spectra for 58 of 62 targets using the SpeX spectrograph \citep{Rayner:2003} on the NASA Infrared Telescope Facility (IRTF) atop Maunakea. Observations were taken between May 2011 and November 2017. Most data were taken as part of programs to characterize the fundamental properties of nearby M dwarfs \citep[e.g.,][]{Mann2013c,Gaidos2014,Terrien2015}. All spectra were taken in SXD mode, providing simultaneous wavelength coverage from 0.9 to 2.5$\mu$m. For 56 of the targets, observations were taken using the 0.3$\times15\arcsec$ slit, which yielded a resolution of $R\simeq2000$. Spectra for two targets (2M2206-20 and 2M2140+16) were taken from \citet{2009ApJ...706..328D} and \citet{Dupuy2012}, which used the 0.5$\times15\arcsec$ and 0.8$\times15\arcsec$ slits (respectively), yielding spectral resolutions of $\simeq$750-1500. 

For Gl 65 and HD 239960 the SpeX slit was aligned to get spectra of both targets simultaneously. For all other targets the binary was unresolved or too poorly resolved to separate in the reduction procedure, and instead the slit was aligned with the parallactic angle to compensate for differential refraction. Each target was nodded between two positions along the slit to remove sky background. Depending on the target brightness and conditions, between 6 and 30 individual exposures were taken following this nodding pattern, with exposure times varying from 8s to 180s. An A0V-type star was observed immediately before or after each target to measure (and remove) telluric lines and flux calibrate the spectrum. The final stacked spectra had S/N of $>100$ per resolving element in the $K$-band for all but the four faintest targets (which had S/N$>50$). 

Basic data reduction was performed with {\tt SpeXTool} package \citep{Cushing2004}. This included flat fielding, sky subtraction, extraction of the one-dimensional spectrum, wavelength calibration, stacking of individual exposures, and merging of individual orders. Telluric lines were removed and the spectrum was flux calibrated using the A0V star observations and the {\tt xtellcor} software package \citep{Vacca2003}. When possible, the same A0V star was used for multiple targets taken near each other in time. 

Three of the four targets lacking SpeX spectra are too warm (earlier than K5) to derive a metallicity from NIR spectra (Gl 792.1, Gl 765.2, and Gl 667), and the third (Gl 54) is too far south to be observed with IRTF. 

\subsection{Adaptive Optics Imaging and Masking}\label{sec:ao}

We analyzed a mix of AO data from our own program with Keck/NIRC2 and archival imaging from the Keck II Telescope, the Canada France Hawaii Telescope (CFHT), the Very Large Telescope (VLT), and the Gemini North Telescope. In general we analyzed all usable images (e.g., non saturated, components resolved) regardless of observing mode and filter. 

For our analysis, we considered a single dataset a collection of observations with a unique combination of filter, target, and epoch. Each combined dataset consisted of a $\Delta m$ (for a given filter), separation, and position angle. 

We separate the observations and reduction by instrument/telescope below. The full list of astrometry and contrast measurements is given in Table~\ref{tab:astrom}, sorted by target and date. 

\begin{deluxetable*}{c l l l l l l}
\tablecaption{Input Astrometry and Photometry \label{tab:astrom} }
\tablehead{
   \colhead{UT Date\tablenotemark{a}} & \colhead{$\rho$} & \colhead{$\theta$} & \colhead{Filter}  & \colhead{$\Delta m$\tablenotemark{b}} & \colhead{Source\tablenotemark{c}} & \colhead{PI\tablenotemark{d}}  \\
    \colhead{(YYYY-MM-DD)} & \colhead{(mas)} & \colhead{(deg)} & \colhead{} & \colhead{(mag)} & \colhead{}  & \colhead{}
}
\startdata
\multicolumn{7}{c}{2M0213+36}  \\
\multicolumn{7}{c}{R.A., Dec = 02:13:20.6, $+$36:48:50}  \\
\hline
2006-11-11 & $215.0\pm5.4$ & $72.0\pm4.0$ & \nodata & \nodata & \citet{Jnn2012} & \\
2007-08-12 & $181.0\pm2.8$ & $56.5\pm2.3$ & \nodata & \nodata & \citet{Jnn2012} & \\
2012-08-29 & $226.0\pm8.2$ & $81.70\pm0.88$ & \nodata & \nodata & \citet{Jnn2014} & \\
2012-11-25 & $217.0\pm4.5$ & $76.10\pm0.82$ & \nodata & \nodata & \citet{Jnn2014} & \\
2014-07-31 & $128.58\pm0.19$ & $47.307\pm0.047$ & K' & $1.481\pm0.014$ & Keck/NIRC2 & Dupuy\\
2015-07-21 & $70.42\pm0.62$ & $331.06\pm0.49$ & K$_{\rm cont}$ & $1.391\pm0.034$ & Keck/NIRC2 & Kraus\\
2015-07-22 & $68.42\pm0.49$ & $332.00\pm0.35$ & K' & $1.379\pm0.066$ & Keck/NIRC2 (NRM) & Kraus\\
2016-11-15 & $59.7\pm3.3$ & $216.4\pm1.1$ & K' & $1.285\pm0.064$ & Keck/NIRC2 & Mann\\
2016-11-15 & $54.17\pm0.39$ & $216.24\pm0.48$ & K' & $1.459\pm0.060$ & Keck/NIRC2 (NRM) & Mann\\
\hline
\multicolumn{7}{c}{Gl 125}  \\
\multicolumn{7}{c}{R.A., Dec = 03:09:30.8, $+$45:43:58}  \\
\hline
1998-10-11 & $581.0\pm3.7$ & $10.4\pm1.2$ & \nodata & \nodata & \citet{Bag2002} & \\
1999-10-27 & $509.4\pm3.2$ & $8.6\pm1.1$ & \nodata & \nodata & \citet{Bag2004} & \\
2000-11-17 & $418.0\pm3.7$ & $6.8\pm1.1$ & \nodata & \nodata & \citet{Bag2006b} & \\
2001-10-04 & $330.0\pm4.3$ & $3.7\pm1.1$ & \nodata & \nodata & \citet{Bag2006b} & \\
2002-09-26 & $222.0\pm3.1$ & $357.2\pm1.1$ & \nodata & \nodata & \citet{Bag2005} & \\
2003-10-16 & $101.0\pm3.1$ & $334.0\pm1.1$ & \nodata & \nodata & \citet{Bag2005} & \\
2003-12-05 & $87.0\pm3.1$ & $327.7\pm1.1$ & \nodata & \nodata & \citet{Bag2005} & \\
2004-10-27 & $83.0\pm3.7$ & $240.5\pm1.1$ & \nodata & \nodata & \citet{Bag2007b} & \\
2005-07-15 & $161.55\pm0.13$ & $214.110\pm0.035$ & K' & $1.166\pm0.062$ & Keck/NIRC2 & Liu\\
2010-09-18 & $340.7\pm4.3$ & $188.20\pm0.87$ & \nodata & \nodata & \citet{Hor2017} & \\
2011-09-10 & $244.6\pm4.3$ & $181.80\pm0.87$ & \nodata & \nodata & \citet{Hor2017} & \\
2015-10-01 & $290.49\pm0.14$ & $25.981\pm0.023$ & K$_{\rm cont}$ & $1.183\pm0.011$ & Keck/NIRC2 & Mann\\
2015-11-18 & $305.58\pm0.14$ & $25.277\pm0.023$ & K$_{\rm cont}$ & $1.187\pm0.010$ & Keck/NIRC2 & Gaidos\\
2016-08-02 & $381.05\pm0.11$ & $22.295\pm0.013$ & K$_{\rm cont}$ & $1.169\pm0.010$ & Keck/NIRC2 & Gaidos\\
2016-09-20 & $394.80\pm0.17$ & $21.856\pm0.014$ & K$_{\rm cont}$ & $1.183\pm0.010$ & Keck/NIRC2 & Mann\\
\hline
\enddata
\tablecomments{Table \ref{tab:astrom} is available in its entirety in the ancillary files with the arXiv submission. A portion is shown here for guidance regarding its form and content.} 
\tablenotetext{a}{Dates from literature points may be off by 1 day owing to inconsistency in reporting UT versus local date. }
\tablenotetext{b}{Errors on $\Delta m$ values are based on the scatter in individual images and are likely underestimated.}
\tablenotetext{c}{Astrometry with source as Keck/NIRC2, CFHT/KIR, VLT/NaCo, or Gemini/NIRI are from this paper. All other measurements list the paper reference.}
\tablenotetext{d}{Principal investigator for AO data analyzed in this paper (from our program or the archive) as it was listed in the image header. }
\end{deluxetable*}

\subsubsection{Keck II/NIRC2 Imaging and Masking}
As part of a long-term monitoring program with Keck II atop Maunakea, between June 2015 and July 2018 we observed 51 of the 55 multiple-star systems analyzed here. All observations were taken using the facility AO imager NIRC2 in the vertical angle mode (fixed angle relative to elevation axis) and the narrow camera ($\approx$10\,mas\,pixel$^{-1}$). Depending on the target brightness and observing conditions, images were usually obtained through either the $K'$ ($\lambda_c=2.124$\um) or narrow $K_{\rm{cont}}$ ($\lambda_c=2.271$\um) filters, and nonredundant aperture masking (NRM) was always taken using the 9-hole mask and $K'$ filter. After acquiring the target and allowing the AO loops to close, we took four to 10 images or 6-8 interferograms (for NRM), adjusting coadds and integration time based on the brightness of the target. As most of our targets are bright, observations were usually taken using the Natural Guide Star (NGS) system \citep{2000PASP..112..315W,2004ApOpt..43.5458V}, only utilizing the Laser Guide Star (LGS) mode for the faintest ($R\gtrsim13$) targets or in poor conditions. In total, our observations provided 155 datasets.

In addition to our own data, we downloaded images from the Keck Observatory Archive (KOA), spanning March 2002 to November 2015, all of which were taken with the NIRC2 imager. Archival data comprised a wide range of observing modes, filters, and cameras, although the majority were taken with the narrow camera using either the $H$- or $K$-band filters. We included nearly all data with clear detections of both binary components independent of the observing setup. We discarded saturated images, those taken with the coronagraph for either of the component stars, and images where the target is completely unresolved. A total of 36 datasets were used from the archive. 

The same data reduction was applied to observations both from our own program and from the archive, following our custom procedure described in \citet{Kraus2016a}. To briefly summarize, we corrected for pixel value nonlinearity in each frame then dark- and flat-corrected it using calibrations taken the same night. In cases where no appropriate darks or flats were taken in the same night, we used a set from the nearest available night. We interpolated over ``dead'' and ``hot'' pixels, which were identified from superflats and superdarks built from data spanning 2006 to 2016. Because flats are rarely taken in narrowband filters, we used superflats built from the nearest (in wavelength) broadband filter where appropriate (e.g., for $K_{\rm{cont}}$ we used $K'$ flats). Pixels with flux levels $>10\sigma$ above the median of the eight adjacent pixels (primarily cosmic rays) were replaced with the median (average of the 4th and 5th ranked). Images were visually inspected as part of identifying the binary location, and a handful ($<1\%$) of images were negatively impacted by our cosmic ray removal (e.g., removal of part of the source). For these, we used the data prior to cosmic-ray rejections.

\subsubsection{CFHT/KIR Imaging}

We obtained data for 34 of our targets from the Canadian Astronomy Data Centre archive, all taken with the 3.6m Canada-France-Hawaii Telescope (CFHT) using the Adaptive Optics Bonnette \citep[AOB, often referred to as PUEO after the Hawaiian owl,][]{1994SPIE.2201..833A} and the KIR infrared camera \citep{1998SPIE.3354..760D}. After removing images where the target was saturated, unresolved, or had poor AO correction, a total of 239 datasets were included. Observations spanned December 1997 to January 2007, covering most of the time PUEO was in use at CFHT (1996 to 2011). Images were taken using a range of filters across $JHK$ bands, but the majority used either the narrowband Br$\gamma$ or [FeII] filters. All data were taken using a 3-5 point dither pattern and included at least two images at each dither location. 

Data reduction for KIR observations followed the same basic steps as our NIRC2 data. We first applied flat-fielding and dark correction using a set of superflats and superdarks built by splitting the datasets into 6 month blocks and combining calibration data within the same time period. As with the NIRC2 data, we identified bad pixels by comparing each image to a set of superflats built from calibration data spanning all downloaded data. To identify cosmic rays, we first stacked consecutive images of each target (at a fixed location, so not including dithers), recording the robust mean and standard deviation at each pixel. Pixels $>10\sigma$ above this robust mean were replaced with the median of the eight surrounding points. Since KIR data were taken in sets of $>5$ images before the object was dithered, this median-filtering was effective for removing nearly all cosmic rays.

\subsubsection{VLT/NaCo Imaging}

We downloaded AO-corrected images from the ESO archive taken with the Nasmyth Adaptive Optics System Near-Infrared Imager and Spectrograph (NAOS-CONICA, or NaCo) instrument on VLT. Data spanned November 2002 to October 2016, with about half of the 72 datasets taken from 2001 to 2005. Based on the program abstracts, $\lesssim$\nicefrac{1}{2} of the observations were meant to use these binaries as astrometric or photometric calibration (e.g., science case is unrelated to M dwarfs or binaries). Data covered 21 of our targets, excluding saturated or otherwise unusable data. Observations were taken with a wide range of filters, cameras, and observing patterns, but the majority were taken using the S13 camera ($\approx$13\,mas\,pixel$^{-1}$) with either broadband $K_s$ and $L$, or narrowband [FeII] and Br$\gamma$ filters, and always followed a 2-4 point dither pattern. 

Basic data reduction was applied to NaCo images following a similar procedure with the KIR and NIRC2 data. We applied flat-fielding and dark corrections to each observation using the standard set of calibrations taken each night as part of the VLT queue. In the case where calibration (dark or flat) images were missing or unusable, we used the nearest (in time) set of calibration images matching the filter (for flats) and exposure setup (for darks). Flats taken in broadband filters were used for flat-fielding narrow-band images at similar wavelengths. We built bad pixel masks using median stacks of all images taken within a night after applying flat and dark corrections. To identify and remove cosmic rays we used the L.A. Cosmic software \citep{2001PASP..113.1420V}.

\subsubsection{Gemini/NIRI Imaging}
We retrieved 36 datasets for 8 of our targets from the Gemini archive, all taken with the AO imager NIRI \citep{2003PASP..115.1388H} on the Frederick C. Gillett Gemini Telescope (Gemini North). All observations were taken between August 2004 and February 2011 with the assistance of the ALTtitude conjugate Adaptive optics for the InfraRed (ALTAIR). Most observations were taken with the f32 camera ($\approx$21\,mas\,pixel$^{-1}$) using broadband $J$-, $H$-, or $K$-band filters. All observations followed a 2-4 point dither pattern and took at least two images at each dither location. 

Data from NIRI were reduced using the same basic methods as for all other adaptive optics data. First, we applied flat and dark corrections to each set of images using the standard calibration images taken as part of the Gemini queue, usually within 24h of the target observations. In most cases, flats taken in broadband filters were used for narrowband flat-fielding. We then identified bad pixels from median filtering of all images within a given night. Observations with a target near or on top of a heavily impacted pixel (identified with the mask) were discarded. We used the L.A. Cosmic software for the identification and removal of cosmic rays. 

\section{Astrometry and photometry}\label{sec:astrometry}

Extracting separations and position angle measurements followed a similar multi-step procedure across all instruments, excluding the NRM data (which is described below). Our method is based largely on that described in \citet{2016ApJ...817...80D} and \citet{Dupuy2017}, which is built on the techniques from \citet{2008ApJ...689..436L} and \citet{2010ApJ...721.1725D}.

We first cross-correlated each image with a model Gaussian PSF to identify the most significant peaks. The cross-correlation peak occasionally centers on instrumental artifacts, often struggling to separate partially overlapping binaries, and can easily identify the wrong source for triple systems. This step was checked by eye and updated as needed. The eye-check phase also allowed us to manually remove data of poor quality: e.g., no or poor AO correction, saturated data, or unresolved systems. We used these centers as the initial guess for the pixel position utilized in the next phase. 

We then fit the PSF centers by either: 1) running {\tt StarFinder}, a routine designed to measure astrometry and photometry from adaptive optics data by deriving a PSF template from the image and iteratively fitting this model to the components \citep[for more details, see][]{2000A&AS..147..335D}, or 2) fitting the binary image with a PSF modeled by three-component elliptical 2D Gaussians using the least-squares minimization routine {\tt MPFIT} \citep{Markwart2009}. Although the results of these two methods generally agreed, {\tt StarFinder} was preferred, as it used a non-parametric and more realistic model of the PSF and worked with mediocre AO correction, provided that the component PSFs were well separated. {\tt StarFinder}, however, failed on the tightest binaries, where it was unable to distinguish two stars and as a result, incorrectly built an extended PSF that fit the blended image. The Gaussian fit was used for these cases where {\tt StarFinder} failed. 

As part of the PSF fit, both methods provided a flux ratio of the PSF normalization factors, which we used to determine the contrast $\Delta m$ in the relevant band. Data from all filters are used for astrometry, although only measurements in the $K$-band ($K_s$, $K$, $K'$, $Br\gamma$, and $K_{\rm{cont}}$) were used in the estimate of \mks\ (see Section~\ref{sec:mags}).

PSF fitting provides pixel-position measurements of each component, but converting these to separation ($\rho$) and position angle (PA) on the sky requires an astrometric calibration of the instrument. For the NIRC2 narrow camera, we used the \citet{Yelda2010} distortion solution for data taken before 2015 Apr 13 UT, and the \citet{2016PASP..128i5004S} solution for data taken after this. These calibrations include a pixel scale and orientation determination of 9.952$\pm$0.002 mas\,pixel$^{-1}$ and $0.252\pm0.009^{\circ}$ for the former, and 9.971$\pm$0.004 mas\,pixel$^{-1}$ and $0.262\pm0.020^{\circ}$ for the latter. For the NIRC2 wide camera we used the solution from Fu et al. (2012, priv. comm.)\footnote{\href{http://homepage.physics.uiowa.edu/~haifu/idl/nirc2wide/}{http://homepage.physics.uiowa.edu/$\sim$haifu/idl/nirc2wide/}}, with a pixel scale of $39.686\pm0.008$ mas\,pixel$^{-1}$ and the same orientation as the narrow camera. For the f/32 camera on NIRI, we used the distortion solution from the Gemini webpage\footnote{\href{http://www.gemini.edu/sciops/instruments/niri/undistort.pro}{http://www.gemini.edu/sciops/instruments/niri/undistort.pro}}. 

For other instruments and cameras, data were always taken following a dither pattern to sample different regions of the CCD distortion pattern. So the RMS between dithered images should reflect errors due to uncorrected distortion (which is included in our errors; see below). For KIR (CFHT/PUEO), we adopted a pixel scale of 34.8$\pm$0.1 mas pixel$^{-1}$ \citep{2003ApJ...589..410S} and an orientation of $0\pm2$\degree \footnote{\href{http://www.cfht.hawaii.edu/Instruments/Detectors/IR/KIR/}{http://www.cfht.hawaii.edu/Instruments/Detectors/IR/KIR/}}. For NaCo, we assumed a pixel scale of 13.24$\pm0.05$ mas pixel$^{-1}$ for the S13 camera \citep{2003A&A...411..157M,2005A&A...435L..13N} and the values given in the ESO documentation for all others\footnote{\href{http://www.eso.org/sci/facilities/paranal/instruments/naco/doc.html}{http://www.eso.org/sci/facilities/paranal/instruments/naco/doc.html}} (with the same error). The rotation taken from the NaCo headers was assumed to be correct to 0.4$^{\circ}$ \citep{Sef2008}. For NIRI observations, we used a pixel scale provided in the Gemini documentation\footnote{\href{http://www.gemini.edu/sciops/instruments/niri/imaging/pixel-scales-fov-and-field-orientation}{http://www.gemini.edu/sciops/instruments/niri/imaging/pixel-scales-fov-and-field-orientation}} for each camera (117.1, 49.9, and 21.9 mas pixel$^{-1}$ for f/6, f/14, and f/32, respectively), with a global uncertainty of 0.05 mas pixel$^{-1}$ on the pixel scale and 0.1$^{\circ}$ on the orientation \citep{2004ApJ...614..235B}. 

Calculation of separations and position angles from non-redundant masking (NRM) observations followed the procedures in the appendix of \citet{Kraus2008} with the aid of the latest version of the ``Sydney'' aperture-masking interferometry code\footnote{\href{https://github.com/mikeireland/idlnrm}{https://github.com/mikeireland/idlnrm}}. To remove systematics, each NRM observation of a science star was paired with that of a single calibrator star taken in the same night with a similar magnitude and airmass. Binary system profiles were then fit to the closure phases to produce estimates of the separation, position angle, and contrast of the binary components. More details on the analysis of masking data can be found in \citet{2006ApJ...650L.131L}, \citet{Kraus2008}, and \citet{2012ApJ...744..120E}.

All data in a single set (same target, filter, and night) were combined into a single measurement (after applying all corrections above), with errors estimated using the RMS in the individual images within a night. This scatter across images was combined with the uncertainty in the orientation and pixel scale in quadrature. We assumed that the pixel scale and orientation uncertainties were completely correlated within a night and filter, so they do not decrease with repeat observations. 

We also corrected separation and position angle measurements for differential atmospheric refraction \citep[DAR, ][]{2010SPIE.7736E..1IL} using filter wavelength information and weather data from the header (for VLT) or from the CFHT weather archive\footnote{\href{http://mkwc.ifa.hawaii.edu/archive/wx/cfht/}{http://mkwc.ifa.hawaii.edu/archive/wx/cfht/}} (for Keck, CFHT, and Gemini). We disregarded the chromatic component of this effect, as the correction is small compared to measurement errors.

\subsection{Literature Astrometry}\label{sec:litas}

To help identify literature astrometry for our targets, we used the fourth catalog of interferometric observations of binary stars \citep[INT4,][]{Hartkopf:2001}. We only used measurements with both a separation and position angle. In cases where the literature data were also available in one of the archives above (i.e., the same dataset used in the reference), we adopted our own measurements over the literature data. We did not utilize contrast measurements from the literature.

In total, we used 597 measurements (each including a separation and position angle) covering 51 of the 55 systems analyzed here. Although we pulled astrometry from 71 different publications, most of the measurements come from $\sim$10 different surveys (which may be spread across numerous publications). For example, 160 points came from {\it HST}, primarily the Fine Guidance Sensors measurements of 14 systems \citep[e.g.,][]{Benedict2016}, and 180 measurements came from speckle observations on the Special Astrophysical Observatory (SAO) 6m \citep[e.g.,][]{Bag2002}, WIYN \citep[e.g.,][]{Hor2017}, or SOAR \citep[e.g.,][]{Tok2017} telescopes. The rest of the measurements are from a mix of surveys focusing on taking many epochs of specific systems to determine orbits \citep[e.g.,][]{Koh2012}, broader surveys (e.g., for multiplicity) that obtain 1-2 epochs on dozens of binaries \citep[e.g.,][]{Jnn2012}, and programs targeting M dwarfs \citep[e.g.,][]{Mason2018}.

For a single system, GJ 1005, the position angle measurements from \citet{Benedict2016} differed significantly from our own astrometry and other literature determinations. The offset is consistent with a sign error in the individual positions assigned to each target before computing the final position angle. An earlier analysis of this same dataset by \citet{1998AJ....116.1440H} gave position angles consistent with ours and discrepant from \citet{Benedict2016}. We opt to use the \citet{Benedict2016} values over \citet{1998AJ....116.1440H} because the former is more precise, but we apply the relevant correction to the \citet{Benedict2016} position angles to correct the sign error. The separations were not affected, and no other system showed a similar discrepancy. 

One complication using older literature astrometry is inhomogeneous reporting of separation and position angle errors. Many references provided separations and position angles without uncertainties or have a discussion of general uncertainties, but do not provide them for individual measurements. A separate problem is references that reported measurement errors only, usually derived from a set of observations of a given target within a night or observing run (e.g., errors due to scatter in the PSF fit or variations in separation and position angle between exposures). Because we combined measurements from multiple instruments and sources, it is critical that we also account for systematic effects, i.e., those that impact all the images in a given set of observations (and hence are likely not reflected in the reported measurement uncertainties). This includes field distortion, the adopted pixel scale and instrument orientation, and DAR. These effects cannot be removed or modeled from individual epochs and can be larger than measurement uncertainties alone, particularly for extremely high precision measurements \citep[e.g.,][]{Lu2009}.

The most robust corrections for systematic were accomplished by observing crowded fields at multiple pointings and orientations. The extracted position of each star can then be compared to an external catalog and/or to repeat measurements with the star on different regions of the detector. Similar methods were used to calibrate a wide range of high-precision adaptive optics systems, including those used in this work \citep{Yelda2010, 2015MNRAS.453.3234P, 2016PASP..128i5004S}. In the absence of such data, observations of binaries with relatively well-determined orbits were often used as a low-order correction to the orientation or pixel scale \citep[e.g.,][]{Tok2015c}. Corrections for these effects, whether derived from observations of binaries or dense fields, have been particularly effective for systems that are stabilized and rarely removed from the telescope. For many other instruments and telescopes however, the corrections can vary with time. In such cases, there is rarely enough data to derive a time-dependent correction, so it is easier to model separation and position angle shifts as an extra error term. This method, i.e., modeling systematic errors from the detector/optics/etc as an additional error term using binary orbits, is the strategy we adopted here.

\begin{figure*}[ht]
\begin{center}
\includegraphics[width=0.32\textwidth]{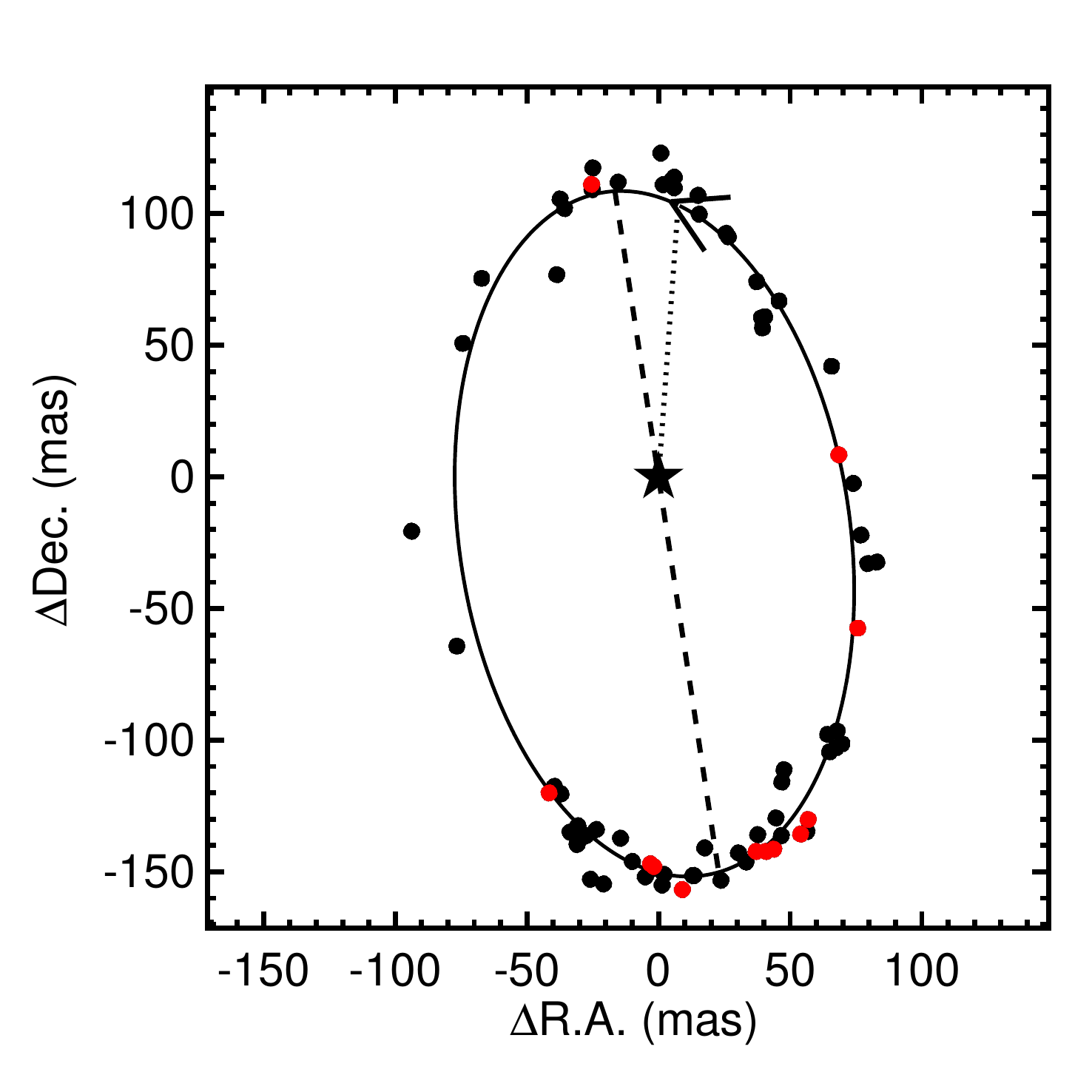}
\includegraphics[width=0.32\textwidth]{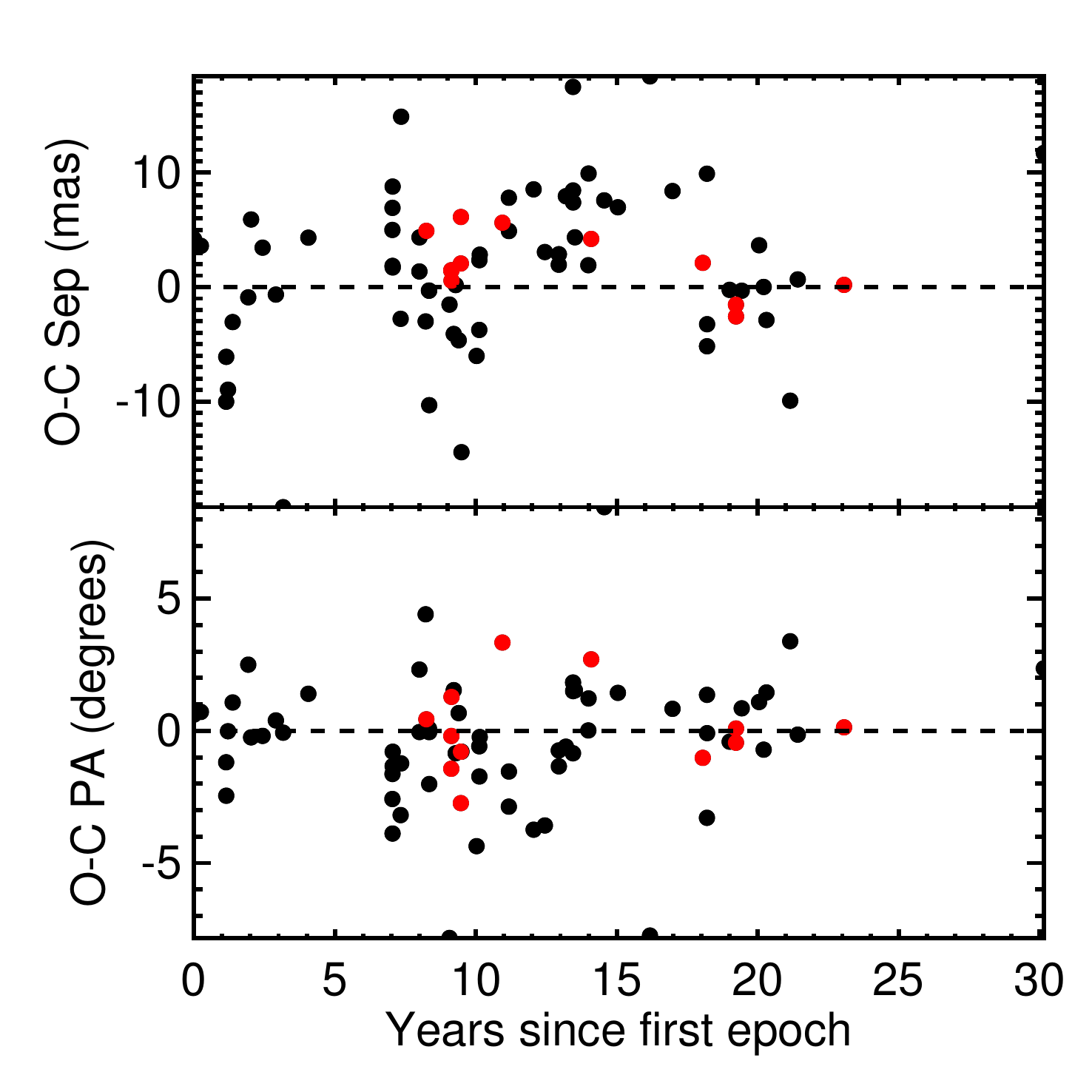}
\includegraphics[width=0.32\textwidth]{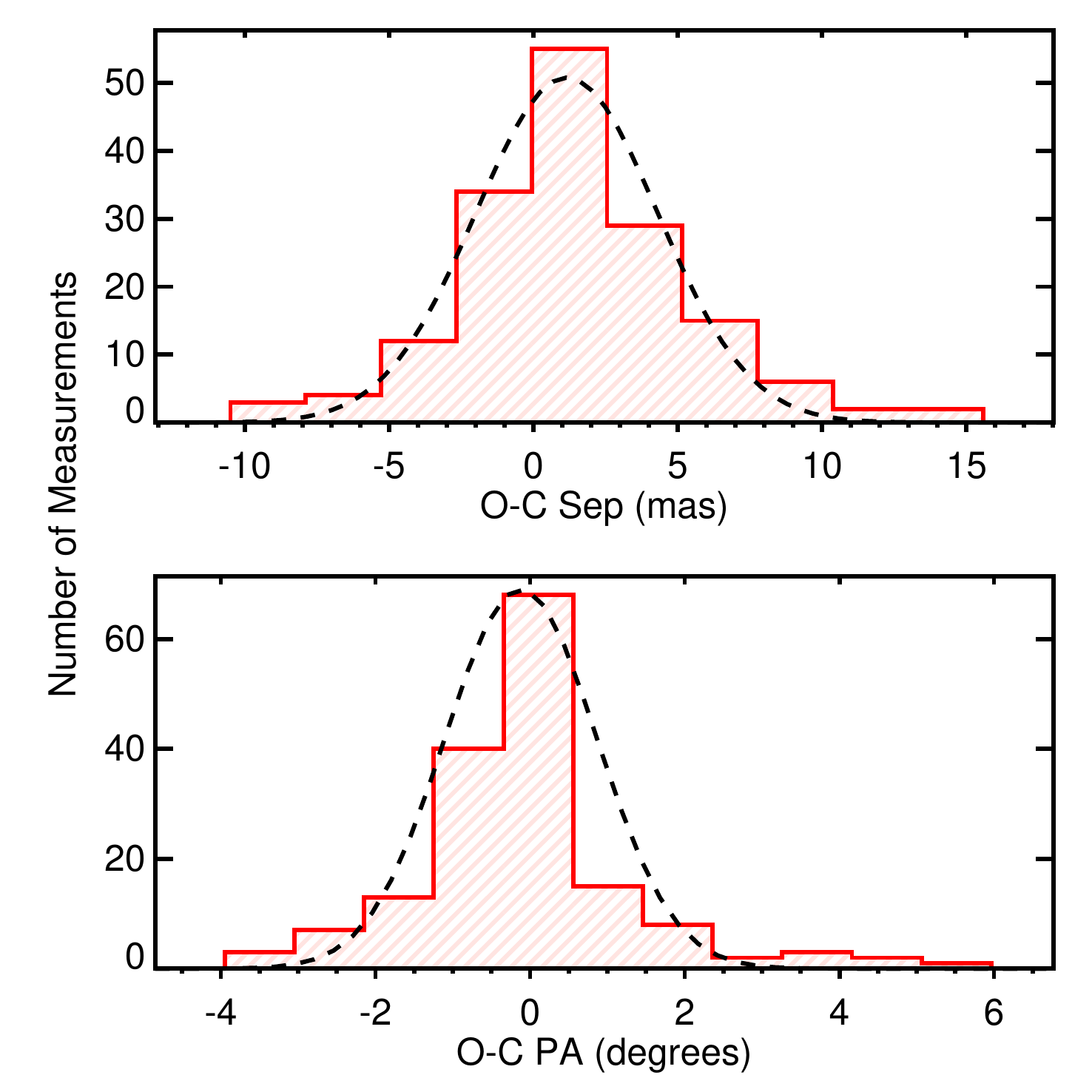}
\caption{Example of our method for assessing any missing error terms in literature astrometry. The left and center panels show the orbit (left) and residuals (center) for an example binary, HR 1331AB. Literature points are shown as circles in the left and center panels. In the left panel, the solid line is the best fit, the dotted line connects periastron passage, with an arrow pointing in the orbital direction, and the dashed line indicates the line of nodes. Measurements taken using the speckle imager on the BTA 6-m are labeled in red. The rightmost panels show a histogram of separation and position angle residuals from measurements taken with the BTA 6-m across 39 different binaries. The total residuals are modeled well by a Gaussian function, which we used to estimate the missing error term after subtracting out reported measurement uncertainties and errors in the orbital parameters. }
\label{fig:literr}
\end{center}
\end{figure*}

We first identified a set of binaries where the orbit can be fit (with $<3\%$ errors on the angular separation) without astrometry from the reference being tested. Literature sources using the same instrument and/or from the same paper series were merged for this comparison. We then fit the orbit of each binary following the method outlined in Section~\ref{sec:orbit}, using only the least-squares method for efficiency. We compared the expected position angle and separation (predicted from the binary orbit) to the measurements from the reference in question across all measurements and binaries included. For a given reference, we typically had tens or hundreds of orbit residual points, from which we computed a reduced $\chi^2$ ($\chi^2_\nu$) for both the separation and position angle, accounting for errors in the orbital parameters. For references with $\chi^2_\nu > 1$, we derived the required missing error term, i.e., the additional error in separation or position angle uncertainty required to yield $\chi^2_\nu = 1$. We show an example of the procedure in Figure~\ref{fig:literr}.

For references where no errors are provided, or for which there is a single uncertainty for all measurements, we adopted our derived uncertainty as the global error for all measurements. For references that report uncertainties for each measurement, we added our value in quadrature with the reported value. The added errors are summarized reference group in Table~\ref{tab:litastrom}, and all literature astrometry used in this paper is listed alongside our own measurements in Table~\ref{tab:astrom}. 

\begin{deluxetable*}{l l l l l l l l l l l l l l}
\tablecaption{Literature Astrometry }
\tablehead{
\colhead{Reference(s)} & \colhead{$\sigma_{\rm{sep}}$} & \colhead{$\sigma_{\rm{P.A.}}$}& \colhead{Median Sep\tablenotemark{a}} & \colhead{Note} \\
\colhead{} & \colhead{(mas)} & \colhead{(deg)} & \colhead{(mas)} & \colhead{} }
\startdata
\multicolumn{4}{c}{Global Error} \\
\hline
  1 & 4.6 & 0.85 &  236 & HST FGS\\
2--9 & 4.3 & 0.87 &  184 & WIYN/DCT DSSI\\
10--18 & 6.4 & 1.7 &  161 & ICCD Speckle\\
19--20 & 10 & 0.72 &  683 & Palomar\\
 21 & 45 & 1.4 &  390 & \\
 22 & 80 & 1.9 & 2080 & \\
23--24 & 9.9 & 1.8 &  208 & CTIO/KPNO USNO Speckle\\
25--32 & 42 & 1.1 & 1060 & Speckle at USNO\\
\hline
\multicolumn{4}{c}{Extra Term} \\
\hline
33--48 & 3.1 & 1.1 &  161 & 6m Speckle\\
49--50 & 3.1 & 0.65 &  224 & Astralux\\
 51 & 4.9 & 0.65 &  168 & \\
52--53 & 9.4 & 0.76 &  365 & \\
 54 & 14 & 1.6 &  334 & \\
 55 & 3.4 & 0.37 &  327 & \\
56--62 & 3.8 & 0.95 &  179 & SOAR Speckle\\
 63 & 5.0 & 0.80 &  342 & \\
64--67 & 5.9 & 1.6 &  123 & Speckle interferometry of binaries\\
\hline
\enddata
\tablenotetext{a}{The median separation for all measurements from a given reference used to estimate the uncertainty.}
\label{tab:litastrom}
\tablecomments{  1=\citet{Benedict2016},  2=\citet{Hor2002a},  3=\citet{Hor2008},  4=\citet{Hor2010},  5=\citet{Hor2011},  6=\citet{Hor2012a},  7=\citet{Hor2015},  8=\citet{Hor2015b},  9=\citet{Hor2017}, 10=\citet{Hrt1992b}, 11=\citet{Hrt1994}, 12=\citet{Hrt1997}, 13=\citet{Hrt2000a}, 14=\citet{McA1987b}, 15=\citet{McA1989}, 16=\citet{McA1990}, 17=\citet{McA1996a}, 18=\citet{McA1997}, 19=\citet{Hel2009}, 20=\citet{Llo2007}, 21=\citet{RDR2015}, 22=\citet{USN1988b}, 23=\citet{Mason2018}, 24=\citet{Msn2009}, 25=\citet{WSI1999a}, 26=\citet{WSI2000a}, 27=\citet{WSI2000b}, 28=\citet{WSI2002}, 29=\citet{WSI2004a}, 30=\citet{WSI2004b}, 31=\citet{WSI2006b}, 32=\citet{WSI2011}, 33=\citet{Bag1991a}, 34=\citet{Bag1994}, 35=\citet{Bag1997a}, 36=\citet{Bag1999a}, 37=\citet{Bag2001}, 38=\citet{Bag2002}, 39=\citet{Bag2002b}, 40=\citet{Bag2004}, 41=\citet{Bag2005}, 42=\citet{Bag2006b}, 43=\citet{Bag2007}, 44=\citet{Bag2007b}, 45=\citet{Bag2013}, 46=\citet{Doc2006i}, 47=\citet{Doc2008d}, 48=\citet{Doc2010h}, 49=\citet{Jnn2012}, 50=\citet{Jnn2014}, 51=\citet{Frv1999}, 52=\citet{Hrt2008}, 53=\citet{Hrt2009}, 54=\citet{Jod2013}, 55=\citet{Koh2012}, 56=\citet{Tok2010}, 57=\citet{Tok2012d}, 58=\citet{Tok2014a}, 59=\citet{Tok2015c}, 60=\citet{Tok2016a}, 61=\citet{Tok2017}, 62=\citet{Tok2018b}, 63=\citet{Sef2008}, 64=\citet{Bla1987}, 65=\citet{Bnu1986}, 66=\citet{McA1983}, 67=\citet{McA1984a}}
\end{deluxetable*}

Because the assumed uncertainties of each reference may impact the orbital fit (and hence the residuals of another reference), this process was done over all references twice, each time adjusting the uncertainties as appropriate. References where we could not test the reported errors (e.g., due to insufficient data) and those with extremely large aded error terms ($>100$\,mas) were not used. No reference yielded a negative term. 

Some earlier studies modeled the extra uncertainty in separation as a fraction \citep[e.g.,][]{Hrt2008,Hor2011,Tok2012d}. This is consistent with expectations for a plate scale errors, which impact wider binaries more than tighter systems. We found a better fit to separation residuals using a single value than a fraction of the separation (the right panel of Figure~\ref{fig:literr} shows one example). This may be because other effects, such as DAR and field distortion, are as important as plate scale errors. In particular, fractional errors tend to underestimate the uncertainty for the smallest separations. However, because of the relatively narrow range of separations considered here, the two methods gave similar results, and our uncertainties were relatively consistent with these earlier studies. \citet{2007AJ....134.1671M}, for example, compared the separation and position angle predictions from the ``Speckle Interferometry at USNO'' paper series to a set of well-characterized orbits and found a scatter of 1.1$^{\circ}$--1.2$^{\circ}$ in position angle and 2.2\%--5.6\% in separation. For the typical separations we used in this series ($\simeq1\arcsec$), this is consistent with our own determination of 1.2$^{\circ}$ and 37\,mas (Table~\ref{tab:litastrom}). To aid with such comparisons, in Table~\ref{tab:litastrom} we included the typical separation from each reference used for our uncertainty estimates.

\subsection{Summary of input astrometry}

In total we measured or gathered 1142 unique datasets (unique filter/night/target combinations), approximately half of which we measured from adaptive optics images (541) and the other half were drawn from the literature (597). Most of the astrometry measurements derived from our analysis came from either Keck/NIRC2 (198) or CFHT/KIR (239), with a smaller contribution from VLT/NaCo (72) and Gemini/NIRI (36). 

\begin{figure}[htb]
\begin{center}
\includegraphics[width=0.46\textwidth]{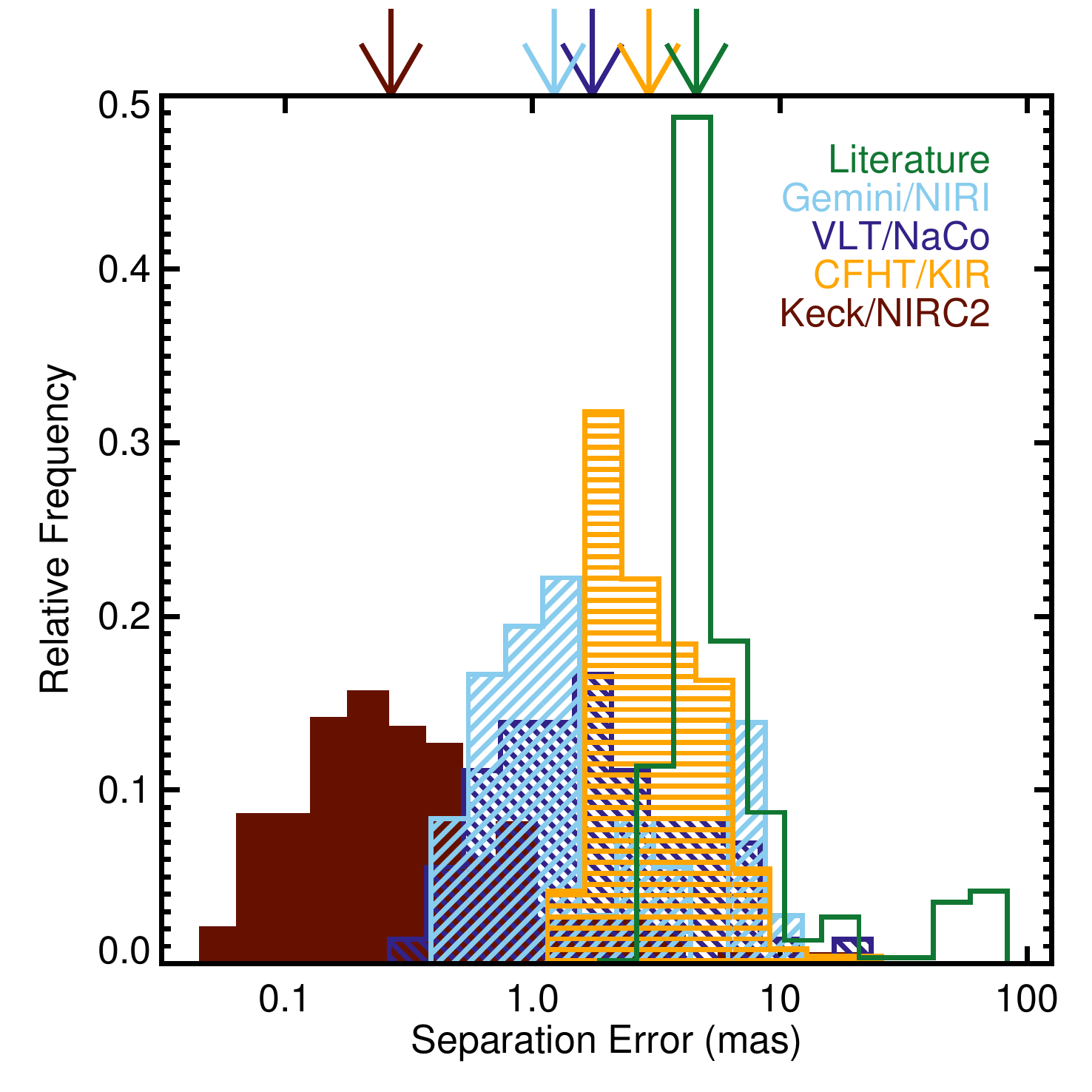}
\caption{Comparison of input errors on separation for all astrometry used in our analysis by source (see Section~\ref{sec:obs}). Arrows on the top X-axis denote the median for each source. Keck/NIRC2 astrometry significantly outperforms other sources, and is the most critical for our analysis. Note that bins are spaced logarithmically to show the full range of separation uncertainties. }
\label{fig:errors}
\end{center}
\end{figure}

Although data from NIRC2 represent only $\lesssim$20\% of the total astrometric measurements, they are critical in constraining orbital parameters. NIRC2 astrometry was typically an order of magnitude more precise than those from the literature, and a factor of 3-8$\times$ more precise than those from KIR, NIRI, and NaCo (Figure~\ref{fig:errors}). In addition to improved Strehl provided by a larger telescope, NIRC2 is rarely removed from the telescope and therefore has a stable and extremely well-characterized distortion solution and pixel scale. Terms we treat as uncertainties for much of the literature astrometry are modeled out for NIRC2 observations. Instruments like NaCo are also capable of achieving astrometry with similar levels of precision \citep[e.g.,][]{Reggiani2016}. However, this requires astrometric calibrators observed in the same run, which were not available for most datasets analyzed here.

We characterized the relative importance of each data source using the total number of unique separation measurements weighted by their uncertainties (1/$\sigma$). Under this metric, the NIRC2 points contributed significantly more orbital information than the literature data (77\% of the total weight from NIRC2 versus 9\% from the literature). Measurements from KIR (7\%) had a comparable total contribution to the literature data, each of which had $\simeq2-3\times$ the weight of measurements from NaCo (4\%) and NIRI (3\%). 

 A comparison based on measurement errors alone significantly underestimates the importance of data sampling and orbital coverage. Literature and archive images tended to be concentrated on the best-characterized systems, while the NIRC2 observations were specifically coordinated to complete orbits and cover under- or unsampled regions of binary orbits. The literature data, however, provide the largest baseline. Over all observations used in our analysis, NIRI data spanned 6.5 yr, compared to 9.1 yr covered by KIR, 13.8 yr by NACO, 16.3 from NIRC2, and 68.9 yr from the literature. The NIRC2 data was also heavily concentrated in a single 3-year window (2015-2018). While a significant fraction of the baseline in the literature astrometry came from a single target (Gl 65), literature data covered 37.2 yr even when this target is excluded. The long baseline provided by the literature astrometry was crucial for analyzing systems with multidecade orbital periods, which included the majority of binaries analyzed here.

\section{Orbit Fitting}\label{sec:orbit}
We fit the astrometry following a Bayesian methodology with Keplerian orbits. Our basic technique is outlined in \citep[][, and references within]{Dupuy2017}, which we summarize here. We used the Monte Carlo Markov Chain (MCMC) software {\tt emcee} \citep{Foreman-Mackey2013}, a Python implementation of the affine-invariant ensemble sampler \citep{goodman2010}. For each system, we explored seven orbital elements: the orbital period ($P$), combined angular semi-major axis ($\alpha_{\rm{ang}}$), eccentricity ($e$), inclination ($i$), argument of periastron ($\omega$), position angle of the line of nodes ($\Omega$), and the position angle at January 1, 2010 00:00:00 UT ($\lambda$). The variable $\lambda$ was fit instead of the usual epoch of periastron passage ($T_0$), because $T_0$ is undefined for circular orbits and multi-valued to aliases of $P$, both of which cause problems for the MCMC exploration. We converted $\lambda$ into $T_0$ after the MCMC chain is complete for reporting purposes. 

We applied non-uniform priors of 1/$P$, 1/$\alpha_{\rm{ang}}$, and $\sin(i)$ to $P$, $\alpha_{\rm{ang}}$, and $i$, respectively. All other parameters evolved under uniform priors. Parameters were limited by physical or definitional constraints, e.g., $P>0$, $0 \le e<1$, and $0\le i \le \pi$, but were given no additional boundaries. A summary of the fit parameters, priors, and limits is given in Table~\ref{tab:priors}. 

\begin{deluxetable}{l l l l l l l l l l l l l l}
\tablecaption{Orbit fit parameters, limits, and priors. \label{tab:priors}} 
\tablehead{
\colhead{Parameter} & \colhead{Limits} & \colhead{Prior} 
}
\startdata
$P$ & (0, $\infty$) & 1/$P$\\
$\alpha_{\rm{ang}}$ & (0, $\infty$) &1/$\alpha_{\rm{ang}}$ \\
$e$ & (0, 1) & uniform\\
$i$ & (0, $\pi$) & $\sin(i)$ \\
$\omega$ & (0, $2\pi$) & uniform\\
$\Omega$ & (0, $2\pi$) & uniform \\
$\lambda$ & (0, $2\pi$) & uniform \\
\enddata
\end{deluxetable}

For each run, walkers were initialized with the best-fit orbit determined using {\tt MPFIT} \citep{Markwart2009} and a spread in starting values based on the {\tt MPFIT} estimated errors. Each MCMC chain was initially run for $10^5$ steps with 100 walkers. We considered a chain converged if the total length was at least 50 times as long as the autocorrelation time \citep{2010CAMCS...5...65G}. Systems that did not converge were run for an additional $10^6$ total steps, which was sufficient for convergence of all systems. We saved every 100 steps in the chain, and the first 10\% of each chain was removed for burn-in. Longer burn-in time did not change the final posterior in any significant way, in part because the initial (least-squares) guesses were always near the final answer from the MCMC. 

Systems of near-equal mass may have the primary and companion confused, both in our own measurements and also those taken from the literature. We identified such measurements by eye during the {\tt MPFIT} stage and manually adjusted the position angles before starting the MCMC run. In total $\simeq$16 measurements were corrected this way, almost all of which were for three systems with contrast ratios close to unity. A more robust solution to this problem would be to feed a double-peaked posterior at the reported value and $\pm$180\degree\ into the likelihood function. However, in all cases the problematic points were obvious by eye, they had reported $\Delta m$ consistent with zero, and a simple 180\degree\ correction completely fixed the orbit. 

\begin{figure*}[htb]
\begin{center}
\includegraphics[width=0.32\textwidth]{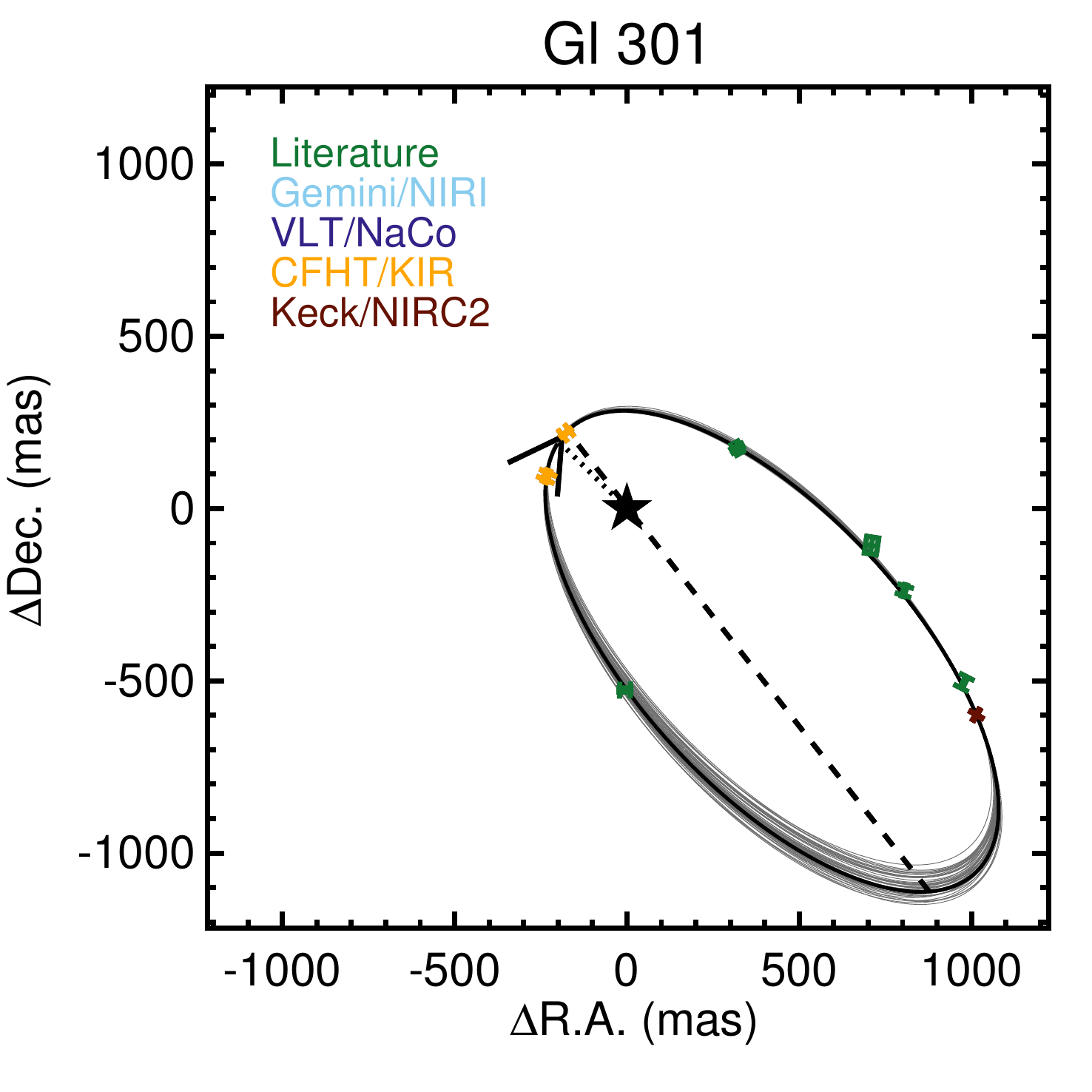}
\includegraphics[width=0.32\textwidth]{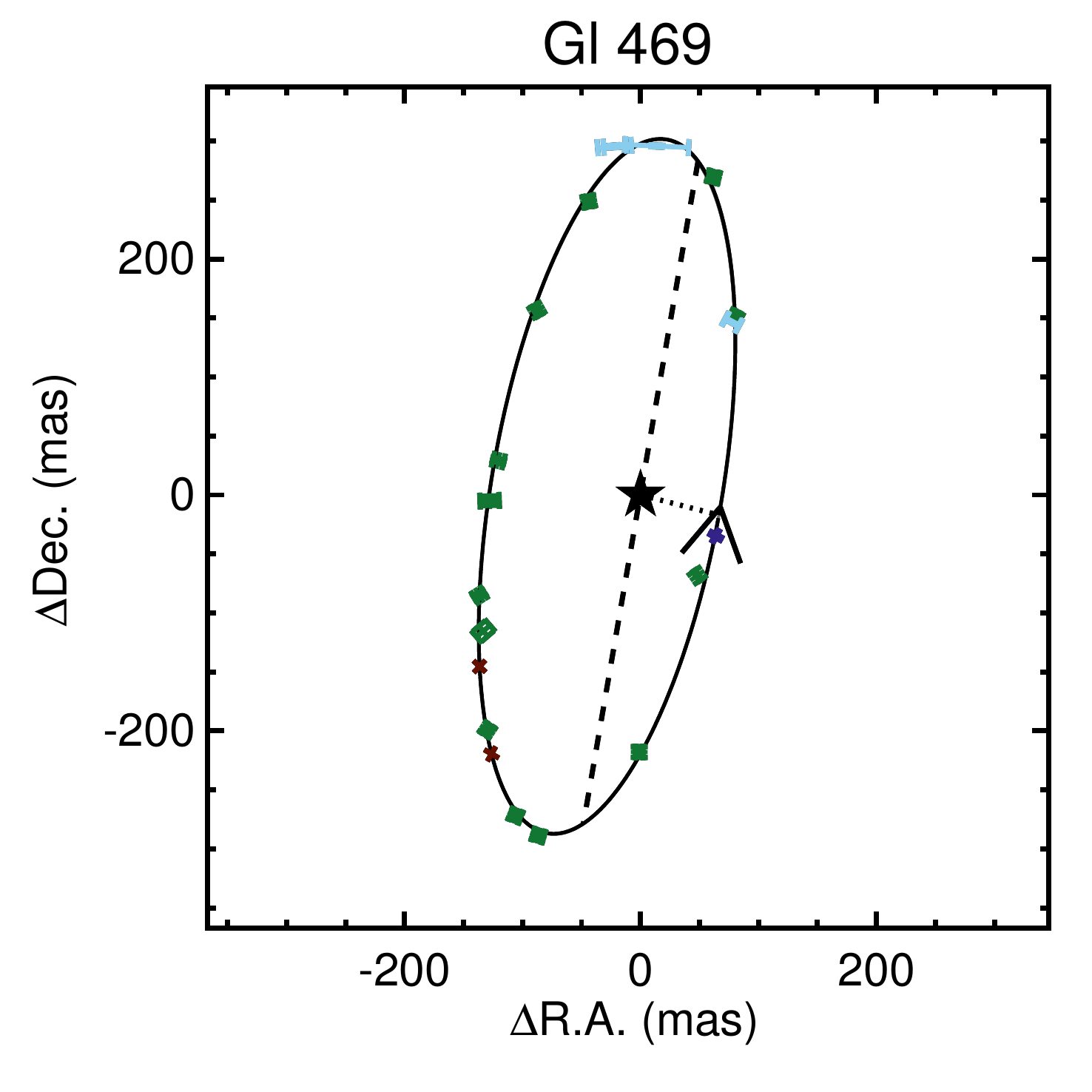}
\includegraphics[width=0.32\textwidth]{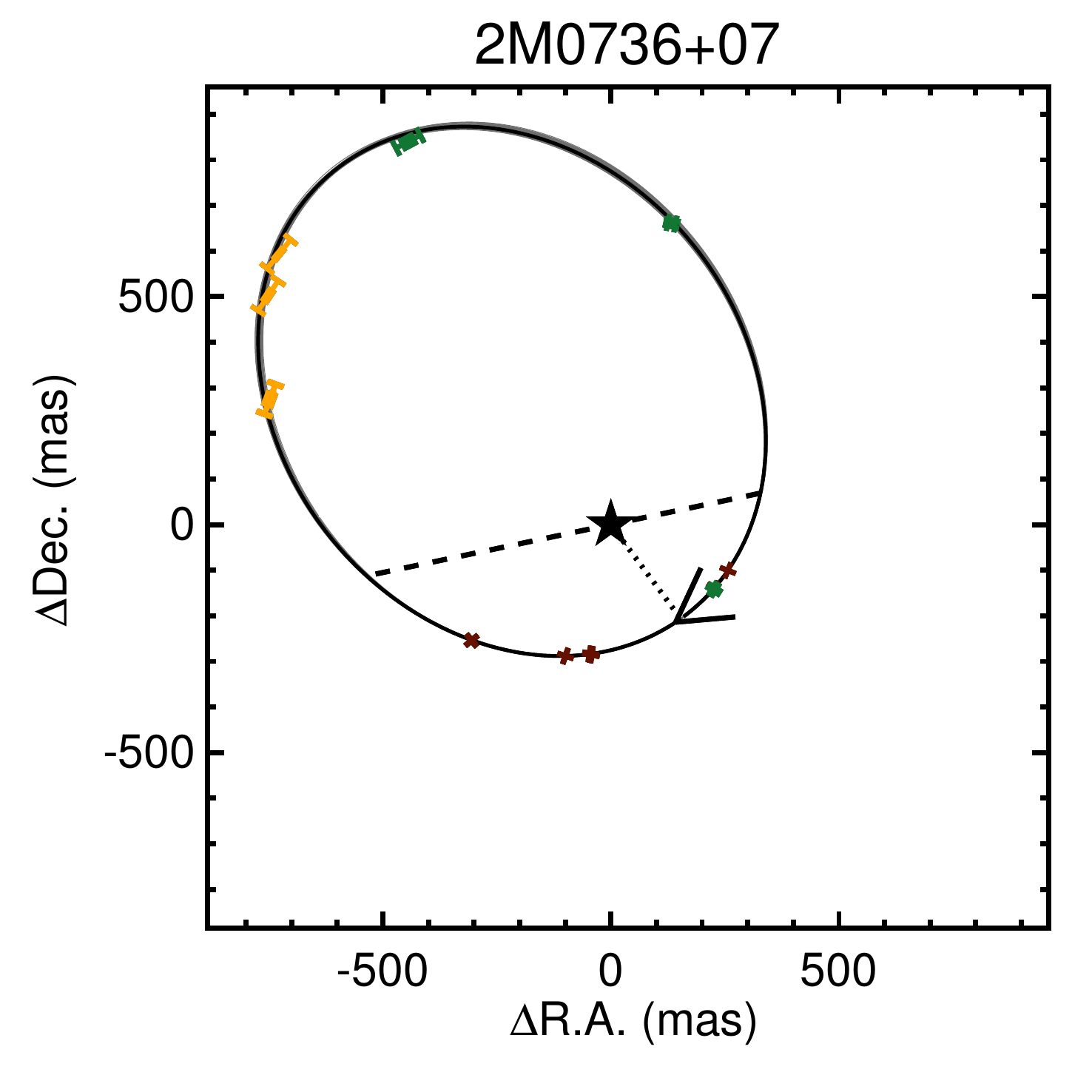}
\includegraphics[width=0.32\textwidth]{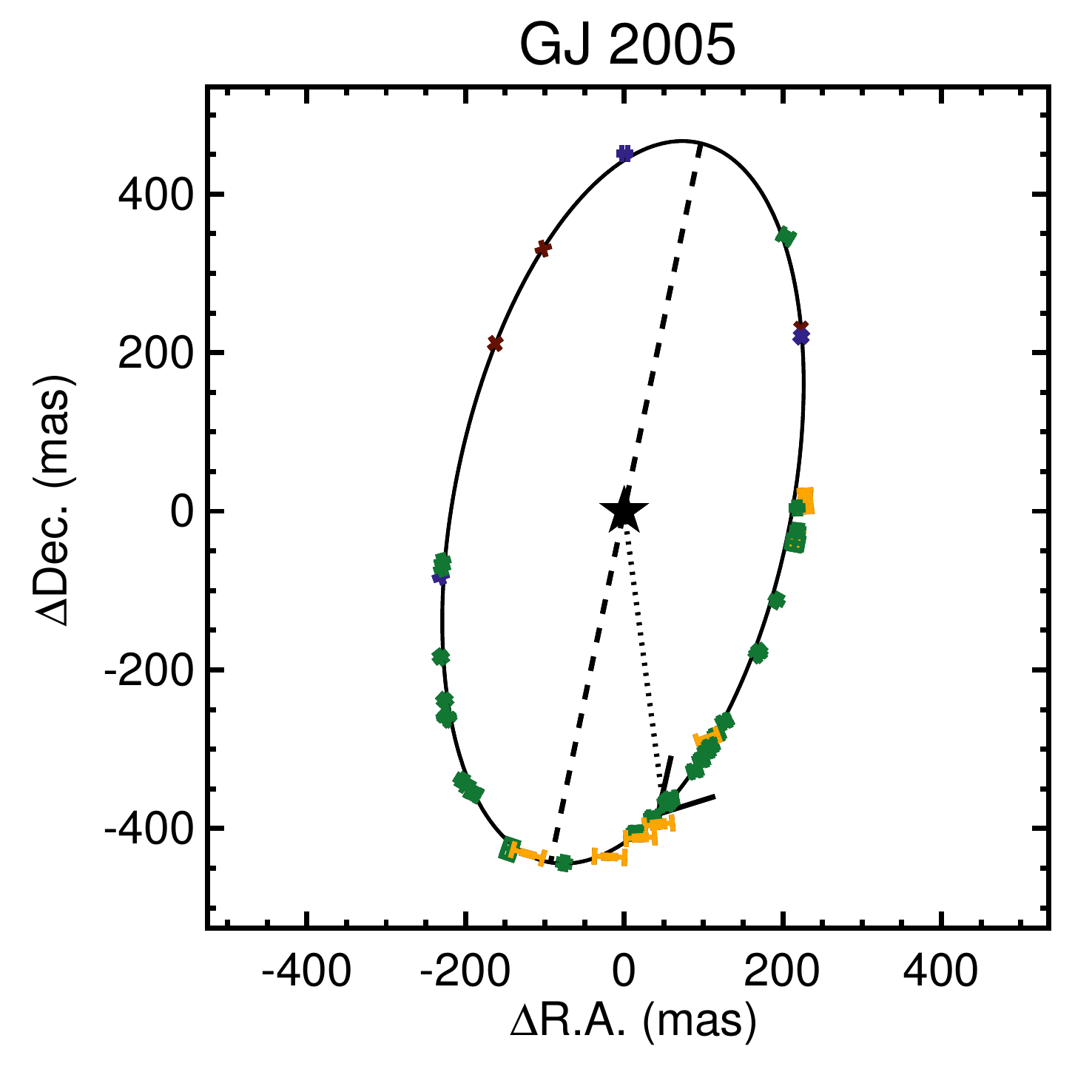}
\includegraphics[width=0.32\textwidth]{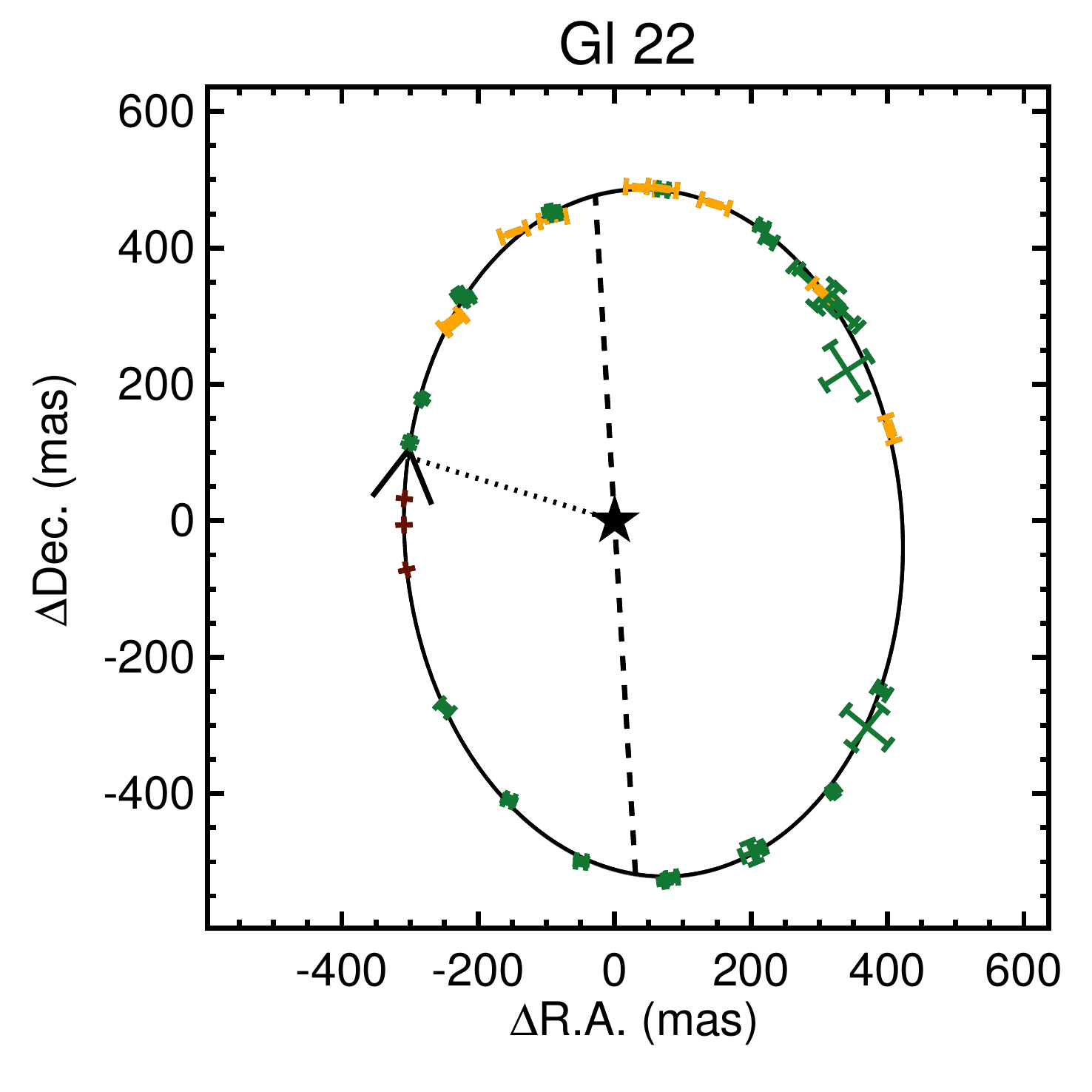}
\includegraphics[width=0.32\textwidth]{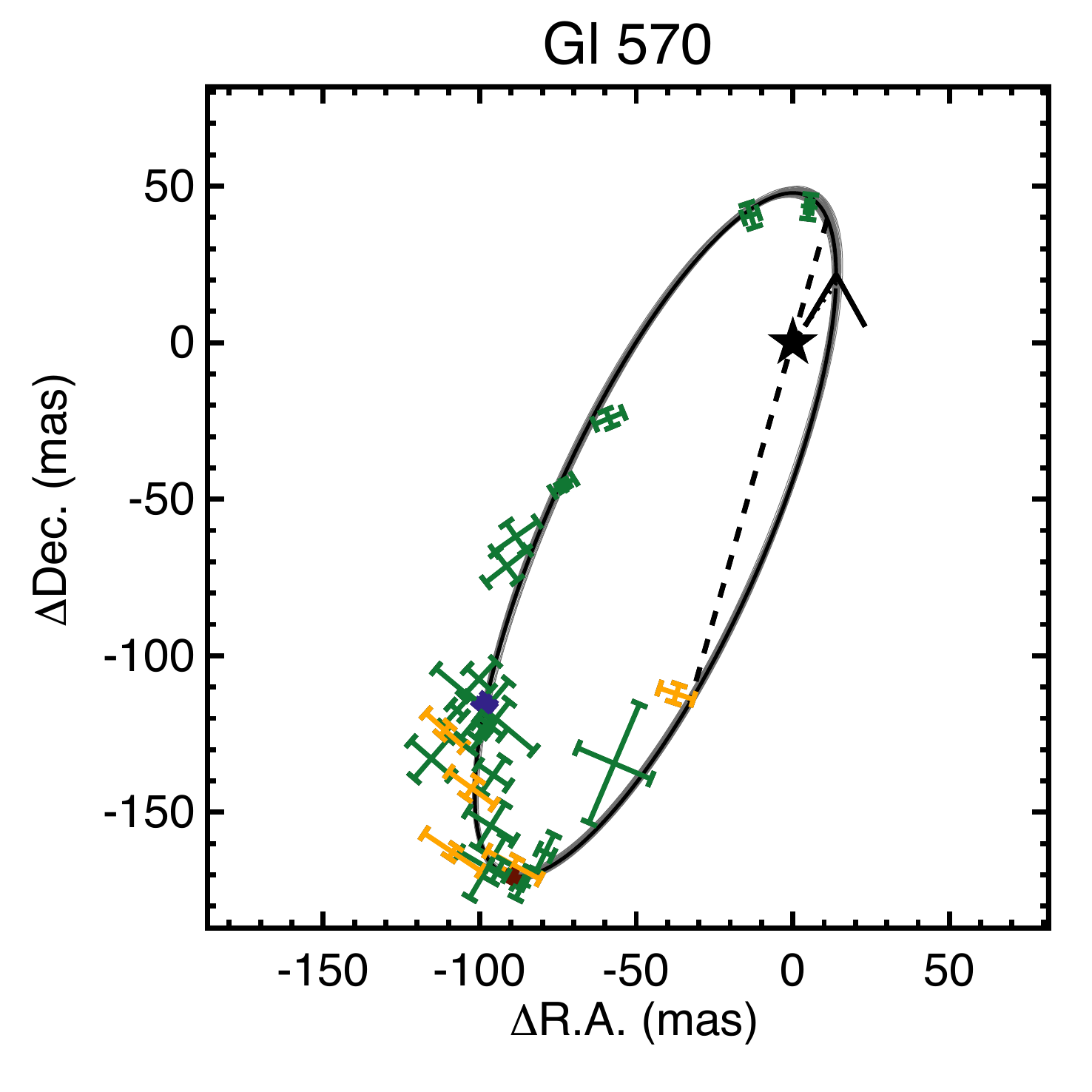}
\includegraphics[width=0.32\textwidth]{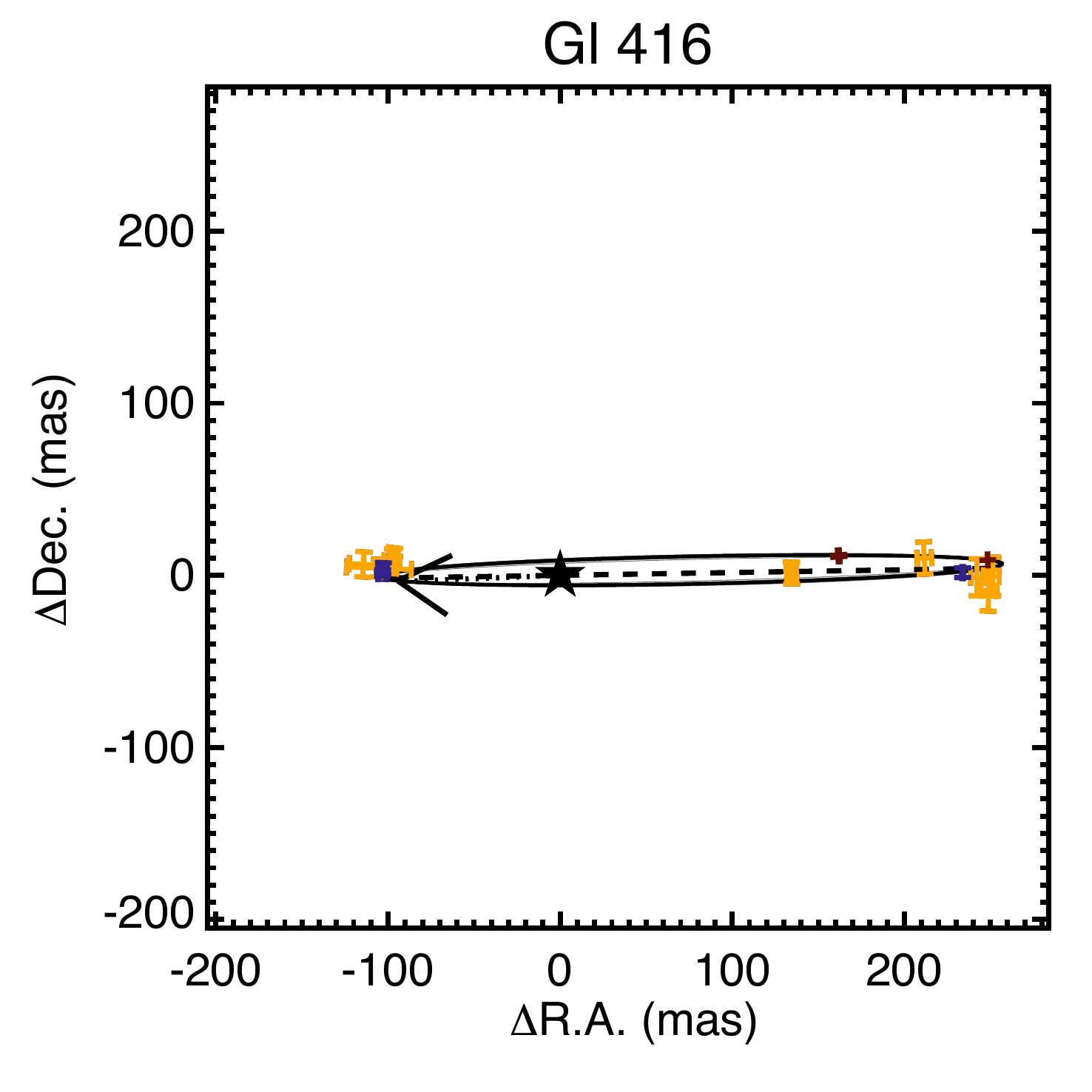}
\includegraphics[width=0.32\textwidth]{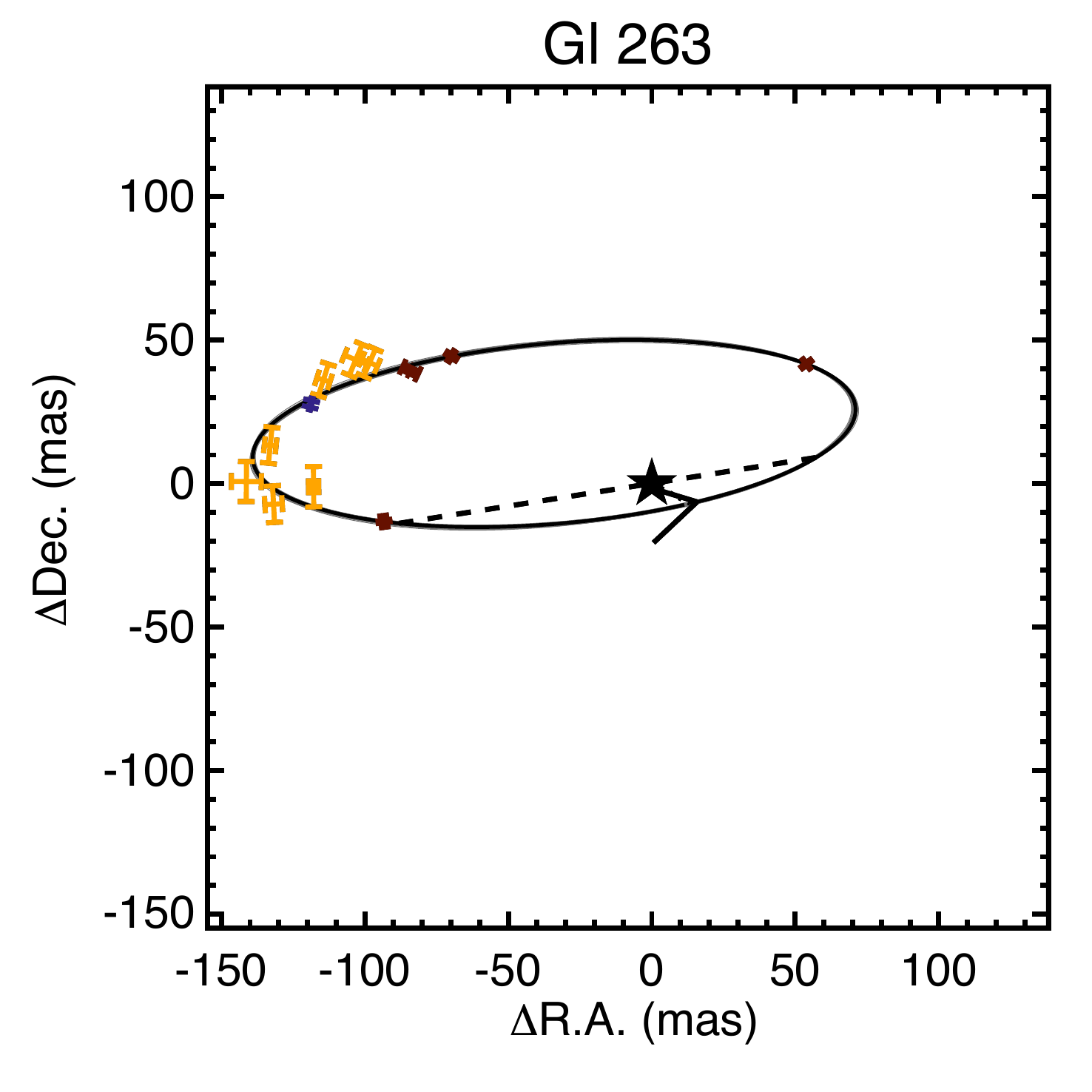}
\includegraphics[width=0.32\textwidth]{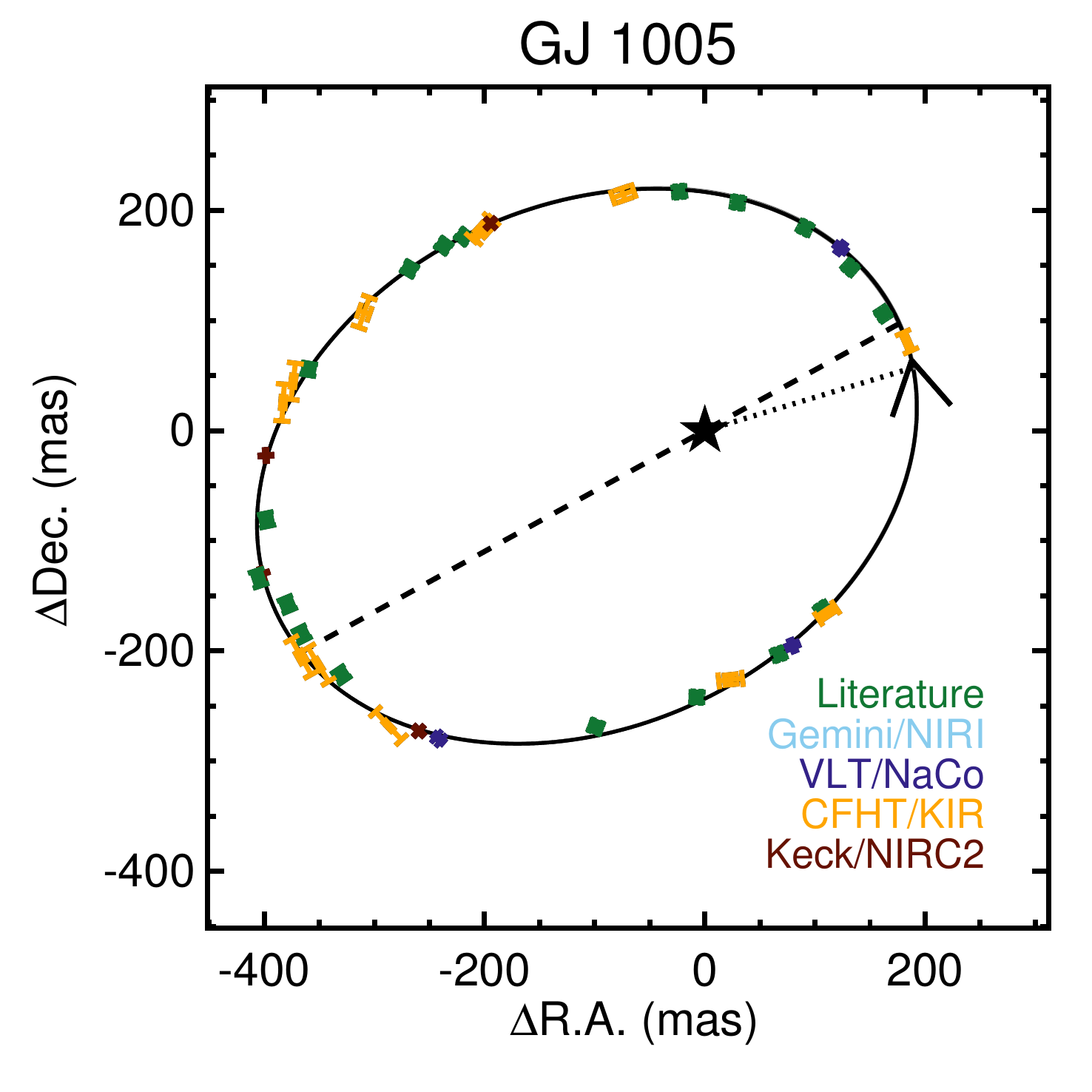}
\caption{Example results of our orbit-fitting procedure. Points are individual measurements of the separation and position angle, color-coded by the astrometry source. Black solid line shows the best-fit (highest-likelihood) orbit. Dark grey lines are drawn by randomly sampling 50 orbit fits from the MCMC chain to display an estimate of the errors. The dotted line connects periastron passage, with an arrow pointing in the orbital direction, and the dashed line indicates the line of nodes. For Gl469AB, high-quality astrometry is available for the full orbit, and the resulting errors on orbital parameters are so small that the grey lines cannot be seen.}
\label{fig:orbits}
\end{center}
\end{figure*}

Overall the quality of our fits was extremely good, with $\chi^2_\nu$ values ranging from 0.1 to 2 and a mean cumulative probability (the probability of getting the $\chi^2$ or smaller given the degrees of freedom) of 63\% across all targets. We show some example orbital fits in Figure~\ref{fig:orbits} and provide the median orbital parameters in Table~\ref{tab:orbits}. Orbits span a wide range in period; the tightest binaries have $P<1$\,yr, while the widest systems have periods of $>50$\,yr. The two systems with $P>50$\,yr (Gl 301 and Gl 277) were also some of the least well characterized. No systems show evidence of period doubling due to limited sampling, an advantage of using data with a mix of tight ($\lesssim$1\,yr) and widely spaced ($>5$\,yr) astrometry.

\begin{deluxetable*}{l l l l l l l l l r r}
\rotate
\tablewidth{\linewidth}
\tablecaption{Orbital Parameters \label{tab:orbits}
}
\tablehead{
    \colhead{Name} & \colhead{$P$} & \colhead{$\alpha_{\rm{ang}}$} & \colhead{$e$} & \colhead{$i$} 
 & \colhead{$\omega$} & \colhead{$\Omega$} & \colhead{$T_0$} & \colhead{$\alpha_{\rm{ang}}^3/P^2$} & \colhead{$\chi^2$/dof} \\
    \colhead{} & \colhead{(yr)} & \colhead{(mas)} & \colhead{} & \colhead{(deg)}
 & \colhead{(deg)} & \colhead{(deg)} & \colhead{MJD} & \colhead{arcsec$^3\,yr^{-2}$} & \colhead{} 
}
\startdata
GJ 1005 & \phantom{0}4.55726$^{+0.00075}_{-0.00074}$ & \phantom{0}312.85$^{+0.50}_{-0.50}$ & 0.36136$^{+0.00097}_{-0.00098}$ & 143.93$^{+0.25}_{-0.24}$ & 345.26$^{+0.62}_{-0.62}$ & \phantom{0}61.23$^{+0.41}_{-0.41}$ & 58172.9$^{+1.9}_{-1.9}$ & (1.4743$\pm$0.0072)x$10^3$ & 85.9/67\\
GJ 2005 & 17.296$^{+0.011}_{-0.011}$ & \phantom{0}463.36$^{+0.57}_{-0.56}$ & 0.02900$^{+0.00093}_{-0.00091}$ & \phantom{0}62.816$^{+0.050}_{-0.051}$ & 143.3$^{+2.4}_{-2.2}$ & \phantom{0}11.798$^{+0.084}_{-0.085}$ & 59158$^{+42}_{-40}$ & (3.326$\pm$0.015)x$10^4$ & 101.6/85\\
Gl 22 & 15.4275$^{+0.0054}_{-0.0054}$ & \phantom{0}510.26$^{+0.74}_{-0.74}$ & 0.1576$^{+0.0013}_{-0.0013}$ & \phantom{0}44.29$^{+0.15}_{-0.15}$ & 104.90$^{+0.53}_{-0.53}$ & 176.75$^{+0.21}_{-0.21}$ & 57447.0$^{+5.1}_{-5.2}$ & (5.582$\pm$0.026)x$10^4$ & 75.5/65\\
Gl 54 & \phantom{0}1.14434$^{+0.00022}_{-0.00022}$ & \phantom{0}126.19$^{+0.39}_{-0.39}$ & 0.1718$^{+0.0024}_{-0.0024}$ & 125.32$^{+0.35}_{-0.35}$ & \phantom{0}47.33$^{+0.92}_{-0.93}$ & \phantom{0}92.04$^{+0.40}_{-0.40}$ & 58542.0$^{+1.2}_{-1.2}$ & (1.534$\pm$0.014)x$10^3$ & 37.9/33\\
GJ 1038 & \phantom{0}5.98$^{+1.02}_{-0.77}$ & \phantom{0}139.7$^{+14.5}_{-9.7}$ & 0.54$^{+0.12}_{-0.11}$ & \phantom{0}72.9$^{+1.9}_{-2.5}$ & 174$^{+13}_{-16}$ & 105.9$^{+1.6}_{-2.7}$ & 58313$^{+41}_{-1976}$ & (7.67$\pm$0.52)x$10^5$ & 1.5/9\\
Gl 65 & 26.359$^{+0.025}_{-0.024}$ & 2049.6$^{+3.4}_{-3.4}$ & 0.6222$^{+0.0015}_{-0.0015}$ & 128.09$^{+0.15}_{-0.15}$ & 283.340$^{+0.085}_{-0.084}$ & 146.29$^{+0.14}_{-0.14}$ & 60591.9$^{+9.4}_{-9.4}$ & (1.2391$\pm$0.0058)x$10^2$ & 96.7/89\\
Gl 84 & 13.328$^{+0.037}_{-0.037}$ & \phantom{0}495.5$^{+1.2}_{-1.2}$ & 0.3771$^{+0.0082}_{-0.0081}$ & \phantom{0}91.771$^{+0.056}_{-0.057}$ & 245.62$^{+0.48}_{-0.48}$ & 102.987$^{+0.022}_{-0.022}$ & 61554$^{+14}_{-14}$ & (6.850$\pm$0.066)x$10^4$ & 21.4/17\\
2M0213+36 & \phantom{0}6.419$^{+0.067}_{-0.071}$ & \phantom{0}161.5$^{+1.3}_{-1.3}$ & 0.4232$^{+0.0039}_{-0.0038}$ & 115.30$^{+0.24}_{-0.24}$ & 207.66$^{+0.72}_{-0.73}$ & \phantom{0}83.73$^{+0.32}_{-0.33}$ & 57604.5$^{+2.8}_{-2.9}$ & (1.023$\pm$0.016)x$10^4$ & 28.8/11\\
Gl 98 & 25.255$^{+0.021}_{-0.021}$ & \phantom{0}559.84$^{+0.66}_{-0.65}$ & 0.2354$^{+0.0013}_{-0.0013}$ & \phantom{0}73.389$^{+0.048}_{-0.047}$ & 231.49$^{+0.21}_{-0.21}$ & 109.116$^{+0.022}_{-0.022}$ & 56417.3$^{+4.6}_{-4.7}$ & (2.7509$\pm$0.0098)x$10^4$ & 87.6/85\\
Gl 99 & 24.015$^{+0.086}_{-0.084}$ & \phantom{0}360.53$^{+0.72}_{-0.71}$ & 0.2087$^{+0.0028}_{-0.0025}$ & \phantom{0}84.605$^{+0.049}_{-0.049}$ & 152.6$^{+1.9}_{-1.9}$ & \phantom{0}98.836$^{+0.087}_{-0.090}$ & 56340$^{+43}_{-43}$ & (8.126$\pm$0.027)x$10^5$ & 23.8/25\\
Gl 125 & 25.67$^{+0.20}_{-0.19}$ & \phantom{0}534.5$^{+2.3}_{-2.3}$ & 0.2271$^{+0.0044}_{-0.0044}$ & \phantom{0}97.186$^{+0.026}_{-0.026}$ & 181.38$^{+0.35}_{-0.36}$ & \phantom{0}13.732$^{+0.082}_{-0.081}$ & 64226$^{+74}_{-73}$ & (2.3173$\pm$0.0098)x$10^4$ & 12.6/23\\
Gl 150.2 & 13.604$^{+0.045}_{-0.046}$ & \phantom{0}254.9$^{+1.3}_{-1.3}$ & 0.268$^{+0.011}_{-0.010}$ & 101.79$^{+0.38}_{-0.38}$ & 250.5$^{+1.2}_{-1.3}$ & 100.68$^{+0.46}_{-0.45}$ & 58570$^{+23}_{-24}$ & (8.95$\pm$0.16)x$10^5$ & 25.4/19\\
Gl 190 & \phantom{0}0.96380$^{+0.00024}_{-0.00024}$ & \phantom{00}99.12$^{+0.85}_{-0.85}$ & 0.2441$^{+0.0093}_{-0.0091}$ & \phantom{0}92.96$^{+0.26}_{-0.25}$ & 186.4$^{+5.7}_{-5.9}$ & \phantom{0}40.42$^{+0.17}_{-0.17}$ & 58534.4$^{+6.5}_{-6.7}$ & (1.048$\pm$0.027)x$10^3$ & 25.5/31\\
GJ 1081 & 11.593$^{+0.034}_{-0.033}$ & \phantom{0}272.9$^{+6.8}_{-6.1}$ & 0.8612$^{+0.0058}_{-0.0055}$ & \phantom{0}97.23$^{+0.35}_{-0.34}$ & 230.5$^{+1.5}_{-1.4}$ & \phantom{0}51.11$^{+0.14}_{-0.14}$ & 57220$^{+18}_{-17}$ & (1.51$\pm$0.10)x$10^4$ & 25.4/19\\
Gl 234 & 16.5777$^{+0.0027}_{-0.0027}$ & 1086.04$^{+0.27}_{-0.28}$ & 0.38229$^{+0.00013}_{-0.00013}$ & \phantom{0}52.910$^{+0.016}_{-0.016}$ & 220.942$^{+0.020}_{-0.020}$ & \phantom{0}30.385$^{+0.030}_{-0.030}$ & 57342.99$^{+0.23}_{-0.23}$ & (4.6612$\pm$0.0033)x$10^3$ & 123.2/105\\
LHS 221 & 13.5943$^{+0.0062}_{-0.0061}$ & \phantom{0}440.9$^{+1.6}_{-1.6}$ & 0.4777$^{+0.0022}_{-0.0022}$ & 109.76$^{+0.13}_{-0.13}$ & \phantom{0}58.59$^{+0.21}_{-0.21}$ & 107.20$^{+0.21}_{-0.21}$ & 59671.2$^{+2.5}_{-2.5}$ & (4.637$\pm$0.053)x$10^4$ & 30.5/49\\
LHS 224 & \phantom{0}3.2860$^{+0.0014}_{-0.0014}$ & \phantom{0}156.38$^{+0.29}_{-0.28}$ & 0.2231$^{+0.0045}_{-0.0044}$ & 131.76$^{+0.52}_{-0.53}$ & \phantom{0}73.73$^{+0.43}_{-0.43}$ & 173.93$^{+0.81}_{-0.84}$ & 58535.0$^{+4.9}_{-4.8}$ & (3.541$\pm$0.020)x$10^4$ & 19.2/27\\
Gl 263 & \phantom{0}3.6205$^{+0.0021}_{-0.0021}$ & \phantom{0}143.8$^{+2.1}_{-2.0}$ & 0.7158$^{+0.0065}_{-0.0064}$ & 103.28$^{+0.36}_{-0.36}$ & 287.52$^{+0.51}_{-0.51}$ & \phantom{0}81.04$^{+0.24}_{-0.24}$ & 58416.0$^{+3.6}_{-3.6}$ & (2.268$\pm$0.099)x$10^4$ & 21.9/17\\
Gl 277 & 53.0$^{+9.2}_{-7.2}$ & 1058$^{+113}_{-84}$ & 0.48$^{+0.14}_{-0.12}$ & \phantom{0}93.53$^{+0.48}_{-0.34}$ & \phantom{0}22$^{+332}_{-17}$ & \phantom{0}10.22$^{+0.26}_{-0.44}$ & 71033$^{+12032}_{-3397}$ & (4.22$\pm$0.16)x$10^4$ & 15.2/13\\
2M0736+07 & 23.768$^{+0.048}_{-0.046}$ & \phantom{0}633.29$^{+0.94}_{-0.93}$ & 0.58621$^{+0.00060}_{-0.00060}$ & \phantom{0}12.3$^{+1.0}_{-1.1}$ & \phantom{0}66.7$^{+5.6}_{-5.3}$ & \phantom{0}77.3$^{+4.9}_{-5.2}$ & 57466.3$^{+2.7}_{-2.7}$ & (4.495$\pm$0.019)x$10^4$ & 16.8/21\\
Gl 301 & 62.2$^{+1.8}_{-1.7}$ & \phantom{0}875$^{+10}_{-10}$ & 0.6778$^{+0.0049}_{-0.0048}$ & \phantom{0}52.31$^{+0.86}_{-0.87}$ & 167.4$^{+1.1}_{-1.1}$ & 142.0$^{+1.1}_{-1.1}$ & 51189$^{+18}_{-19}$ & (1.737$\pm$0.049)x$10^4$ & 20.1/13\\
Gl 310 & 23.48$^{+0.14}_{-0.14}$ & \phantom{0}552.6$^{+5.6}_{-5.4}$ & 0.6976$^{+0.0073}_{-0.0074}$ & 122.06$^{+0.57}_{-0.56}$ & 246.72$^{+0.36}_{-0.37}$ & \phantom{0}49.62$^{+0.39}_{-0.38}$ & 58432.5$^{+5.2}_{-5.3}$ & (3.06$\pm$0.12)x$10^4$ & 21.5/11\\
Gl 330 & 32.69$^{+0.42}_{-0.40}$ & \phantom{0}582$^{+14}_{-12}$ & 0.8301$^{+0.0071}_{-0.0070}$ & 105.78$^{+0.51}_{-0.49}$ & 309.0$^{+1.5}_{-1.5}$ & \phantom{0}38.63$^{+0.50}_{-0.51}$ & 64663$^{+160}_{-157}$ & (1.85$\pm$0.13)x$10^4$ & 20.2/17\\
\enddata
\tablecomments{Table \ref{tab:orbits} is available in its entirety in the ancillary files with the arXiv submission. A portion is shown here for guidance regarding its form and content.
}
\end{deluxetable*}

As a test of our sensitivity to the assumed priors, we reran the five systems with the fewest astrometry measurements (those most sensitive to prior assumptions) with uniform priors on all parameters. Otherwise, these fits were identical. The resulting orbital parameters agree with those from the fits including the prescribed priors to better than $1\sigma$, suggesting insensitivity to our input priors. 

Our orbital fits made heavy use of literature astrometry, many of which had no reported errors. Our method for assigning or correcting errors assumed that all measurements have a common missing error term per source (Section~\ref{sec:litas}). It is more likely that errors depend on the separation and contrast ratio, as well as quantities that were not consistently reported, like weather, setup, and observational strategy. Further, this technique assumed an uncorrelated error term. In the case of an erroneous pixel scale or imperfectly aligned instrument, all measurements from a common instrument err in the same direction. In practice, it is difficult to correct for these effects without access to the actual images. The data suggest that this does not impact our results; the final $\chi^2_\nu$ values for the best-fit orbits shows no correlation with the fraction of astrometry from the literature versus our own measurements, and astrometry from our own measurements agrees well with the literature data. 

As an additional test, we tried refitting six binaries with the most literature data twice, first doubling the error term added to the literature points, then halving it. In all cases, overall parameters and errors did not change significantly (although the final $\chi^2_\nu$ values changed). The main reason for this is that our measurements (particularly those from NIRC2) are far more precise and hence dictate the final solution, even in cases where most of the individual measurements are from the literature. In the case of halving errors, the MCMC landed on a similar solution, but with smaller parameter uncertainties and increased $\chi^2$ values. We conclude that our treatment of literature errors does not significantly impact the final orbital fits and that our assigned errors are reasonable.

\section{Stellar Parameters}\label{sec:params}

\subsection{Parallaxes}\label{sec:plx}

Parallaxes for 59 of the 62 systems were drawn from the literature, with the remaining three from MEarth astrometry (detailed below). To avoid complications from astrometric motion impacting the measured parallax, we used parallax determinations that accounted for centroid motion of the binary, or parallax measurements for a nearby associated companion or primary star where possible. Parallaxes from one of these two categories account for nearly half the sample (29 systems). For the other 30 systems with literature parallaxes, we adopted the most precise parallax available excluding values from {\it Gaia} DR2. While the most precise parallax is not necessarily the most accurate, the majority of systems had only one precise ($<5\%$) parallax in the literature.

Many studies used the weighted mean of all available parallaxes \citep[e.g.,][]{Winters2015} to reduce overall uncertainties. However, excluding the 29 cases above, there are only a few systems where the weighted mean would significantly improve the parallax. Gl 125, as a typical example, has a parallax determination of 63.45$\pm$1.94\,mas from \citet{van-Leeuwen:2007yq}, and 77.2$\pm$11.6\,mas from \citet{van-Altena1995}. Using the weighted mean of these two is 63.82$\pm$1.91\,mas, a negligible improvement from simply adopting the \citet{van-Leeuwen:2007yq} value. More importantly, the weighted mean is only applicable if the parallax measurements are independent. For binaries, the parallax astrometry may be sampling the same systematics due to centroid motion of the unresolved binary. 

For 22 of the systems, we drew parallaxes from the new reduction of Hipparcos data \citep{van-Leeuwen:2007yq}. We used parallaxes from \citet{Dupuy2017} for the seven overlapping binaries, and from \citet{Benedict2016} for 13. For four systems we pulled parallaxes from the general catalogue of trigonometric parallaxes \citep{van-Altena1995}, and three were taken from the Tycho-Gaia astrometric solution \citep[TGAS or {\it Gaia} DR1,][]{gaiadr1}.

About half (29) of our targets do not have entries in the second data release of {\it Gaia} \citep[DR2, ][]{GaiaDr2,Gaia-Collaboration:2018aa}. They were likely excluded because centroid shifts from orbital motion prevented a five-parameter (single-star) solution (a requirement to be included in DR2). We also found significant differences between the {\it Gaia} DR2 values and earlier measurements (including from TGAS) even when measurements were available. Many wide triples or higher-order systems in {\it Gaia} DR2 (where the wider star is easily resolved) have significantly different parallaxes reported for each set of stars. For example, GJ 2069AC has a {\it Gaia} parallax of 60.237$\pm$0.080\,mas, while GJ 2069BD has a {\it Gaia} parallax of 62.02$\pm$0.21\,mas, a difference of 1.8\,mas (7.9$\sigma$). While both AC and BD components are binaries, GJ 2069AC is an eclipsing binary, and too tight to be have detectable astrometric motion. Orbital motion in GJ 2069BD is likely impacting the parallax measurement or uncertainties, an issue that should be resolved in future {\it Gaia} data releases that will include fits for orbital motion. We found no such issues with our other parallax sources.

We adopted {\it Gaia} DR2 parallaxes for five systems, GJ 1245AC, GJ 277AC, Gl 570BC, Gl 667AB, and HIP 111685AB. In each case we used the parallax of their wider common-proper-motion companion. The wider associated stars are not known to harbor another unresolved star, and hence they should not be impacted by the same binarity issue. Gl 695BC, GJ 2005BC, Gl 22AC, and 2M1047+40 also have nearby associated stars. However, GJ 2005A has no entry in {\it Gaia} DR2, the {\it HST} parallax for Gl 695BC is more precise than the {\it Gaia} DR2 value for Gl 695A, GJ 22B does not pass the cuts on {\it Gaia} astrometry suggested in \citet{GaiaDr2} and \citet{Dr2_HR}, and the wide companion to 2M1047+40 is itself a tight binary \citep[LP 213-67AB,][]{Dupuy2017} with a large reported excess astrometric noise in {\it Gaia} \citep[a sign of binarity, ][]{2018RNAAS...2...20E}.

For three systems we derived new parallaxes using MEarth astrometry \citep{Nutzman2008}. Updated parallaxes were measured following the procedure from \citet{2014ApJ...784..156D}. The only difference was that we used $\simeq$two additional years of data, which helps average out systematic errors arising from centroid motion due to the binary orbit, and significantly reduces the overall uncertainties. 

The remaining five systems had parallaxes from a range of other literature sources, each containing just one system in our sample. All adopted parallaxes and references are listed in Table~\ref{tab:sample}.

\subsection{Metallicity}\label{sec:feh}

We estimated [Fe/H] using our SpeX spectra and the empirical relations from \citet{Mann2013a} for K5-M6 dwarfs, and \citet{Mann2014} for M6-M9 dwarfs. These relations  are based on the strength of atomic lines (primarily Na, Ca, and K features) in the optical or NIR \citep[e.g.,][]{RojasAyala:2010,Terrien:2012lr}, empirically calibrated using wide binaries containing a solar-type primary and an M-dwarf companion \citep[e.g.,][]{Bonfils:2005,Johnson2009,Neves2012}. The calibrations were based on the assumption that components of such binaries have similar or identical metallicities \citep[e.g.,][]{2015ApJ...801L..10T}. Similar methods have been used extensively to assign metallicities across the M dwarf sequence \citep[e.g.,][]{Terrien2015,Muirhead2015,Dressing2017,2018ApJ...853...30V,2018ApJ...854..145M}. Final adopted [Fe/H] values are given in Table~\ref{tab:sample}. Errors account for Poisson noise in the spectrum, but because of the relatively high SNR of the spectra, final errors on [Fe/H] are generally dominated by the uncertainties in the calibration itself, conservatively estimated to be 0.08~dex \citep{Mann2013a,Mann2014}. However, we estimated that we can measure relative [Fe/H] values (one M dwarf compared to another) to 0.04~dex. 

For all but two systems (Gl 65 and HD 239960), our NIR spectra are for the combined flux of the binary components. \citet{Mann2014} explored the issue of measuring metallicities of binaries with unresolved data by combining spectra of single-stars with equal metallicities and reapplying the same calibration. The bias introduced is negligible ($\lesssim0.02$~dex) when compared to overall uncertainties. The additional scatter is smaller than the measurement uncertainties, and can be explained entirely by Poisson noise introduced in the addition of component spectra. This may be more complicated for nearly or marginally resolved systems, where the narrow slit (0.3\arcsec) is preferentially including light from one star. However, repeating the tests of \citet{Mann2014} and adding a random flux weighting to the fainter star produced only a small increase in the uncertainties (0.01-0.03\,dex).

Two systems (2M2140+16, and 2M2206-20) have SpeX spectra taken with a wider slit, yielding lower spectral resolution. The bands in \citet{Mann2014} are defined using a homogeneous dataset taken with the narrow (0.3\arcsec) slit, so this difference may impact the derived [Fe/H]. We tested this by convolving a set of single-star SpeX spectra taken with the 0.3\arcsec\ slit with a Gaussian to put them at the appropriate lower resolution. The median of the derived [Fe/H] values changed by $<0.01$~dex, but the change varies between targets. Based on the resulting scatter, we estimate the errors on [Fe/H] from the lower-resolution spectra to be 0.12~dex on a Solar scale and 0.08~dex on a relative scale. These systems are marked separately in Table~\ref{tab:sample}. 

Two of the systems in our sample are L dwarfs (2M0746+20 and 2M1017+13). These are most likely above the hydrogen-burning limit, and hence were included in our analysis. However, the \citet{Mann2014} method contained no L dwarf calibrators. Our derived [Fe/H] were extrapolations of the \citet{Mann2014} calibration. The \citet{Mann2014} calibration has only a weak dependence on spectral type, but we still advise treating the assigned values with skepticism until an L dwarf calibration becomes available. 

Three targets (Gl 792.1, Gl 765.2, and Gl 667) are too warm (earlier than K5) for the calibration of \citet{Mann2013a}. For Gl 667, we adopted the [Fe/H] from \citet{2014ApJ...791...54G} for the associated M dwarf companion Gl 667C. [Fe/H] measurements from \citet{2014ApJ...791...54G} are determined in the same way as applied to other targets as explained above. For the other two, we took [Fe/H] values from \citet{2011A&A...530A.138C} and \citet{2010A&ARv..18...67T}, respectively. These [Fe/H] measurements are not necessarily on the same scale as those from \citet{Mann2013a}, which are calibrated against abundances of Sun-like stars from \citet{2015ApJ...805..126B,2016ApJS..225...32B}. Given reported variations in [Fe/H] for these stars, as well as [Fe/H] determination differences \citep{2014AJ....148...54H,2016ApJS..226....4H} we adopted conservative 0.08~dex uncertainties on both systems. For the other target lacking a SpeX spectrum (Gl 54), we derived [Fe/H] using the optical calibration of \citet{Mann2013a} and a moderate-resolution optical spectrum taken from \citet{Gaidos2014}.

\subsection{$K_S$-magnitudes}\label{sec:mags}

To determine $K_S$ magnitudes for each component, we required both unresolved (total) $K_S$ for each system and the contrast ($\Delta K_S$) for each component. We adopted unresolved $K_S$ magnitudes from the Two Micron All Sky Survey \citep[2MASS,][]{Skrutskie2006}. Some of the brightest stars in our sample are near or beyond saturation in 2MASS. For these targets we recalculated $K_S$ magnitudes using available optical and NIR spectra, following the method of \citet{Mann2015a} and \citet{Mann2015b}, using available optical spectra from \citet{Gaidos2014}. Synthetic magnitudes were broadly consistent (mean difference of 0.003$\pm$0.002\,mag) with 2MASS $K_S$ magnitudes (and at similar precision) for fainter targets ($K_S>7$). We only updated $K_S$ magnitudes for bright systems where our synthetic photometry differed from the 2MASS value by more than 2$\sigma$ or the 2MASS photometry was saturated (five systems). We mark these systems in Table~\ref{tab:sample}. 

Reddening and extinction are expected to be $\simeq$0 for all targets, as the most distant system is at 35\,pc, while the Local Bubble (a region of near-zero extinction) extends to $\simeq$70\,pc \citep{2009MNRAS.397.1286A}. Hence, we did not apply any extinction correction to the adopted $K_S$ values.

To compute $\Delta K_S$, we used component contrast measurements from our AO data (Section~\ref{sec:astrometry}). We utilize any contrast taken with a filter centered in the $K$ band, which included $K_s$, $K$, and $K'$ ($K_p$ or $K$-prime), as well as narrowband filters $Br\gamma$ and $K_c$ ($K_{\rm{cont}}$ or $K$-continuum). While all targets considered here had at least one measurement in one of these filters, none of the response functions used were a perfect match to 2MASS $K_S$. We transformed each $K$-band contrast into 2MASS contrasts ($\Delta K_S$) using corrections derived from flux-calibrated spectra as detailed in Appendix~\ref{sec:a1}. These corrections were generally small ($\lesssim0.1$ magnitudes).

\begin{figure}[t]
\begin{center}
\includegraphics[width=0.47\textwidth]{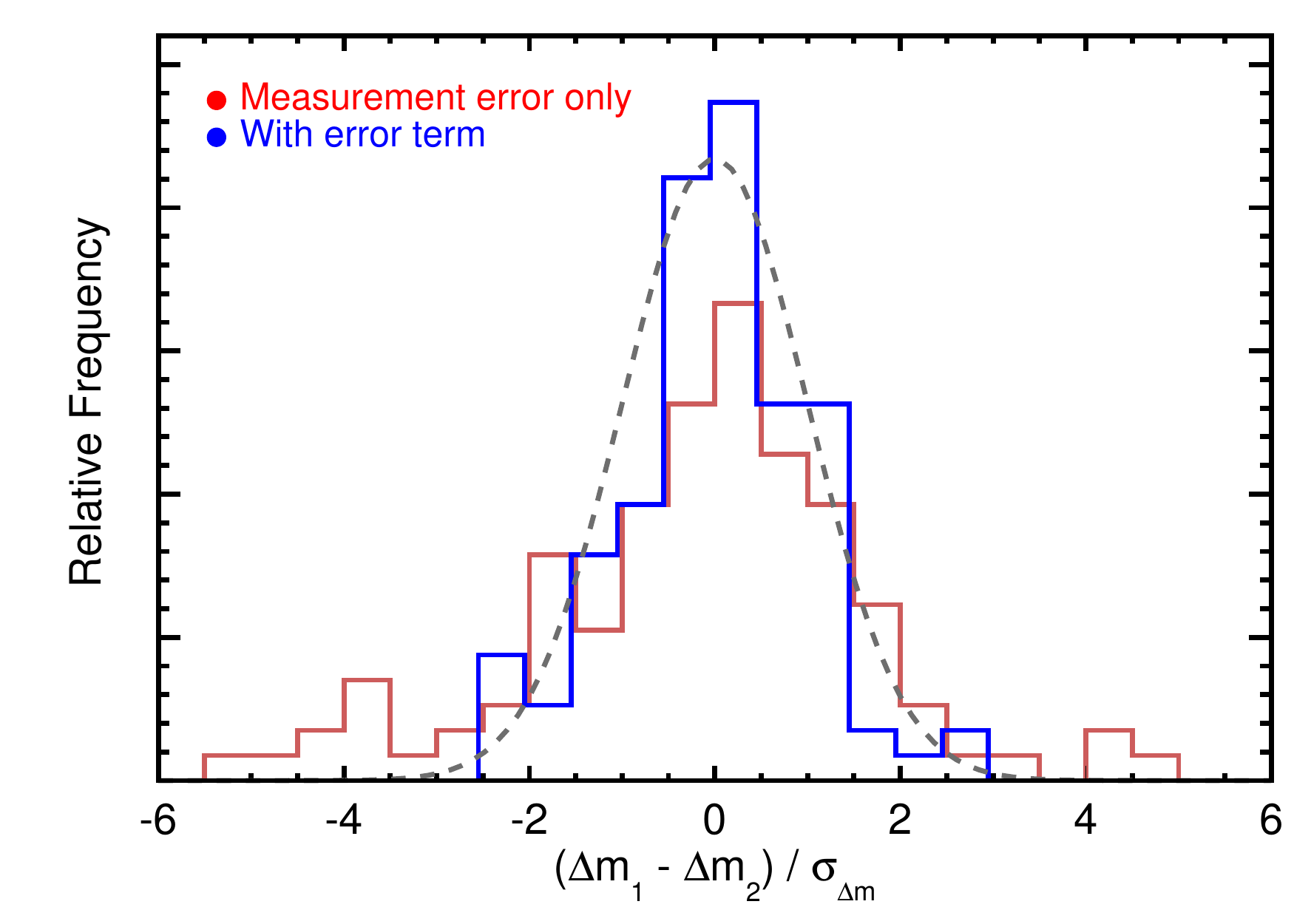}
\caption{Distribution of contrast ratio differences (in units of standard deviations) for data taken on the same target, and with the same filter and instrument, but in different nights. The red line is before adding the missing error term, while the blue line shows the distribution after adding this. The grey dashed line shows the expected Normal distribution. The histograms are offset slightly for clarity, although identical bins are used as input.}
\label{fig:magnitudes}
\end{center}
\end{figure}

After converting all contrast measurements to $\Delta K_S$, we combed multiple measurements using the robust weighted mean. Errors on contrasts for each dataset were taken to be the RMS in flux measurements among consecutive images. These errors may be underestimated because of imperfect PSF modeling, flat-fielding, uncorrected nonlinearities in the detector, as well as intrinsic variability of the star. To test for this, we compared $\Delta K_X$ measurements of the same star using the same filter and instrument but on different nights (Figure~\ref{fig:magnitudes}). The comparison suggested a missing error term of 0.016\,magnitudes for NIRC2, 0.02 for KIR and NaCo, and 0.03 for NIRI. We did not split this into separate error terms per filter; many filters do not have enough multi-epoch data on their own, and a single error term across all filters for a given instrument gave a reasonable fit. We included this term as an additional error term common to all measurements in our final computation of $\Delta K_S$.

For GJ 2005BC and Gl 900BC, the 2MASS PSF included flux from the A component. In both cases, we used our AO data to measure $\Delta K_S$ between all three components. The total $K_S$ magnitudes given in Table~\ref{tab:sample} already have the A components removed. 

\section{The Mass-Luminosity Relation}\label{sec:relation}

\subsection{Methodology}\label{sec:methods}

For main-sequence stars, the mass-luminosity relation traditionally takes the form
\begin{equation}\label{eqn:ml}
\frac{L_*}{L_\odot} = C \left(\frac{M_*}{M_\odot}\right)^\alpha,
\end{equation}
where $\alpha$ depends on the dominant energy transport mechanism (e.g., radiative versus convective) and internal structure of the star \citep{2004sipp.book.....H}.

\begin{figure*}[htb]
\begin{center}
\includegraphics[width=\textwidth]{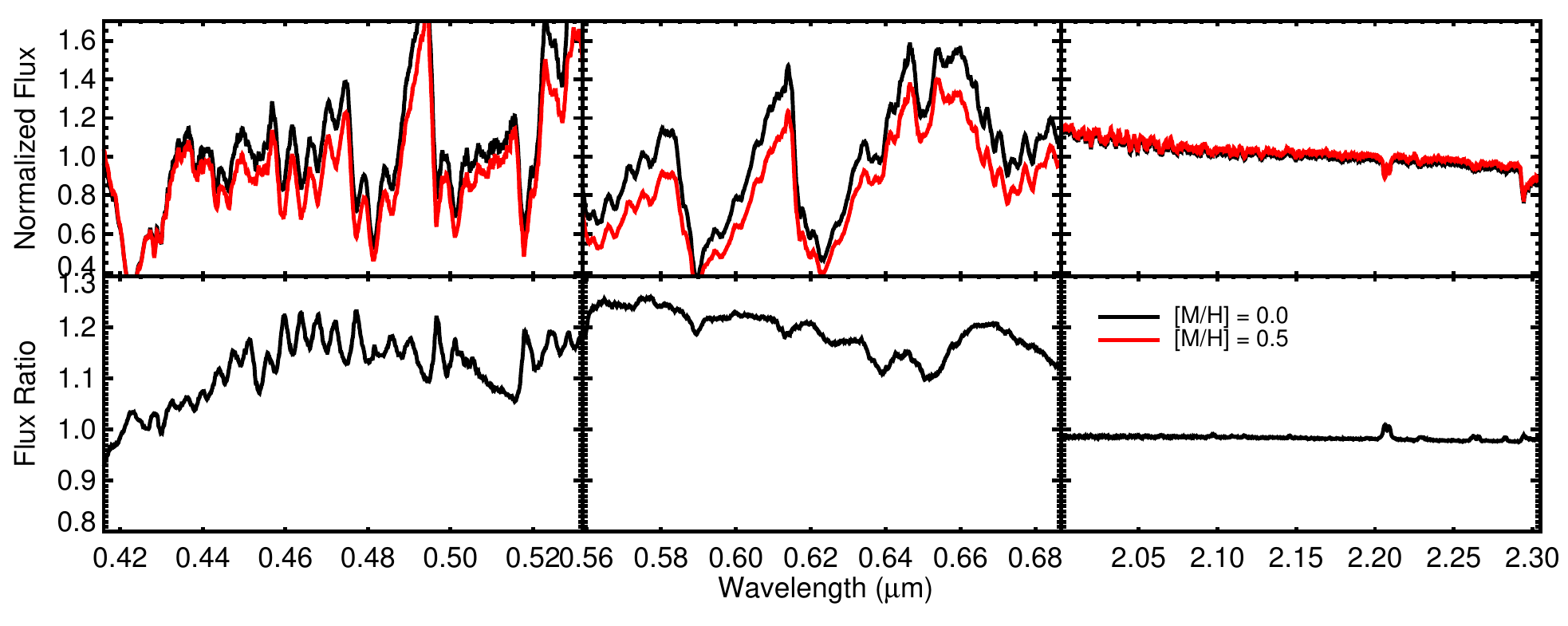}
\caption{Effect of changes in [M/H] on a model spectrum at \teff=3200\,K, $\log(g)=5$ in $g$- (left), $r$ (middle), and $K$-band (right). The top panel shows [M/H]=0 (black) and [M/H]=+0.5 (red) spectra from the CFIST BT-SETTL models \citep{Allard2012}. The bottom panel shows the ratio of the two, highlighting how small an effect [M/H] has in the $K$-band compared to optical regions. The one feature that stands out in the $K$-band is the Na doublet, which is a commonly used as a metallicity diagnostic for dwarfs \citep{RojasAyala:2010,Terrien:2012lr,Newton:2014}, and a gravity diagnostic for pre-main-sequence stars \citep[e.g.,][]{Schlieder2012}. }
\label{fig:metal}
\end{center}
\end{figure*}

We rewrite Equation~\ref{eqn:ml} in terms of \mks\ instead of $L_*$. Absolute magnitudes are more easily measured than overall luminosity, and avoid introducing errors from uncertain bolometric corrections or the need to take flux-calibrated spectra in order to measure the bolometric flux directly. Switching to \mks\ also mitigates effects of abundance differences. The $K$-band is heavily dominated by metal-insensitive CO and H$_2$O molecular absorption bands. Optical bands are dominated by much stronger molecular bands (e.g. TiO, CO, CaH, MgH, and VO) that are sensitive to both [Fe/H] and [$\alpha$/Fe] \citep[Figure~\ref{fig:metal}, also see][]{Woolf:2006uq,Lepine:2007fk,Mann2013a}. 

Our sample encompassed almost an order of magnitude in mass and hence a range of underlying stellar physics. No single power law is expected to fit over the full sequence. Instead, we assumed that $\alpha$ depends on \mks, which we approximated as a polynomial. This yields an \mmk\ relation of the form
\begin{equation}\label{eqn:mmk}
\log_{10} \left( \frac{M_*}{M_\odot} \right) = \sum_{i=0}^{n} a_i(M_{K_S}-zp)^i,
\end{equation}
where $a_i$ are the fit coefficients. The order of the fit ($n$) was determined using the Bayesian Information Criterion (BIC). The constant $zp$ is a zero-point (or anchor) magnitude, which is defined to be 7.5. This approximately corresponded to the logarithmic average mass of stars in our sample. The zero-point was effectively a coordinate shift, was not constrained by the fit, and did not impact the final result (a test fit with no zero-point gave consistent results). However, a value representative of the sample helped reduce the number of significant figures required for the $a_i$ values and improved fit convergence time. 

The true relation between between $\alpha$ and \mks\ is likely more complicated than Equation~\ref{eqn:mmk}, and may depend on other astrophysical parameters (e.g., activity). We explore the impact of using this model in Section~\ref{sec:test}, and the role of [Fe/H] on the relation in Section~\ref{sec:metal}. More complicated astrophysical effects are included as an additional error term (discussed below).

For the left-hand side of Equation~\ref{eqn:mmk}, we computed the total dynamical mass (\mdyn) for each binary. To this end, we combined the orbital period ($P$) and total angular semi-major axis ($\alpha_{\rm{ang}}$) from our fits to the orbital parameters (Section~\ref{sec:orbit}) with the parallax determinations ($\pi$, Section~\ref{sec:plx}) following a rewritten form of Kepler's laws:
\begin{equation}\label{eqn:mass}
M_{\rm{tot,dyn}} = M_{1}+M_{2} = \frac{(\alpha_{\rm{ang}}/\pi)^3}{P^2},
\end{equation}
where $P$ is in years, $\alpha_{\rm{ang}}$ and $\pi$ are in arcseconds, and \mtot\ is in solar masses. 

Equation~\ref{eqn:mass} provides only the total mass of a given binary system, as opposed to individual/component masses used in earlier work on the \mmk\ relation. Thus, when fitting for the $a_i$ coefficients we performed the comparison between the predicted (\mpred, from the \mmk\ relation) and dynamical total mass (\mtot, from Equation~\ref{eqn:mass}) for each system. For this, we rewrote Equation~\ref{eqn:mmk} to obtain an expression for the total mass predicted by the relation (\mpred):
\begin{eqnarray}\label{eqn:mmk2}
M_{\rm{tot,pre}} &&= 10^{\sum_{i=0}^{n} a_i(M_{K_S,2}-zp)^i} \nonumber \\
&& + 10^{\sum_{i=0}^{n} a_i(M_{K_S,1}-zp)^i}, 
\end{eqnarray}
where \mks$_{,1}$ and \mks$_{,2}$ are the primary and companion absolute $K_S$-band magnitudes derived from our measured $\Delta K_S$ and unresolved $K_S$ magnitudes (Section~\ref{sec:mags}). Note that while the \mmk\ relation is designed to make predictions for the masses of single stars from their \mks\ magnitudes, because we have resolved magnitudes we can combine predictions for the individual mass of each component into a prediction for \mpred, which can be compared directly to \mdyn. In this way we could solve for the $a_i$ coefficients in the \mmk\ relation without using individual masses or mass ratios. We also note that Equation~\ref{eqn:mmk2} could be modified for arbitrarily higher-order star systems, providing individual \mks\ magnitudes and the total mass of the system is known. 

We fit for the $a_i$ terms in Equation~\ref{eqn:mmk2} using the MCMC code {\tt emcee}, which accounts for the strong covariance between coefficients and provides a robust estimate of the uncertainties on the derived relation by exploring a wide range of allowed fits. Each coefficient was allowed to evolve under uniform priors without limits, and was initialized with the best-fit value derived from {\tt MPFIT}. We ran the MCMC chain with 500 walkers for $10^6$ steps after a burn-in of 50,000 steps. We ran separate MCMC chains testing values of $n$ (fit order) from three to seven. Initial $a_i$ values were taken from a least-squared fit using {\tt MPFIT}. 

Errors on \mdyn\ and \mpred\ values are correlated to each other owing to a common parallax. \mdyn\ estimates scale with the cube of the parallax (Equation~\ref{eqn:mass}). As a result, the parallax was a major source of uncertainty on \mdyn\ for many systems. Similarly, our component $K_S$ magnitudes had relatively small errors (0.016-0.06\,mag), so \mks\ errors tended to be dominated by the parallax. Because this correlation is usually along (parallel to) the direction of the \mmk\ relation (a greater distance increases both \mdyn\ and \mpred), it can tighten the fit if properly taken into account (when compared to assuming uncorrelated errors). 
 
We wanted the MCMC to explore the full `ellipse' representing the correlation between \mks\ and $M_*$ for each binary. To this end, we treated the distance of each system as a free parameter, letting each evolve under a prior from the observed parallaxes. As input, the MCMC was provided $a\arcsec^3/P^2$ and $K_S$ (with uncertainties) for each system, from which \mdyn\ and \mks\ were calculated using the common parallax. We converted \mks\ into a \mpred\ for each binary, which we compared to the corresponding \mdyn\ values within the likelihood function. Thus, the MCMC is forced to explore the range of possible parallaxes consistent with the input Gaussian uncertainties, while both \mdyn\ and \mpred\ shifted in a correlated way owing to changes in the (shared) parallax. Since the orbital information provides no direct constraint on the distances, this method effectively forced the MCMC to explore a distribution along the input prior. 

\begin{figure}[htb]
\begin{center}
\includegraphics[width=0.47\textwidth]{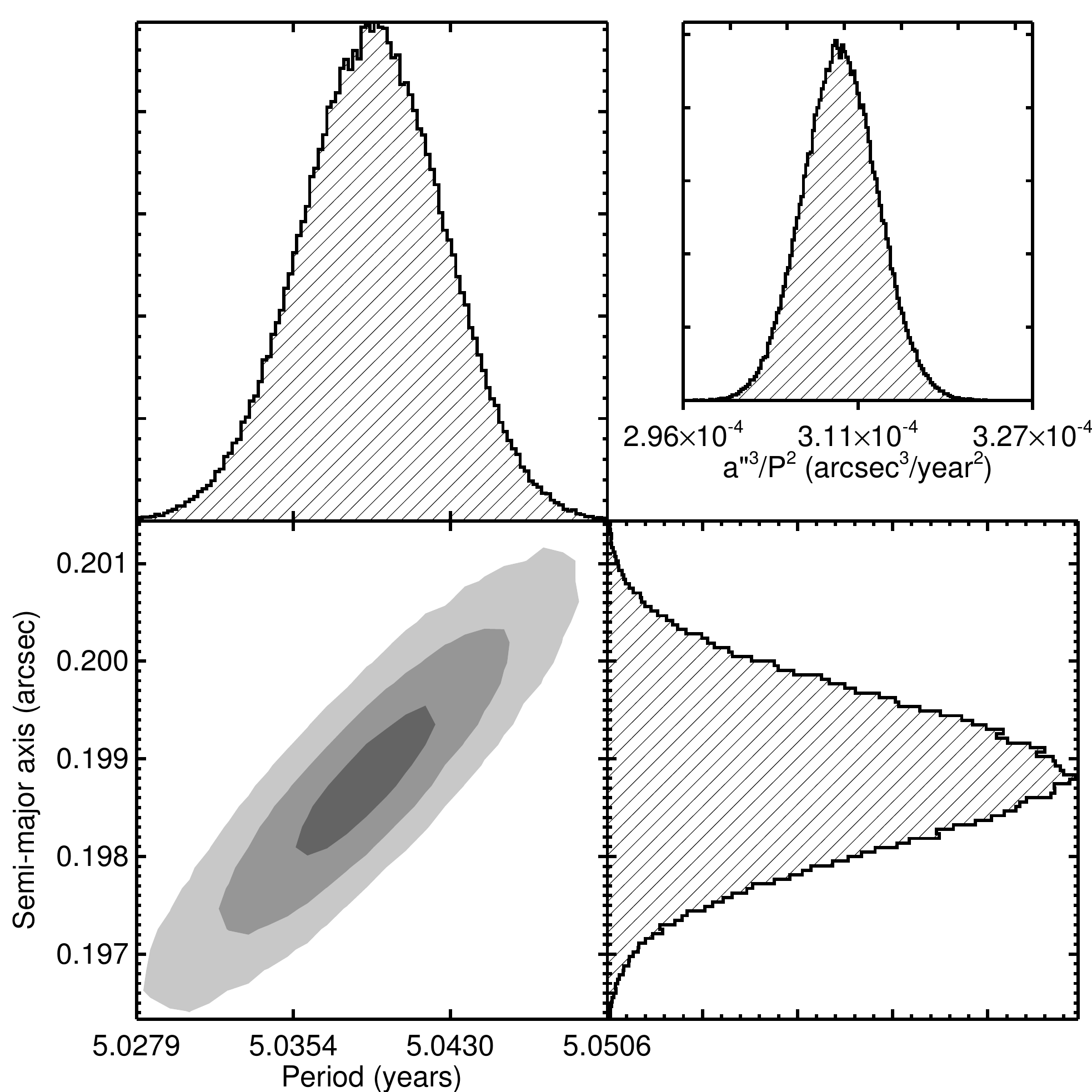}
\caption{Example joint posterior on semi-major axis and orbital period (bottom left) for the system LHS 6167. Grey regions show 1, 2, and 3$\sigma$ (from darkest to lightest) of the points. The histograms above and to the right show the one-dimensional distributions of each parameter. The parameter fed into the \mmk\ fit is $\alpha_{\rm{ang}}^3/P^2$ (in arcsec$^3$ per year$^2$), which is shown in the top right inset. }
\label{fig:correlated}
\end{center}
\end{figure}

For computational efficiency, we assumed Gaussian errors on $\alpha_{\rm{ang}}^3/P^2$. Although $\alpha_{\rm{ang}}$ and $P$ were often correlated and non-Gaussian, posteriors of $\alpha_{\rm{ang}}^3/P^2$ were all well described by a Gaussian (Figure~\ref{fig:correlated}). 

For main-sequence dwarfs at fixed metallicity, more massive stars should always be brighter. Thus, we required that the resulting fit have a negative derivative (higher \mks\ always gives a smaller \mpred) over the full range of input objects considered. We tested running without this constraint, and found similar results over most of the parameter range considered. The major difference was near the edges of the input sample. Without the negative derivative constraint, the fit could become double valued where there were few points.

 We specifically explore the role of [Fe/H] on the relation in Section~\ref{sec:metal}, but other astrophysical effects, such as detailed abundances and activity/rotation/magnetic fields, are not explicitly modeled in our fit, and hence may increase the overall scatter in the \mmk\ relation. We modeled these effects as an additional dimensionless parameter, $\sigma_e$. In addition to missing astrophysical variance, additional variation modeled by $\sigma_e$ could come from underestimated uncertainties in our input parallaxes (e.g., owing to uncorrected orbital motion on the astrometry) or orbital parameters. In either case, it is critical to include $\sigma_e$ as a free parameter to avoid underestimating the final uncertainties in the final relation. We implemented $\sigma_e$ as a fractional uncertainty in the total mass, added to the measurement uncertainties (from the orbit and parallax errors) in quadrature. We also tested included $\sigma_e$ as an additional uncertainty in the parallax (broadening the priors), or in the assigned $K$-band magnitudes. We discuss the differences between these implementations in Section~\ref{sec:res}.

To briefly summarize, each step of the MCMC chain included the following components: 
\begin{itemize}
\item We assumed an \mmk\ relation following Equation~\ref{eqn:mmk}. The first iteration used seed guesses for the $a_i$ coefficients from a least-squared fit. 
\item From the measured $\Delta K_S$, unresolved $K_S$ magnitudes, and input parallaxes, we computed \mks\ for each of the 124 stars, as well as uncertainties arising from errors in $\Delta K_S$ and $K_S$.
\item We applied the \mmk\ relation from the first step to compute 124 individual mass estimates.
\item We summed the component masses in each binary, providing predictions for the total masses (\mpred) of each of the 62 systems and corresponding uncertainties.
\item To handle any missing uncertainties or intrinsic variation in the \mmk\ relation, we inflated uncertainties on \mpred\ by a fraction, $\sigma_e$. $\sigma_e$ was treated as a free parameter and was initially set to $\simeq$0. 
\item Using the input orbital parameters ($\alpha_{\rm{ang}}^3/P^2$) for each system and the same parallaxes used for computing each \mks, we computed a total dynamical mass (\mdyn) for each binary system (see Equation~\ref{eqn:mass}). 
\item We calculated the likelihood, which is the $\chi^2$ difference between all predicted total masses (from the relation) and total dynamical masses multiplied by uniform priors for $a_i$ and $\sigma_e$ and the Gaussian priors on each parallax.
\item Based on the log-likelihood, {\tt emcee} adjusted the $a_i$ coefficients, $\sigma_e$, and parallaxes for all systems and repeated the process.
\end{itemize}
We emphasize that the comparison was done completely in total mass; the fitting method included no assumptions about the mass ratio, nor were mass ratios needed to fit the $a_i$ coefficients or $\sigma_e$. The fit was done using just $\pi$, $\alpha_{\rm{ang}}^3/P^2$, $K_S$, and $\Delta_{K_S}$ for each system. We address potential biases from using total masses (instead of individual masses) in Sections~\ref{sec:totalvsind} and \ref{sec:test}. 

\subsection{Results and Uncertainties}\label{sec:res}

\begin{figure*}[p]
\begin{center}
\includegraphics[width=0.9\textwidth]{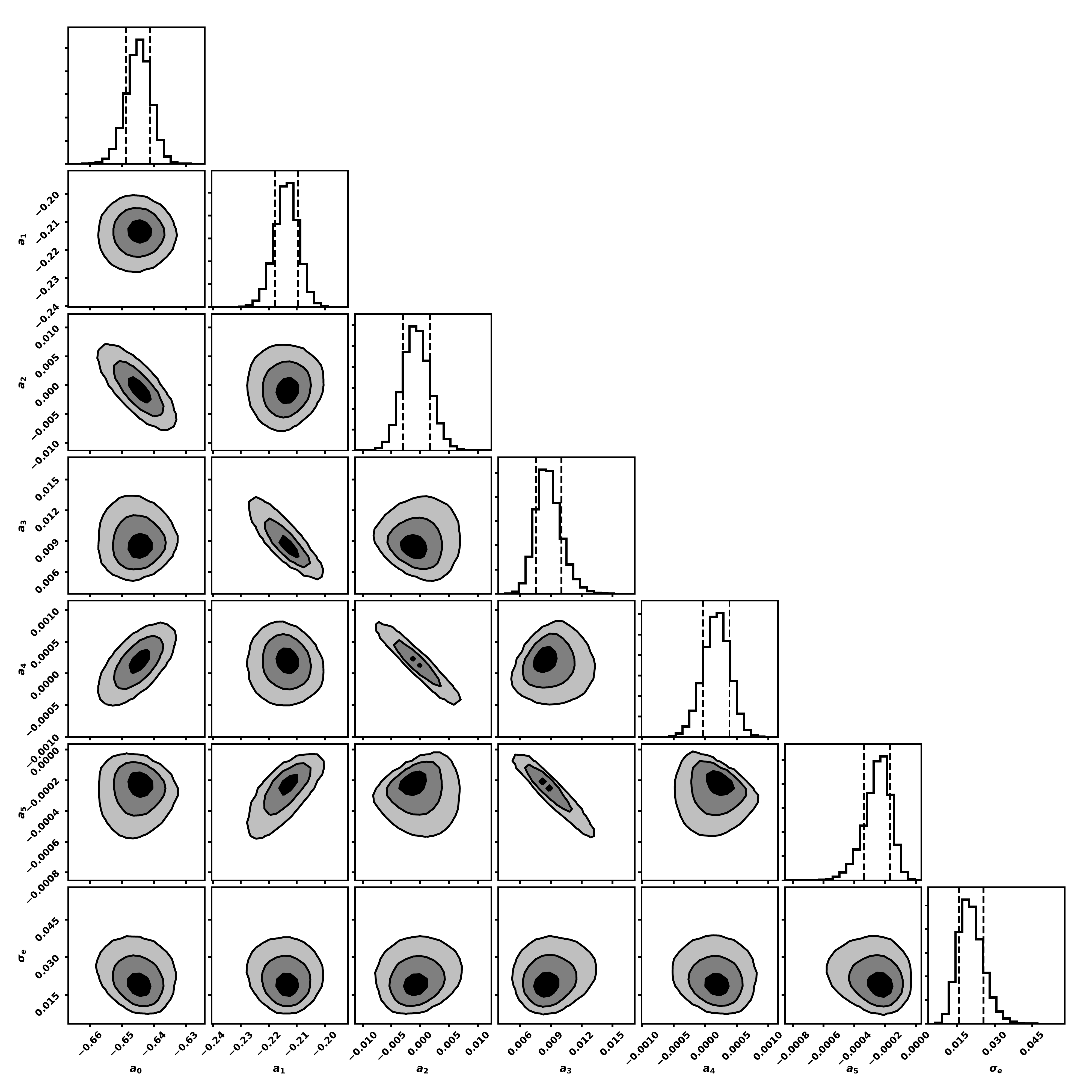}
\caption{Posterior projections for the $a_i$ values derived from our MCMC fit to Equation~\ref{eqn:mmk2} as well as the additional error term $\sigma_e$. Contours denote the 1, 2, and 3$\sigma$ confidence intervals, and the dashed lines in the histogram mark 1$\sigma$. The $\sigma_e$ parameter represents the fractional error in the total mass, added to account for intrinsic variation in the relation or underestimated uncertainties in the input masses. Figure was generated using {\tt corner.py} \citep{corner}.}
\label{fig:fitpost}
\end{center}
\end{figure*}

\begin{figure*}[htp]
\begin{center}
\includegraphics[width=\textwidth]{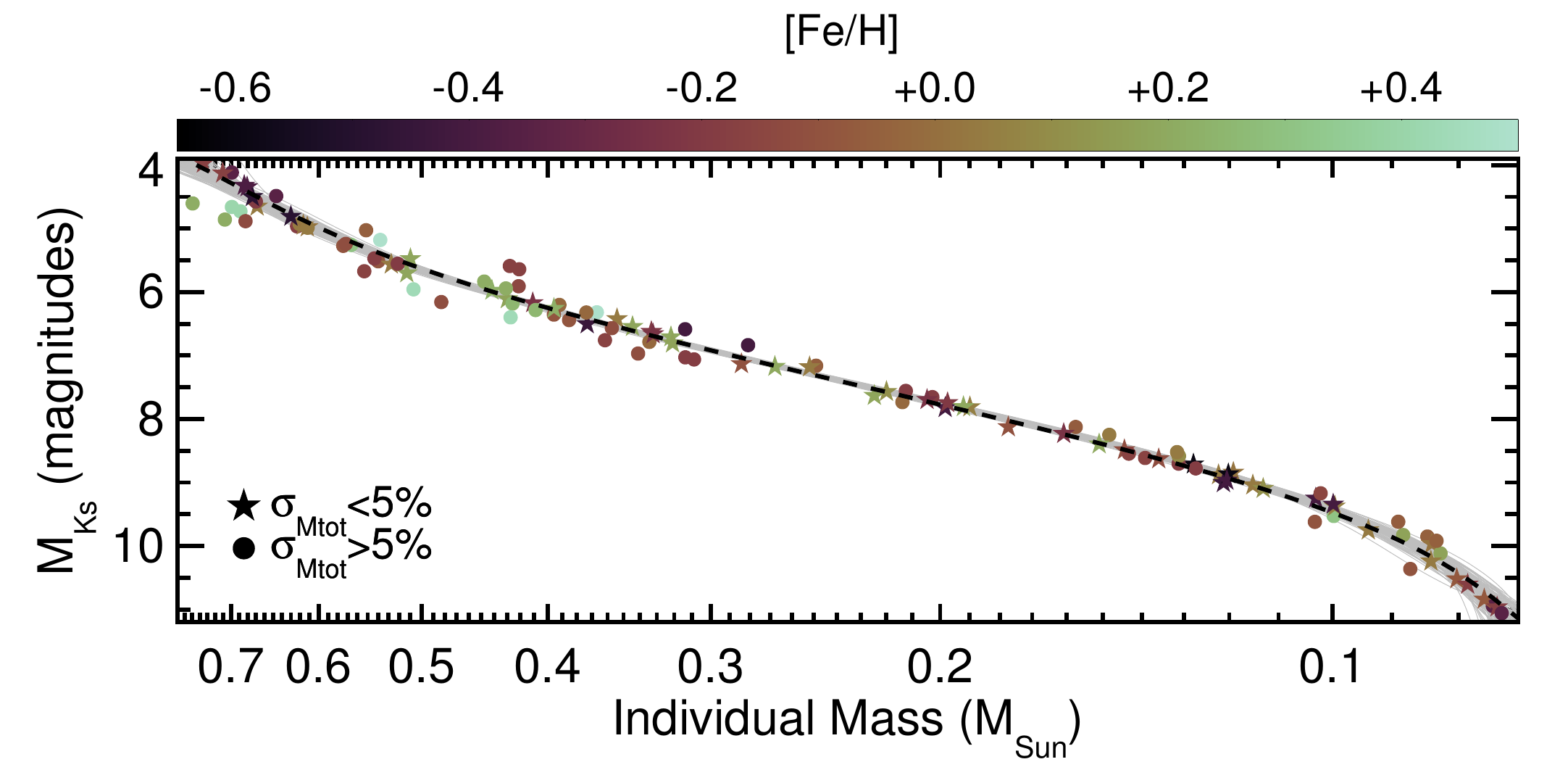}
\caption{Absolute $K_S$-band magnitude as a function of mass for targets in our sample. Stars indicate systems with total dynamical mass uncertainty $<5\%$ while those with larger uncertainties are shown as circles. All points are color-coded by their estimated metallicity. The black dashed line indicates the best-fit (highest-likelihood) relation from our MCMC analysis. To provide an estimate of the scatter in the relation as a function of mass, we show 100 randomly selected fits from the MCMC chain in grey. Note that our orbit fits only provide \mdyn; we used the mass ratios derived from the best-fit \mmk\ relation here, and this figure should be considered for display purposes only. Figure~\ref{fig:m_m} shows the comparison between \mpred\ (from our \mmk\ relation) and \mdyn\ (from Equation~\ref{eqn:mass}), which is more reflective of how the MCMC fit is done. }
\label{fig:relation}
\end{center}
\end{figure*}

\begin{figure}[ht]
\begin{center}
\includegraphics[width=0.48\textwidth]{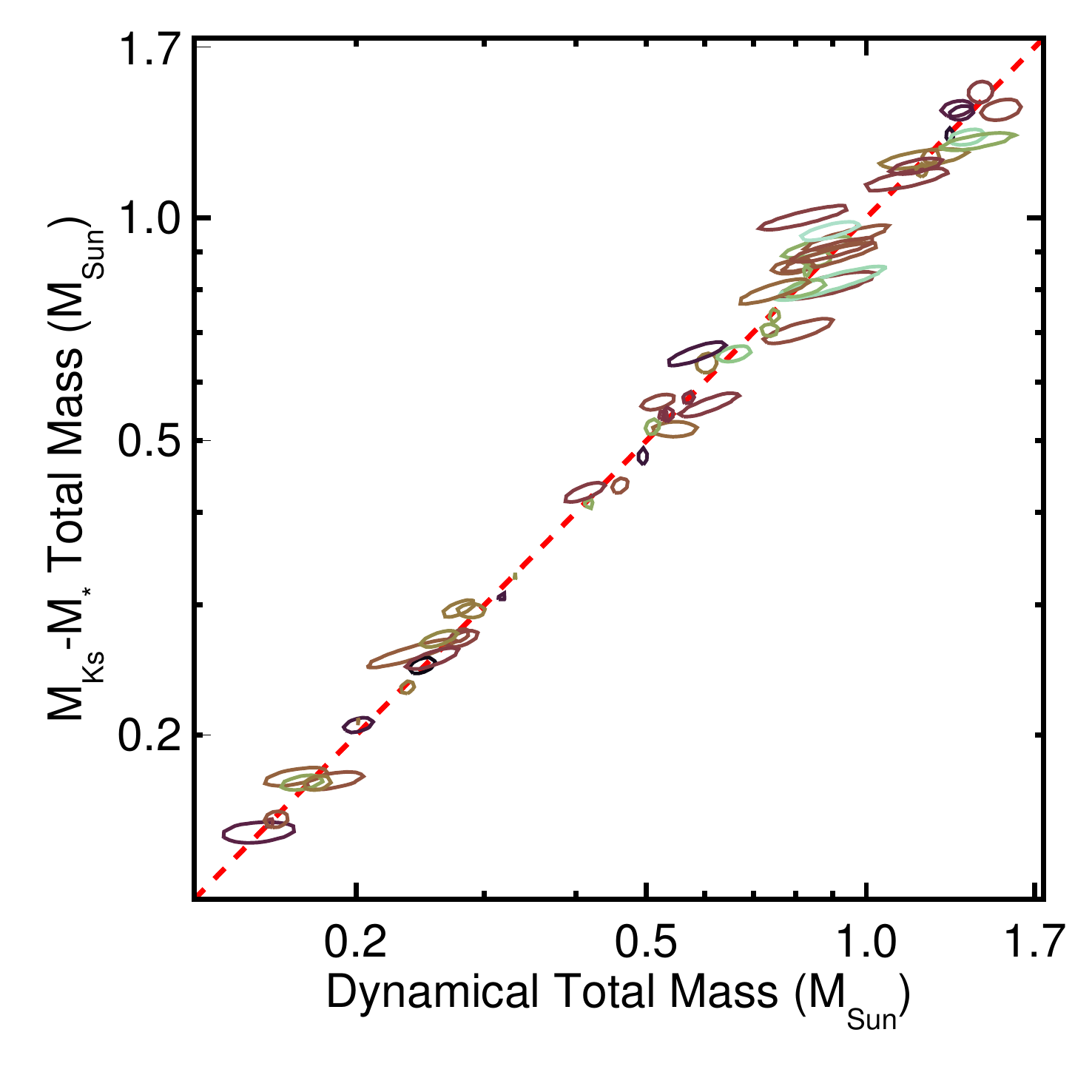}
\caption{Predicted total (system) mass from the \mmk\ relation (\mpred) as a function of the total dynamical mass determined from the orbital fits (\mdyn). Ellipses represent $\simeq1\sigma$ distribution of values for each point, accounting for parallax errors common to both the predicted and dynamical mass. Predicted masses (Y-axis) also account for errors arising from $\sigma_e$ and correlated uncertainties in the $a_i$ values. Color-coding by [Fe/H] matches that of Figure~\ref{fig:relation}.}
\label{fig:m_m}
\end{center}
\end{figure}

We show the resulting posteriors for the polynomial coefficients ($a_i$) in Figure~\ref{fig:fitpost}. The final fit was tightly constrained over the full sequence, which we show for individual masses in Figure~\ref{fig:relation}, and combined masses in Figure~\ref{fig:m_m}. For the former figure, we have assumed a mass ratio for each system from the \mmk\ relation, i.e., the ratio of the two predicted masses given their individual \mks\ (see Sections~\ref{sec:plx} and \ref{sec:mags} for more details). These ratios were not used in the fit and have strongly correlated uncertainties given a common parallax and total $K_S$. Thus, we only used these mass ratios for displaying the relation. The latter figure (Figure~\ref{fig:m_m}) is a more realistic representation of how the MCMC fit for \mmk\ was done (i.e., comparing \mdyn\ to \mpred).

Coefficients for powers of the same parity (even or odd) were strongly correlated to each other. This was expected, as a decrease in the slope (from linear) is best explained using an odd power, and increases with an even power. Coefficients for powers with even parity (with the exception of the $a_0$ term) were generally centered around zero. This also was expected in the context of the shape of the relation seen in Figure~\ref{fig:relation}, and our requirement that the mass always decrease with decreasing luminosity. A power with even parity will prefer to turn upward at low masses. We investigated this further by redoing the fit with no even powers ($a_0$ was retained), but exploring higher-order odd powers. The resulting fit was significantly worse, with a $\sigma_e$ value twice as large as fits with the same number of free parameters including even powers. The resulting fits also showed significant systematic deviations from the empirical data for $0.3M_\odot<M_*<0.5M_\odot$. We opted to include the even orders for all analyses despite their near-zero values. 

\begin{figure}[h]
\begin{center}
\includegraphics[width=0.48\textwidth]{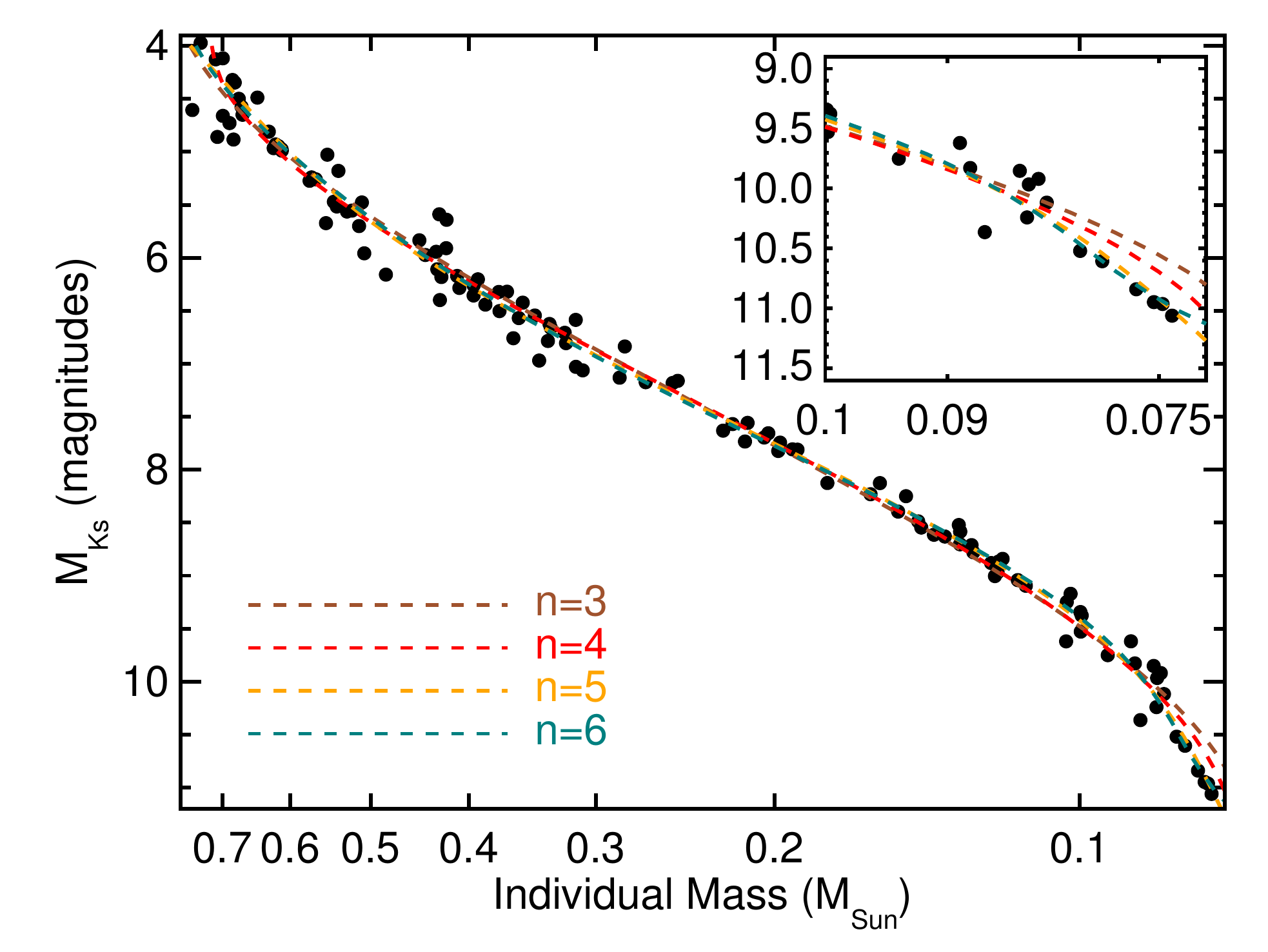}
\caption{Median of the mass posteriors at each \mks\ using fits of varying order ($n$) to Equation~\ref{eqn:mmk2} compared to the empirical values for the lowest-mass stars in our sample (black points). The inset shows objects with $M<0.1M_\odot$, where the disagreement between different orders is largest. A high-order ($n=\order$) is required to reproduce objects below 0.085$M_\odot$ as the relation becomes increasing nonlinear. The systematic offset seen between the low-mass sample and the best-fit relation can be seen in the coefficient posteriors as well as the best-fit relation (i.e., the distribution of fits are systematically high). }
\label{fig:order}
\end{center}
\end{figure}

We adopted $n=\order$ as the preferred solution based on both the BIC values and visual inspection of the residuals. Lower-order fits did a reasonable job fitting most of the sequence but poorly reproduced the masses of objects with $M_*<0.085M_\odot$. In this regime, the relation becomes increasingly nonlinear. The result is that lower-order fits systematically underestimated masses of the coolest objects in our sample (Figure~\ref{fig:order}) and overestimated $\sigma_e$ to compensate. Higher-order ($n>5$) fits explain the data well but were not justified statistically (e.g., marginal decrease in $\sigma_e$) and showed slope changes outside the calibration sample that are not expected by theoretical considerations. 

The three different implementations of $\sigma_e$ were broadly consistent with each other. For example, implementing $\sigma_e$ as broadening on the parallax priors yielded an extra error $\gtrsim$\nicefrac{1}{3} that of implementing it on the final mass. Since the parallax term is cubed in the total mass calculation (Equation~\ref{eqn:mass}) these are functionally equivalent (although parameter correlation forces a slightly larger error in the parallax). However, we found that implementing $\sigma_e$ as an error on the total mass best explained the data. Taking $\sigma_e$ as an error on the $K$-band magnitude led to a $\chi^2_\nu$ value too low when just considering stars below 0.25$M_\odot$, and too high for stars above 0.5$M_\odot$, while applying $\chi^2_\nu$ as an error on the total mass yielded $\chi^2_\nu$ closer to one over the whole mass range. Taking $\sigma_e$ as an error on $M_*$ is also easily implemented when applying the relation. 

We list the best-fit (highest-likelihood) coefficient values in Table~\ref{tab:coeff}, as well as the median value of $\sigma_e$ and BIC values for each fit. We also provided trimmed posteriors for each coefficient and $\sigma_e$ for our suggested relations.\footnote{\href{https://github.com/awmann/M_-M_K-/tree/master/resources}{https://github.com/awmann/M\_-M\_K-/tree/master/resources}} Fits using $n=4$ and $n=6$ are included in Table~\ref{tab:coeff} for reference, although we suggest only using the $n=\order$ relation. 

\begin{deluxetable*}{l c c c c c c c c c c }
\tablecaption{Best-fit Coefficients for Equations~\ref{eqn:mmk2} and \ref{eqn:mmk3}}
\tablehead{
\colhead{$n$} & \colhead{$a_0$} & \colhead{$a_1$} & \colhead{$a_2$} & \colhead{$a_3$} & \colhead{$a_4$} & \colhead{$a_5$} & \colhead{$a_6$} & \colhead{$f$} & \colhead{$\sigma_e$} & \colhead{BIC}
}
\startdata
4 &   -0.649 &   -0.202 & $    5.16\rm{x}10^{-3}$ & $    4.91\rm{x}10^{-3}$ & $   -3.54\rm{x}10^{-4}$ & \nodata & \nodata & \nodata &  0.025 & 90\\
5 &   -0.642 &   -0.208 & $   -8.43\rm{x}10^{-4}$ & $    7.87\rm{x}10^{-3}$ & $    1.42\rm{x}10^{-4}$ & $   -2.13\rm{x}10^{-4}$ & \nodata & \nodata &  0.020 & 86\\
6 &   -0.642 &   -0.209 & $   -5.11\rm{x}10^{-3}$ & $    7.25\rm{x}10^{-3}$ & $    1.45\rm{x}10^{-3}$ & $   -1.30\rm{x}10^{-4}$ & $   -7.61\rm{x}10^{-5}$ & \nodata &  0.020 & 89\\
\hline
4 &   -0.643 &   -0.199 & $    7.36\rm{x}10^{-4}$ & $    4.45\rm{x}10^{-3}$ & $   -7.69\rm{x}10^{-5}$ & \nodata & \nodata &  0.0076 &  0.026 & 92\\
5 &   -0.647 &   -0.207 & $   -6.53\rm{x}10^{-4}$ & $    7.13\rm{x}10^{-3}$ & $    1.84\rm{x}10^{-4}$ & $   -1.60\rm{x}10^{-4}$ & \nodata & -0.0035 &  0.021 & 88\\
6 &   -0.644 &   -0.221 & $   -5.51\rm{x}10^{-3}$ & $    1.13\rm{x}10^{-2}$ & $    1.18\rm{x}10^{-3}$ & $   -4.25\rm{x}10^{-4}$ & $   -4.71\rm{x}10^{-5}$ & -0.0010 &  0.020 & 93
\enddata
\label{tab:coeff}
\tablecomments{Fits follow the form: $\log_{10}\left( \frac{M_*}{M_\odot} \right) = \sum_{i=0}^{n} a_i(M_{K_S}-zp)^i$, where $zp\equiv7.5$. The $n=5$ fit is preferred, while the others are listed for reference. }
\end{deluxetable*}

To estimate the uncertainty in our \mmk\ relation, we computed the standard deviation in the derived masses for a fixed \mks\ across all MCMC fits, adding errors from $\sigma_e$ in quadrature. This accounts for (correlated) uncertainties in the $a_i$ coefficients in addition to intrinsic scatter in the \mmk\ relation as characterized by $\sigma_e$. We list uncertainties as a function of \mks\ in Table~\ref{tab:err}. Including all sources of uncertainty, the relation is precise to $\simeq2\%$ over most of the mass range, exceeding $3\%$ near the edges where there are fewer binaries to constrain the fit. 

\begin{deluxetable}{l l | l l l l l}
\tablecaption{Error in \mmk\ Relation}
\tablehead{
 \colhead{$M_{K_S}$} & \colhead{$M_*$} & \colhead{SpT$^a$} & \colhead{$\sigma_{M_*}^b$} & \colhead{$\sigma_{M_*}^b$}  \\
 \colhead{(mag)} & \colhead{$M_\odot$} & \colhead{} & \colhead{$M_\odot$} & \colhead{\%} 
}
\startdata
\multicolumn{7}{c}{No [Fe/H] term ($f=0$), 5th order} \\
\hline
 4.0 & 0.754 & K4.5 & 0.028 &  3.7\\
 4.5 & 0.6739 & K7.0 & 0.016 &  2.4\\
 5.0 & 0.6020 & M0.0 & 0.015 &  2.4\\
 5.5 & 0.5255 & M1.5 & 0.012 &  2.3\\
 6.0 & 0.4440 & M2.5 & 0.0099 &  2.2\\
 6.5 & 0.3630 & M3.0 & 0.0081 &  2.2\\
 7.0 & 0.2890 & M3.5 & 0.0064 &  2.2\\
 8.0 & 0.1776 & M4.5 & 0.0039 &  2.2\\
 8.5 & 0.1411 & M5.0 & 0.0032 &  2.2\\
 9.0 & 0.1153 & M6.0 & 0.0026 &  2.3\\
 9.5 & 0.0977 & M6.5 & 0.0023 &  2.4\\
10.0 & 0.0863 & M7.5 & 0.0022 &  2.6\\
10.5 & 0.0791 & M9.0 & 0.0021 &  2.6\\
11.0 & 0.0742 & L1.0 & 0.0024 &  3.2\\
\hline
\multicolumn{7}{c}{[Fe/H] term ($f$), 5th order} \\
\hline
 4.0 & 0.753 & K4.5 & 0.029 &  3.9\\
 4.5 & 0.6734 & K7.0 & 0.017 &  2.5\\
 5.0 & 0.6017 & M0.0 & 0.015 &  2.5\\
 5.5 & 0.5255 & M1.5 & 0.012 &  2.4\\
 6.0 & 0.4441 & M2.5 & 0.010 &  2.3\\
 6.5 & 0.3630 & M3.0 & 0.0082 &  2.3\\
 7.0 & 0.2889 & M3.5 & 0.0065 &  2.3\\
 8.0 & 0.1775 & M4.5 & 0.0040 &  2.3\\
 8.5 & 0.1411 & M5.0 & 0.0033 &  2.3\\
 9.0 & 0.1152 & M6.0 & 0.0027 &  2.4\\
 9.5 & 0.0977 & M6.5 & 0.0024 &  2.5\\
10.0 & 0.0863 & M7.5 & 0.0023 &  2.6\\
10.5 & 0.0791 & M9.0 & 0.0021 &  2.7\\
11.0 & 0.0742 & L1.0 & 0.0025 &  3.3\\
\enddata
\label{tab:err}
\tablecomments{This table assumes \mks\ (and [Fe/H]) are known perfectly. Total errors on $M_*$ should take into account errors in the measured parameters and the relation. }
\tablenotetext{$a$}{Spectral types are given for reference, but are extremely rough because of a significant dependence on metallicity and the spectral typing scale. It is not recommended to use this table as a means to compute \mks\ or $M_*$ from a spectral type or vice versa. }
\tablenotetext{$b$}{The uncertainty in the resulting $M_*$ at a given $M_{K_S}$ accounting intrinsic scatter as characterized by $\sigma_e$.}
\end{deluxetable}

Since $\sigma_e$ (intrinsic scatter in the relation) represents the major source of uncertainty over most of the mass range, we also tried to estimate the intrinsic scatter using a more traditional $\chi^2$ approach. For this test, we adopted the best-fit (highest-likelihood) coefficient parameters given in Table~\ref{tab:coeff} for $n=\order$. We applied the relation to each of the 124 component stars in our binary sample to compute their predicted individual masses, then summed the masses of each component in a system to obtain total masses (\mpred) for each of the 62 binary systems. We compared this to the dynamical total masses (\mdyn), computing a $\chi^2_\nu$ value over all 62 systems. Our $\chi^2_\nu$ computation accounted errors in $K_S$ magnitudes, parallaxes, and orbital parameters. The final $\chi^2_\nu$ from this comparison was 1.7. Adding a missing error term of 1.6\% in the output $M_*$ values from the relation yields $\chi^2_\nu \simeq 1$, somewhat smaller than our $\sigma_e$ estimates from the MCMC analysis. The difference arises because the $\chi^2$ method fails to fully account for correlations between \mks\ and \mdyn. Adopting a larger 5\% uncertainty yielded a $\chi^2_\nu=0.6$, which has a $<0.2\%$ chance of occurring given the number of degrees of freedom. This rules out a significantly larger intrinsic scatter in our fit and confirms that our 2-3\% uncertainties are consistent with the data. 

Some systems land $>10\%$ outside the relation in Figures~\ref{fig:relation} and \ref{fig:m_m}, however, all of these targets have similarly large ($>10\%$) uncertainties in \mdyn. If we restrict our sample to the 47 binaries with uncertainties on \mdyn$<10\%$, the RMS for the fit residuals is only 4.3\%. Similarly, the RMS is 2.6\% for the 28 binaries with $<5\%$ mass uncertainties and 2.0\% for the 13 systems with $<2\%$ total mass uncertainties, confirming the 2-3\% precision for the derived \mmk\ relation. 

While the values in Table~\ref{tab:err} can be used to estimate mass uncertainties arising from using our given \mmk\ relation, a more robust method would be to use the full fit posteriors. This can be important in regions of the fit where the posteriors are asymmetric around the best-fit (e.g., between 0.2$M_\odot$  and 0.3$M_\odot$ the best-fit sits below the median, see Figures~\ref{fig:relation}). To aid with using our relation and computing appropriate uncertainties, we included the fit posteriors and a simple code that provides output $M_*$ posteriors given a $K_S$, distance, and associated uncertainties.\footnote{\href{https://github.com/awmann/M_-M_K-}{https://github.com/awmann/M\_-M\_K-}} The program combines the scatter in the coefficients (accounting for correlations between $a_i$ values) with the median value of $\sigma_e$ to produce a realistic $M_*$ posterior including any asymmetry. We note that while the relation is precise to 2-3\%, because of parallax and $K_S$ magnitude uncertainties, the final uncertainties on $M_*$ are usually 3-4\% for stars with {\it Gaia}-precision distances.

\subsection{Testing for biases in the \mmk\ relation}\label{sec:totalvsind}
In Section~\ref{sec:methods} we outlined the methodology and mathematical framework for fitting the \mmk\ relation using \mtot\ instead of individual masses. Because the relation is meant to be used on single stars, it is useful to explore what potential biases are introduced when using \mtot\ to fit the \mmk\ relation, and especially how it might impact our overall uncertainties. 

To this end, we generated a set of synthetic binaries with component (individual) masses assigned according to an input \mmk\ relation, and tested how well we can recover the assumed relation using only \mtot\ and our framework from Section~\ref{sec:methods}. First, we generated a random set of 124 \mks\ values matching the overall distribution of our input sample, then assigned masses to each system assuming our best-fit \mmk\ relation for $n=\order$.  We converted this set of synthetic stars into synthetic binaries by splitting the sample and randomly matching a star from one half with one from the other half. 

To assign \mtot\ for each system, we summed the assigned masses for each component. We then randomly assigned each system a parallax between 30 and 200\,mas (matching our calibration sample), which enabled us to convert the assigned \mks\ values for each system to an unresolved $K_S$ and $\Delta K_S$ and \mtot\  into $\alpha_{\rm{ang}}^3/P^2$ by inverting Equation~\ref{eqn:mass}. At this phase, each binary has the required set of information that went into our MCMC framework, specifically $K_S$, $\Delta K_S$, $\alpha_{\rm{ang}}^3/P^2$, and $\pi$. 

To keep our input errors consistent with our binary calibration sample, we drew uncertainties for each parameter ($K_S$, $\Delta K_S$, $\alpha_{\rm{ang}}^3/P^2$, and $\pi$) and system by randomly sampling errors from our observed sample. Errors on $\alpha_{\rm{ang}}^3/P^2$ and $\pi$ were treated as fractional Gaussian uncertainties, while errors on $K_S$ and $\Delta K_S$ were taken as Gaussian errors in magnitudes. Synthetic binaries were sometimes assigned a large error in both $\alpha_{\rm{ang}}^3/P^2$ {\it and} $\pi$, yielding total mass uncertainties greater than 20\%. Such systems would have not passed our selection criteria (Section~\ref{sec:targets}); so, in these cases we redrew uncertainties for the system. 

All parameters were then randomly perturbed by their assigned uncertainties (assuming Gaussian errors). To replicate the effects of intrinsic scatter, we then perturbed the $\alpha_{\rm{ang}}^3/P^2$ by 2.0\% (median value of $\sigma_e$). Note that these perturbations changed the assigned \mtot\ and \mks; for the purposes of this test we can consider the original values the true \mtot\ and \mks\ (they follow the input \mmk\ perfectly), while the perturbed values are observed (imperfect measurements with realistic errors). 

We ran the 62 synthetic binaries through our MCMC framework, exactly as we did with the real sample (Section~\ref{sec:methods}). The only change was that we ran the MCMC chain for only 10,000 steps for computational efficiency. We repeated this process 100 times, each time generating a new binary sample and re-running our MCMC analysis. After each run, we saved the best-fit $a_i$ values and median $\sigma_e$. The shorter chain meant that not all fits passed our requirements for convergence, but we are mostly interested in the best-fit values and not a full exploration of the uncertainties that require a long chain. 

\begin{figure}[h]
\begin{center}
\includegraphics[width=0.48\textwidth]{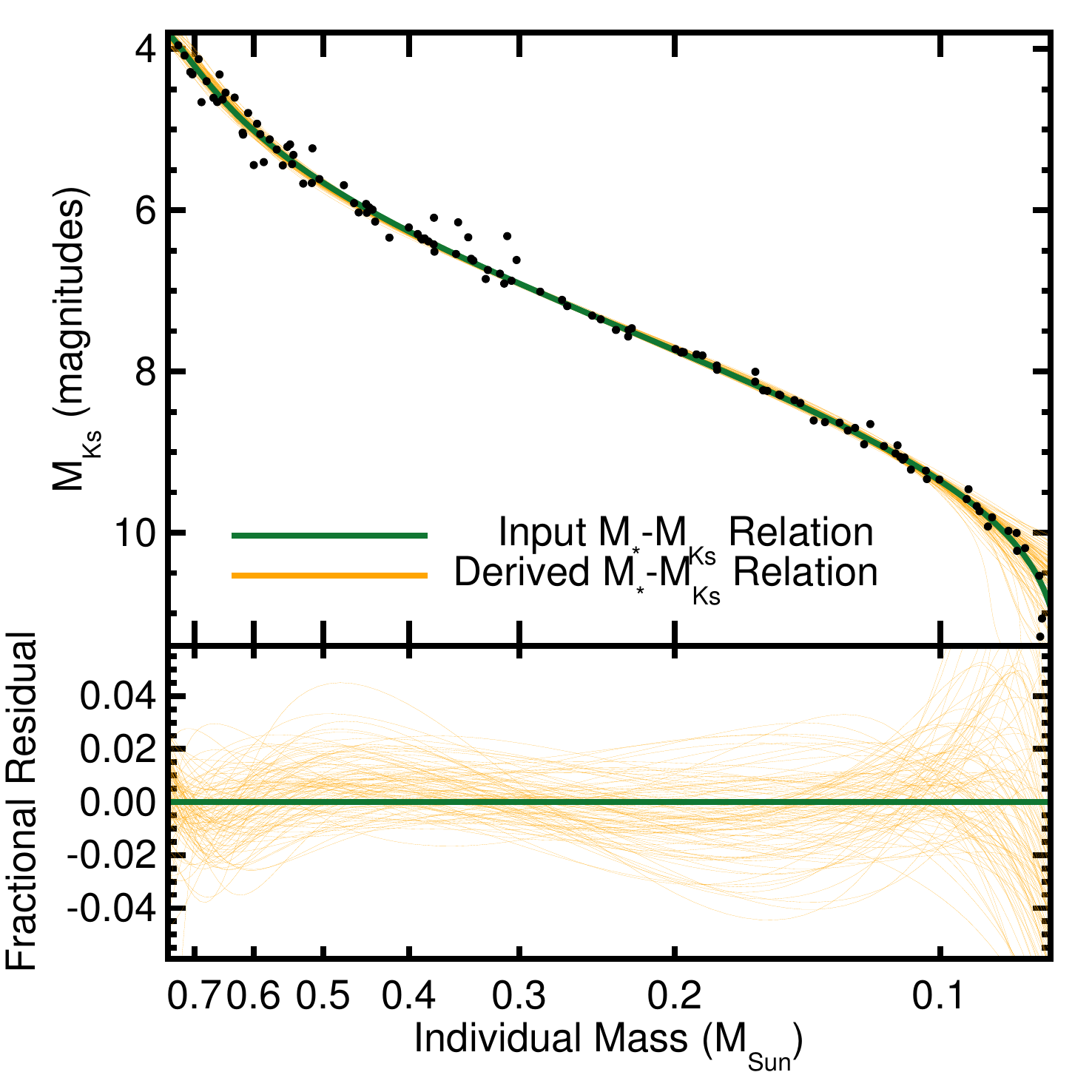}
\caption{Input \mmk\ relation (green) compared to the recovered relations (orange). The recovered \mmk\ fits are the result of running our MCMC fit on a set of synthetic binaries using only their total masses, but their individual components follow the input \mmk\ relation. The black points are one of the 100 randomly generated sets of synthetic binaries, which are shown for reference (each blue fit will use a different set of synthetic binaries). }
\label{fig:inject}
\end{center}
\end{figure}

We show the resulting distribution of \mmk\ relations using our synthetic binary sample alongside the input (true) relation in Figure~\ref{fig:inject}. The fits using the synthetic binaries followed input distribution closely in all cases. The range of solutions deviated from the input by $\simeq\sigma_e$, as expected, with the exception of $<0.1M_\odot$ where there were a wider range of possible solutions. However, the larger scatter is reflected in our MCMC fit to the calibration binaries (Figure~\ref{fig:relation}) and is well described by our adopted uncertainties ($\simeq3\%$) in the very low-mass regime. Since the fits using synthetic binaries used no information about the individual component masses, the consistency between the input and output \mmk\ relation in this test confirms our use of total masses does not significantly bias our result.

\begin{figure}[h]
\begin{center}
\includegraphics[width=0.48\textwidth]{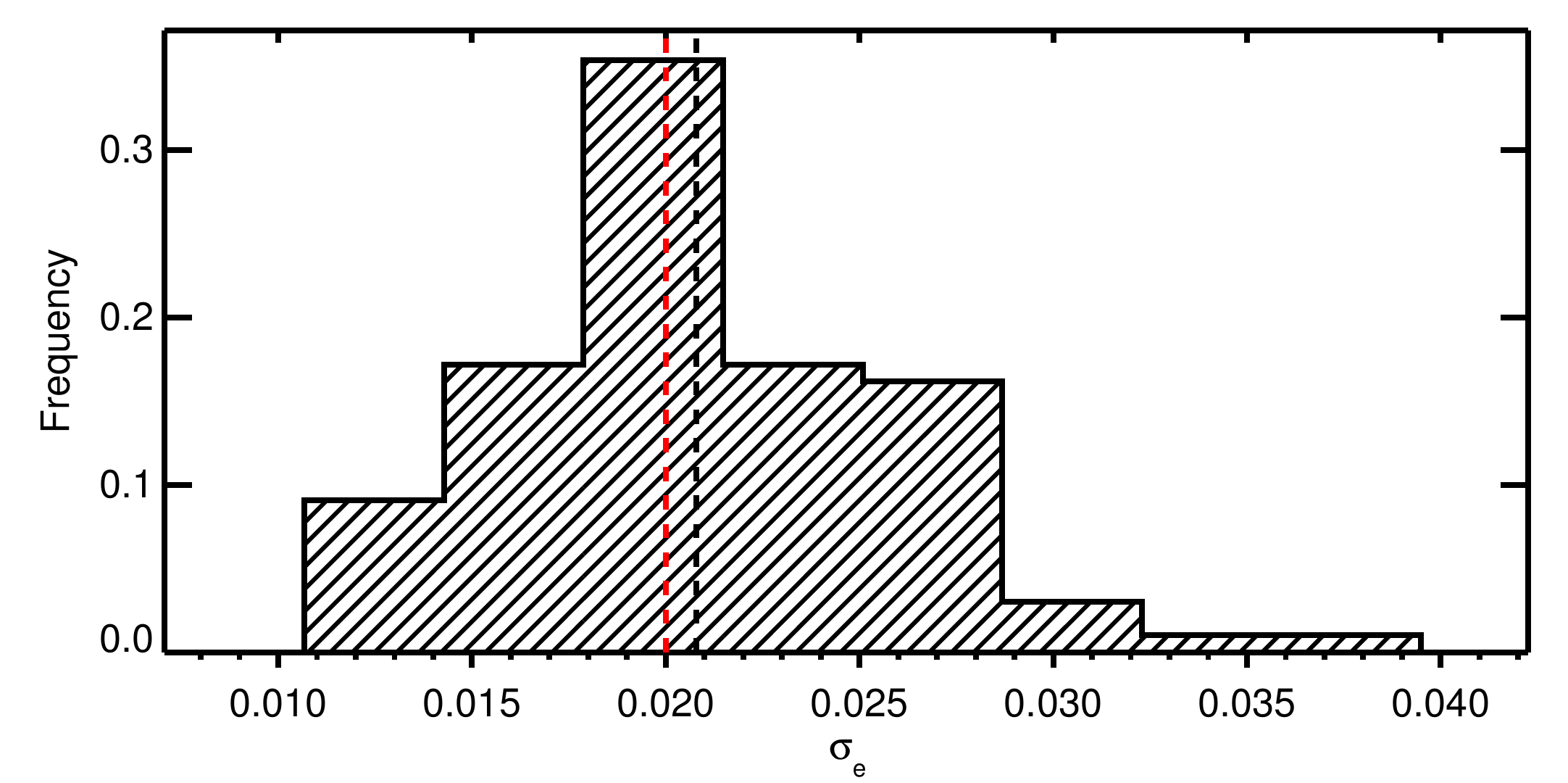}
\caption{Input $\sigma_e$ (red dashed line) compared to the distribution of values determined from 100 different sets of synthetic binaries (black histogram). The black dashed line indicates the median of the recovered $\sigma_e$ values. }
\label{fig:sigetest}
\end{center}
\end{figure}

We show how well we recovered the input $\sigma_e$ in Figure~\ref{fig:sigetest}. The median of recovered $\sigma_e$ values was slightly higher than our input value, although the two were consistent given the range of possible recovered values. Since $\sigma_e$ was the dominant source of uncertainty over most of the relation, this confirms that our overall errors are reasonable despite our use of total masses. 

The above test assumed that our functional form for the \mmk\ (Equation~\ref{eqn:mmk}) is perfect. However, the real \mmk\ relation is unknown; our assumption was that we could use an exponential to approximate this unknown relation with a smooth relation between $\alpha$ and \mks. To test the impact of these assumptions, we used the same method of generating synthetic binaries, but instead assign the true masses using a different formula. We then tested how well we could recover the assumed \mmk\ using the functional form given in Equation~\ref{eqn:mmk}. 

For this test, we assumed a fictitious \mmk\ relation that follows a piecewise function of the form
\[ M_* = 
\begin{cases}\label{eqn:piece}
      10^{-0.136*M_{\rm{K_S}}+0.36} & 3.5 < M_{\rm{K_S}} \le 5.0 \\
       10^{-0.16*M_{\rm{K_S}}+0.48}  & 5.0 < M_{\rm{K_S}} \le 8.0 \\
      10^{-0.11*M_{\rm{K_S}}+0.08}   & 8.0 < M_{\rm{K_S}} \le 11.5,
   \end{cases}
\]
where $M_*$ is given in Solar masses and \mks\ in magnitudes. Equation~\ref{eqn:piece} was partially motivated by the form of \citet{Hen1993}, adjusted to meet the boundary conditions of our sample. However, we highlight that the goal is not to assign a formula that is accurate, but rather to assign one that is plausible but {\it different} from the exponential we assumed when fitting the relation. A piecewise equation is useful for this purpose because the breaks owing to transitions at \mks=5 and \mks=8 might not be obvious when looking at the distribution of {\it total} masses, and are harder to approximate with a polynomial or exponential. A piecewise function is also a useful test of potential astrophysical breaks in the true relation, such as the fully-convective boundary. Thus, a piecewise can be taken as a worst-case but still plausible scenario for the \mmk\ relation, and hence represents a strong test of our method. 

\begin{figure}[h]
\begin{center}
\includegraphics[width=0.48\textwidth]{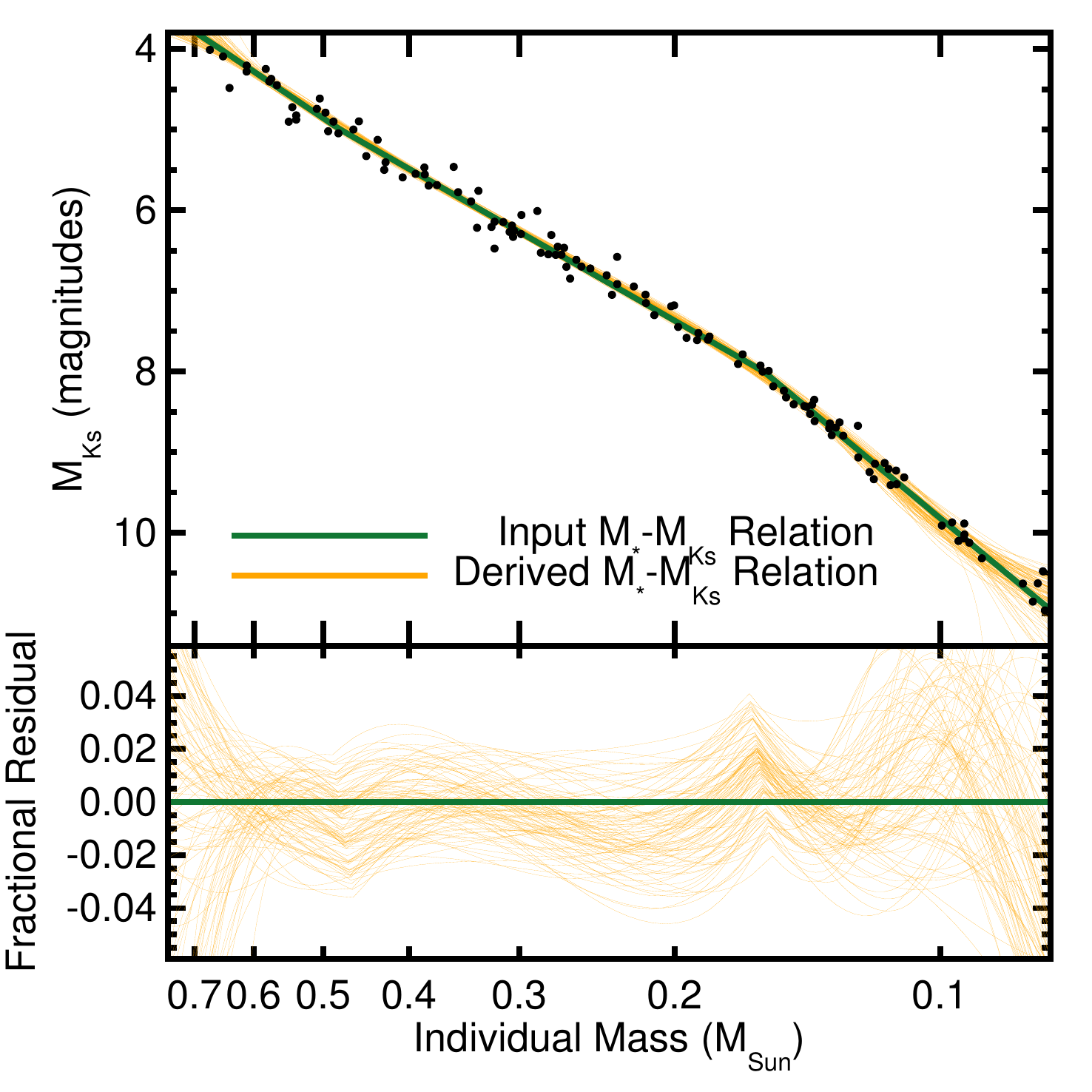}
\caption{Same as Figure~\ref{fig:inject}, but assuming a piecewise input \mmk\ relation (Equation~\ref{eqn:piece}). The fit (orange lines) assumes the \mmk\ follows Equation~\ref{eqn:mmk}. Note that the black points are a randomly selected set of synthetic binaries, and do not necessarily match any values from Figure~\ref{fig:sigetest} or the real binary calibration sample. }
\label{fig:piece}
\end{center}
\end{figure}

In Figure~\ref{fig:piece} we show the result of repeating our synthetic binary test using an input \mmk\ relation from Equation~\ref{eqn:piece}. The transitions between different sections of the piecewise equation are clear in the residuals, where the input and output relations show discontinuities and larger discrepancies. If there is a sharp astrophysical break (e.g., fully versus partially convective stars) the resulting relation would systematically miss masses right at the transition, but the relation would be unaffected just above or below the break. Further, the divergence is never significantly larger than our uncertainties. 

We can adjust the coefficients in Equation~\ref{eqn:piece} to make the breaks sharper. In these cases the deviations between our input and output increase at the breaks (although elsewhere the fit still follows the input relation). However, in these cases our derived value of $\sigma_e$ increases proportionately. Even if we assume physically unrealistic breaks, our final uncertainties would still capture these deviations. Thus, all tests confirm that our output relation and assigned uncertainties are reasonable and the use of total, instead of individual, masses has no significant impact. 

 \subsection{Testing the relation on individual dynamical masses}\label{sec:test}
 
Our tests with synthetic binaries in Section~\ref{sec:totalvsind} confirmed that our use of total masses and our assumed function form for the \mmk\ relation have no significant impact on our relation. However, it is still useful to perform a completely independent test our \mmk\ relation using dynamically measured individual masses. Further, a comparison between our relation and literature mass determinations would be useful to confirm or refute our assigned uncertainties, and may reveal the origin of $\sigma_e$. To this end, we utilized two samples of binaries with precisely determined {\it individual} masses from the literature, each of which are completely independent of our calibration sample: 1) astrometric binaries or triples with radial velocities or absolute astrometry not included in our binary sample and 2) M dwarf - M dwarf eclipsing binaries.
  
We drew (1) from \citet{Hen1993}, \citet{Delfosse2000}, and \citet{Benedict2016}, excluding those in our calibration sample (Section~\ref{sec:targets}). We added one target, GJ 2005A, for which the BC pair was analyzed in this paper, but \citet{Sef2008} provides a mass determination for the A component. For (2), we drew eclipsing binaries from the compilations of \citet{Hartman2011} and \citet{Parsons2018}, restricting the sample to double-lined eclipsing binaries, those with individual mass estimates better than 5\%, and systems with a parallax from the second {\it Gaia} data release passing the {\it Gaia} quality cuts given in Appendix C of \citet{GaiaDr2}. We removed 19e-3-08413AB because the distance is too large ($\gg500$\,pc) to assume zero reddening, and excluded systems from \citet{Kraus2011} because they have no reported flux or luminosity ratios, which are needed to estimate $\Delta K_S$ (detailed below). Lastly, we removed PTFEB132+19AB \citep{2017ApJ...845...72K} because it is young ($\simeq$650\,Myr). In total this gives us individual masses for 29 stars with which to test our \mmk\ relation. 

A significant advantage of eclipsing binaries is that we could adopt the much more precise parallaxes from {\it Gaia}, as they all have orbits that are too tight to show detectible centroid motion. For the astrometric binaries/triples, we drew parallaxes either from {\it Gaia} DR2 parallaxes of nearby companions (e.g., for Gl 644ABC we used the {\it Gaia} parallax form the wider companion Gl 643), or from sources that accounted for the high-order nature of the system. 

As with binaries analyzed in this paper, unresolved $K_S$ values were taken from 2MASS. We used adaptive optics data from VLT/NaCo to derive $\Delta K$ for Gl 866AC-B, and our own Keck/NIRC2 measurements for Gl 644A-BC and GJ 2005A-B-C (all three are resolved) just as was done for binaries analyzed in this paper. For other systems, including unresolved components of the triples (Gl 866A-C and Gl 644B-C), the literature only provided contrast ratios in optical bands. We converted these to $\Delta K_S$ using the synthetic and observed magnitudes given in \citet{Mann2015b}, following a procedure analogous to that outlined in \citep{2017ApJ...845...72K}. To briefly summarize, we found the combination of two single-star spectral templates that reproduced both the unresolved spectral energy distribution (Gaia and 2MASS photometry) and the measured contrast(s) from the literature reference (usually $V$, $R$, or $Kepler$), then computed a $\Delta K_S$ value for the best-fit template combination. This is similar to our conversion of $\Delta K_X$ to $\Delta K_S$ detailed in Appendix~\ref{sec:a1}. 

Correction from optical to NIR contrasts depends on metallicity \citep[e.g., ][]{Schlaufman2010} and precise metallicities of these systems were not known. Instead, we assumed all systems were $-0.6<$[Fe/H]$<+0.4$, and we adopted uncertainties that encompass the range of values owing to unknown metallicity. Errors introduced from these contrast conversions were extremely small for nearly equal mass systems (because $\Delta\,K_S\simeq0$), but became large ($>0.1$\,mag) for systems with mass ratios $\lesssim$0.6. 

Table~\ref{tab:ind} lists all systems with their adopted parallaxes, $M_*$ estimates, corresponding references, and our derived $K_S$ magnitudes.

\begin{deluxetable*}{l l l l l l l l l l}
\tablecaption{Targets with Individual Masses }
\tablehead{
 \colhead{Name} & \colhead{$M_*$} & \colhead{$K_S$} & \colhead{$\pi$} & \colhead{Type\tablenotemark{a}} & \colhead{$M_*$ Ref} & \colhead{$\pi$ Ref}   \\
 \colhead{} & \colhead{($M_\odot$)} & \colhead{(mag)} & \colhead{(mas)} & \colhead{} & \colhead{} & \colhead{} 
}
\startdata
HAT-TR-318-007A  & 0.448$\pm$0.011 & 11.509$\pm$ 0.041 &   8.345$\pm$  0.076 & EB &  1 &  2 \\
HAT-TR-318-007B  & 0.2721$\pm$0.0042 & 12.459$\pm$ 0.062 &   8.345$\pm$  0.076 & EB &  1 &  2 \\
NGTS J0522-2507A  & 0.1739$\pm$0.0015 & 11.798$\pm$ 0.055 &  18.378$\pm$  0.072 & EB &  3 &  2 \\
NGTS J0522-2507B  & 0.1742$\pm$0.0019 & 11.798$\pm$ 0.056 &  18.378$\pm$  0.072 & EB &  3 &  2 \\
HATS 551-027A     & 0.2440$\pm$0.0030 & 10.401$\pm$ 0.071 &  25.484$\pm$  0.061 & EB &  4 &  2 \\
HATS 551-027B     & 0.1790$\pm$0.0015 & 10.852$\pm$ 0.080 &  25.484$\pm$  0.061 & EB &  4 &  2 \\
1RXS J1547+4508A  & 0.2576$\pm$0.0085 &  8.967$\pm$ 0.023 &  45.120$\pm$  0.035 & EB &  5 &  2 \\
1RXS J1547+4508B  & 0.2585$\pm$0.0080 &  8.967$\pm$ 0.026 &  45.120$\pm$  0.035 & EB &  5 &  2 \\
Kepler-16A       & 0.6897$\pm$0.0035 &  9.060$\pm$ 0.042 &  13.289$\pm$  0.027 & EB &  6 &  2 \\
Kepler-16B       & 0.20255$\pm$0.00066 &  12.11$\pm$  0.23 &  13.289$\pm$  0.027 & EB &  6 &  2 \\
LSPM J1112+7626A & 0.3946$\pm$0.0023 & 10.180$\pm$ 0.060 &  17.616$\pm$  0.051 & EB &  7 &  2 \\
LSPM J1112+7626B & 0.2745$\pm$0.0012 & 10.910$\pm$ 0.090 &  17.616$\pm$  0.051 & EB &  7 &  2 \\
NSVS 01031772A    & 0.5428$\pm$0.0027 &  9.420$\pm$ 0.050 &  16.480$\pm$  0.030 & EB &  8 &  2 \\
NSVS 01031772B    & 0.4982$\pm$0.0025 &  9.650$\pm$ 0.060 &  16.480$\pm$  0.030 & EB &  8 &  2 \\
YY GemA          & 0.5992$\pm$0.0047 &  5.960$\pm$ 0.045 &  66.232$\pm$  0.051 & EB &  9 &  2 \\
YY GemB          & 0.5992$\pm$0.0047 &  6.010$\pm$ 0.051 &  66.232$\pm$  0.051 & EB &  9 &  2 \\
LSPM J0337+6910A & 0.375$\pm$0.016 &  9.470$\pm$ 0.071 &  26.907$\pm$  0.041 & EB & 10 &  2 \\
LSPM J0337+6910B & 0.280$\pm$0.015 & 10.048$\pm$ 0.093 &  26.907$\pm$  0.041 & EB & 10 &  2 \\
GU BooA          & 0.6160$\pm$0.0070 & 10.911$\pm$ 0.046 &   6.147$\pm$  0.016 & EB & 11 &  2 \\
GU BooB          & 0.6000$\pm$0.0060 & 11.041$\pm$ 0.061 &   6.147$\pm$  0.016 & EB & 11 &  2 \\
GJ 2069A         & 0.42940$\pm$0.00100 &  7.230$\pm$ 0.041 &  60.138$\pm$  0.092 & EB & 12 &  2 \\
GJ 2069C         & 0.3950$\pm$0.0018 &  7.490$\pm$ 0.054 &  60.138$\pm$  0.092 & EB & 12 &  2 \\
GJ 2005A         & 0.115$\pm$0.010 &  8.714$\pm$ 0.060 &  128.5$\pm$   1.5 & Astr & 13 & 14 \\
Gl 866B          & 0.1145$\pm$0.0012 &  6.593$\pm$ 0.041 &  293.60$\pm$   0.90 & Astr & 15 & 15 \\
Gl 866A          & 0.1187$\pm$0.0011 &  6.557$\pm$ 0.072 &  293.60$\pm$   0.90 & Astr & 15 & 15 \\
Gl 866C          & 0.09300$\pm$0.00080 &  7.127$\pm$ 0.080 &  293.60$\pm$   0.90 & Astr & 15 & 15 \\
Gl 644A          & 0.4155$\pm$0.0057 &  5.350$\pm$ 0.041 &  153.92$\pm$   0.13 & Astr & 15 &  2\tablenotemark{b}\\
Gl 644B          & 0.3466$\pm$0.0047 &  5.610$\pm$ 0.071 &  153.92$\pm$   0.13 & Astr & 15 &  2\tablenotemark{b}\\
Gl 644C          & 0.3143$\pm$0.0040 &  5.890$\pm$ 0.092 &  153.92$\pm$   0.13 & Astr & 15 &  2\tablenotemark{b}\\
\enddata
\tablecomments{References: 1 = \citet{Hartman2018}, 2 = \citet{GaiaDr2}, 3 = \citet{Casewell2018}, 4 = \citet{Zhou2015}, 5 = \citet{Hartman2011}, 6 = \citet{Doyle2011}, 7 = \citet{Irwin2011}, 8 = \citet{LopezMorales2006}, 9 = \citet{Torres2002}, 10 = \citet{Irwin2009}, 11 = \citet{Lopez2005}, 12 = \citet{Wilson2017}, 13 = \citet{Sef2008}, 14 = \citet{Benedict2016}, 15 = \citet{Sgr2000}}
\label{tab:ind}
\tablenotetext{a}{Eclipsing (EB) or astrometric (Astr) binary/triple. }
\tablenotetext{b}{Parallax from companion star Gl 643.}
\end{deluxetable*}

\begin{figure}[th]
\begin{center}
\includegraphics[width=0.47\textwidth]{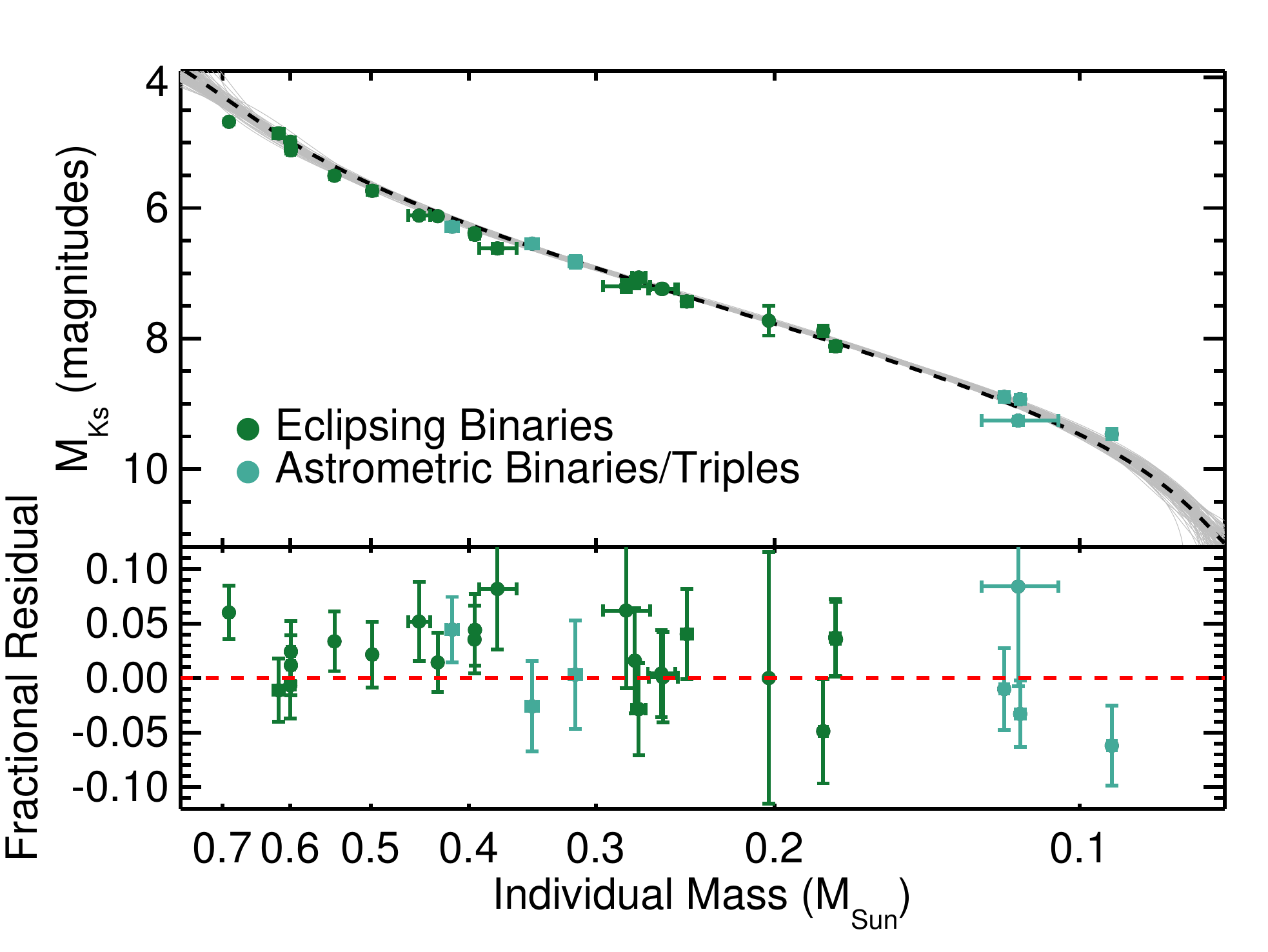}
\caption{Top: $M_*$ and \mks\ for M dwarfs with individual dynamical masses from the literature (points) compared to our derived \mmk\ relation (dashed line). The grey region shows 100 randomly selected fits from the MCMC. Bottom: fractional difference between the empirical mass and the mass predicted by our \mmk\ relation. Points are color-coded by type (eclipsing or astrometric binaries). Errors in the bottom panel account for uncertainty in the \mmk\ relation as well as measurement uncertainties in $K_S$ magnitudes, parallaxes, and the literature dynamical masses. }
\label{fig:ind}
\end{center}
\end{figure}

We show the \mks\ values versus literature dynamical masses for these systems in Figure~\ref{fig:ind} compared to the prediction from our \mmk\ relation. Our result follows the literature individual masses extremely well. To quantify this and test our previously estimated precision on the \mmk, we calculated a predicted mass for each star using the $n=5$ relation as given in Section~\ref{sec:res} and the \mks\ value from our adopted $K_S$ magnitudes and parallaxes. This is exactly as the procedure would be applied to single stars in the field. The output masses account for uncertainty in the relation (including intrinsic scatter characterized by $\sigma_e$) as well as uncertainties in $K_S$ magnitudes and measured parallaxes. 

Of the 29 stars, only one had a literature mass $>2\sigma$ off from the \mmk\ predicted value (Kepler-16A), and the $\chi^2_\nu$ of predicted and dynamical masses was 1.02 ($\chi^2=29.6$). The RMS of the residuals was 3.6\%, although this is driven primarily by the points with the largest errors in $M_*$ or $K_S$. If we restrict the sample to targets with 3\% uncertainties in $M_*$ the RMS is only 2.8\%, in agreement with our estimated uncertainties in the \mmk\ relation. 

Figure~\ref{fig:ind} also suggests a small ($\simeq$2\%) systematic offset, such that literature dynamical masses for eclipsing binaries are preferentially higher than those predicted from the \mmk\ relation. This difference is comparable in size to $\sigma_e$ and our overall precision and hence is within systematic uncertainties (further verified by a $\chi^2_\nu$ near unity). Assuming the offset is astrophysical, it is consistent with a scenario where low-mass eclipsing binaries are inflated compared to single stars owing to increased activity \citep[e.g.,][]{MacDonald2012, Feiden2013,Feiden2014a,Somers2017}. Higher activity levels may inhibit convection, increasing the radius and decreasing the fusion rate and hence overall luminosity at fixed mass. For activity levels expected in most low-mass EBs, \citet{Feiden2016} models suggest a difference of 1-3\% in $M_*$ for a fixed \mks\ over $0.1M_\odot<M_*<0.6M_\odot$, consistent with the offset seen here. 

\subsection{The role of metallicity}\label{sec:metal}

We explored the effects of [Fe/H] on the \mmk\ relation using version 1 of the Mesa Isochrones and Stellar Tracks \citep[MIST,][]{MIST0,MIST1} and an updated version of the Dartmouth Stellar Evolution Database \citep[DSEP,][]{Dotter2008}. The updates to DSEP have been previously detailed in \citet{Feiden2013,Feiden2014a}, and \citet{Muirhead2014}, with additional information on the updates for low-mass stars in \citet{Mann2015b}. MIST models use ATLAS/SYNTHE model atmospheres \citep{2004astro.ph..5087C} with updated TiO opacities for late-type stars that should improve performance. DSEP uses PHOENIX \cite{Hauschildt1999,1999ApJ...525..871H} models, and have been used widely for studies of late-type dwarfs \citep[e.g.,][]{Boyajian2012,Bell2015,Kesseli2018}. While other model grids \citep[e.g.,YaPSI, PARSEC, Lyon][]{Spada2013,Chen2014,BHAC15} show similar agreement with empirical studies of low-mass stars, we leave a more detailed comparison between the full range of model grids and our empirical masses for future analysis, and we restrict our model comparison here to just effects from metallicity. 

\begin{figure*}[htp]
\begin{center}
\includegraphics[width=0.47\textwidth]{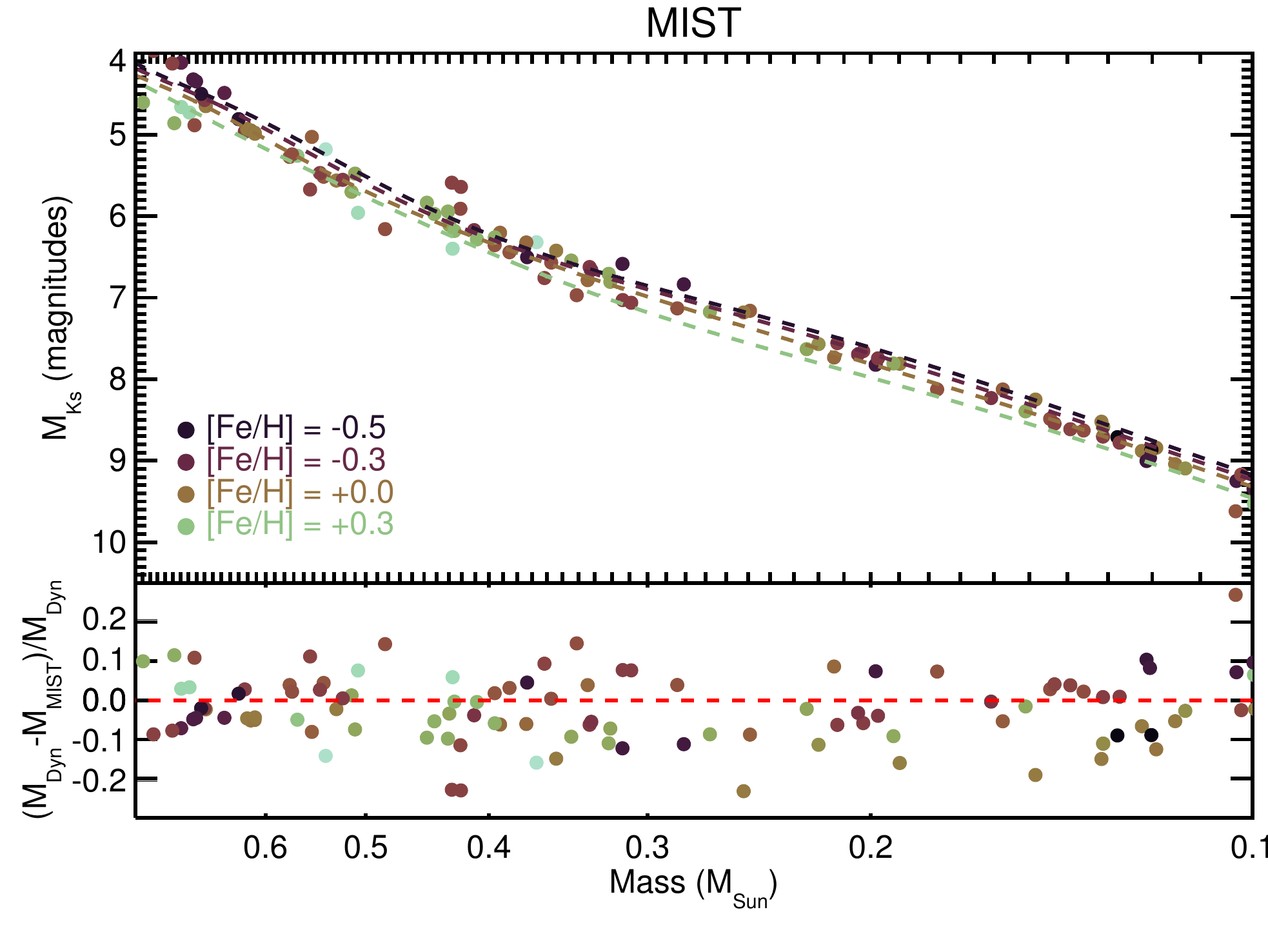}
\includegraphics[width=0.47\textwidth]{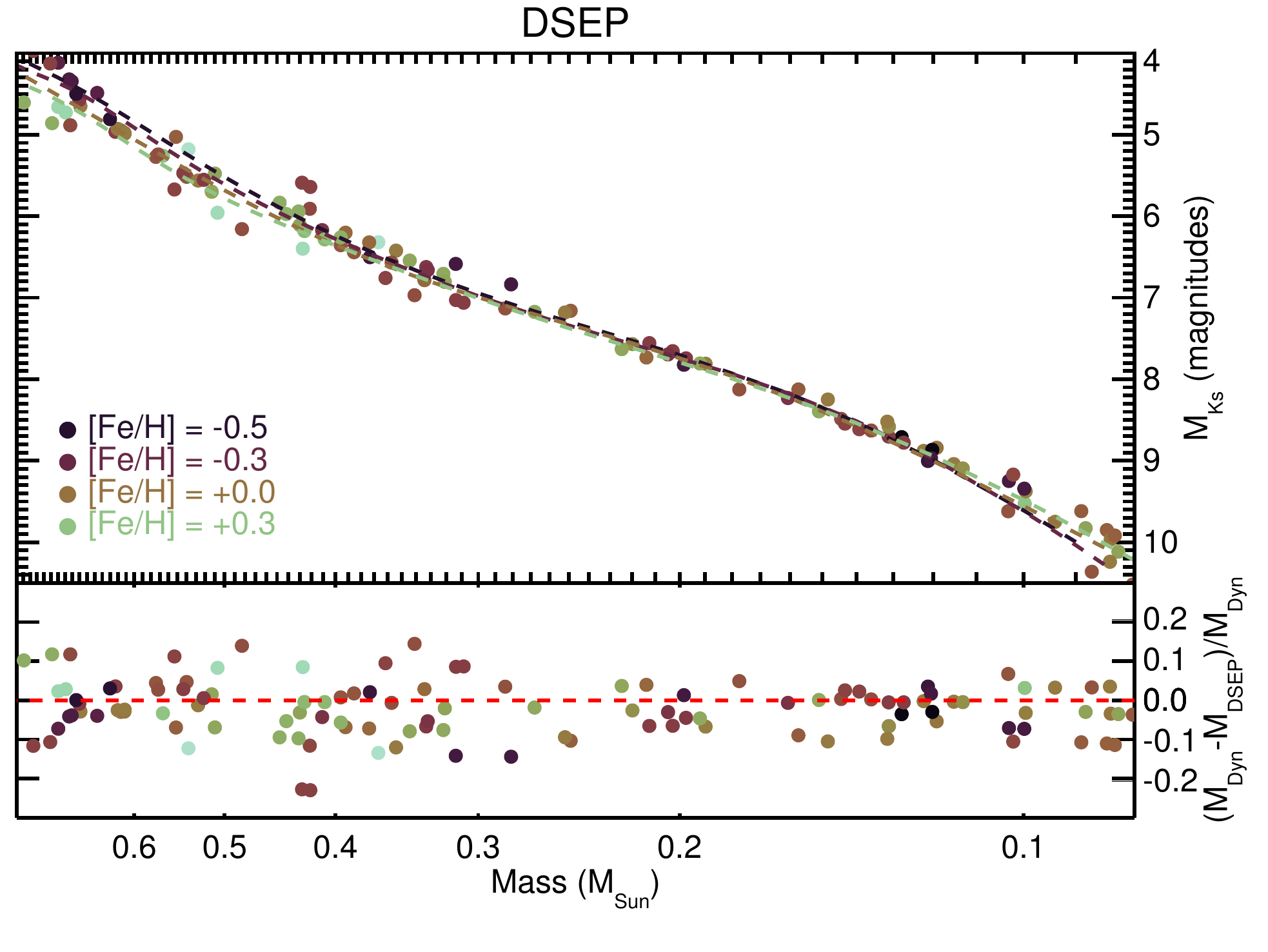}
\caption{\mks\ as a function of $M_*$ using MIST (left) and DSEP (right) tracks of different metallicities (dashed lines) compared to empirical mass determinations (points). Color-coding by metallicity is the same for the points and lines, and matches the color scale of Figure~\ref{fig:relation}. Owing to the limits of the model grids, the plots cut at 0.1$M_\odot$ and 0.085$M_\odot$ for MIST and DSEP, respectively. The bottom panel shows the fractional residual between the model and empirical determinations. This model masses for the residuals were estimated by interpolating over the model grid using the [Fe/H] and \mks\ for a given target. As- with Figure~\ref{fig:relation}, we have converted our dynamical total masses to individual masses using the predicted mass ratios from the \mmk\ for display purposes. }
\label{fig:mk_metal}
\end{center}
\end{figure*}

We show the expected \mks\ tracks from MIST and DSEP for $-0.5<$[Fe/H]$<+0.3$ in Figure~\ref{fig:mk_metal} alongside our empirical determinations. MIST models do not extend below 0.1$M_\odot$, while DSEP goes down to 0.085$M_\odot$. For this comparison we assumed a fixed age of 5\,Gyr, although the choice of age from 1-10\,Gyr makes a negligible difference for the mass range shown (Figure~\ref{fig:age}). 

Metal-rich stars are expected to be less luminous for a fixed $M_*$, whereas the opposite trend is seen for a fixed \teff\ and most color selections. Higher metal abundance increases the opacity, causing the stellar radius to increase at a fixed $M_*$ and surface temperature to decrease. Decreasing surface temperature also decreases the core temperature because it shifts the star to a different adiabat, which reduces nuclear reaction rates and overall luminosity. The trend is weaker (or even reversed) in the $K$-band, because although the overall luminosity is lower at higher [Fe/H], most of the increases in opacity are in the optical, causing a larger fraction of the total luminosity to escape at NIR wavelengths. This difference as a function of wavelength can be seen in Figure~\ref{fig:metal}, where the metal-rich spectrum sits below the solar-metallicity one at optical wavelengths, but above it in the NIR. The MIST and DSEP models likely have some difference in their treatment of one or both of these competing effects, as the DSEP models show a reduced impact of [Fe/H] on the \mmk\ relation as with decreasing stellar mass (likely because of increasing opacity in the optical with decreasing surface temperature), while MIST models show a similarly large impact over the full mass range considered here. 

Based on our dynamical masses, there is a slight trend for metal-rich stars above 0.4$M_\odot$ to land below the median sequence, as expected from both model grids. However, many metal-poor stars also land below the sequence, and there is no obvious trend below 0.4$M_\odot$. Further, the largest metal-rich outlier in the high-mass region (Gl 99) had a relatively poor mass determination (12\%) and is consistent with the solar-metallicity sequence. 

\begin{figure}[htp]
\begin{center}
\includegraphics[width=0.47\textwidth]{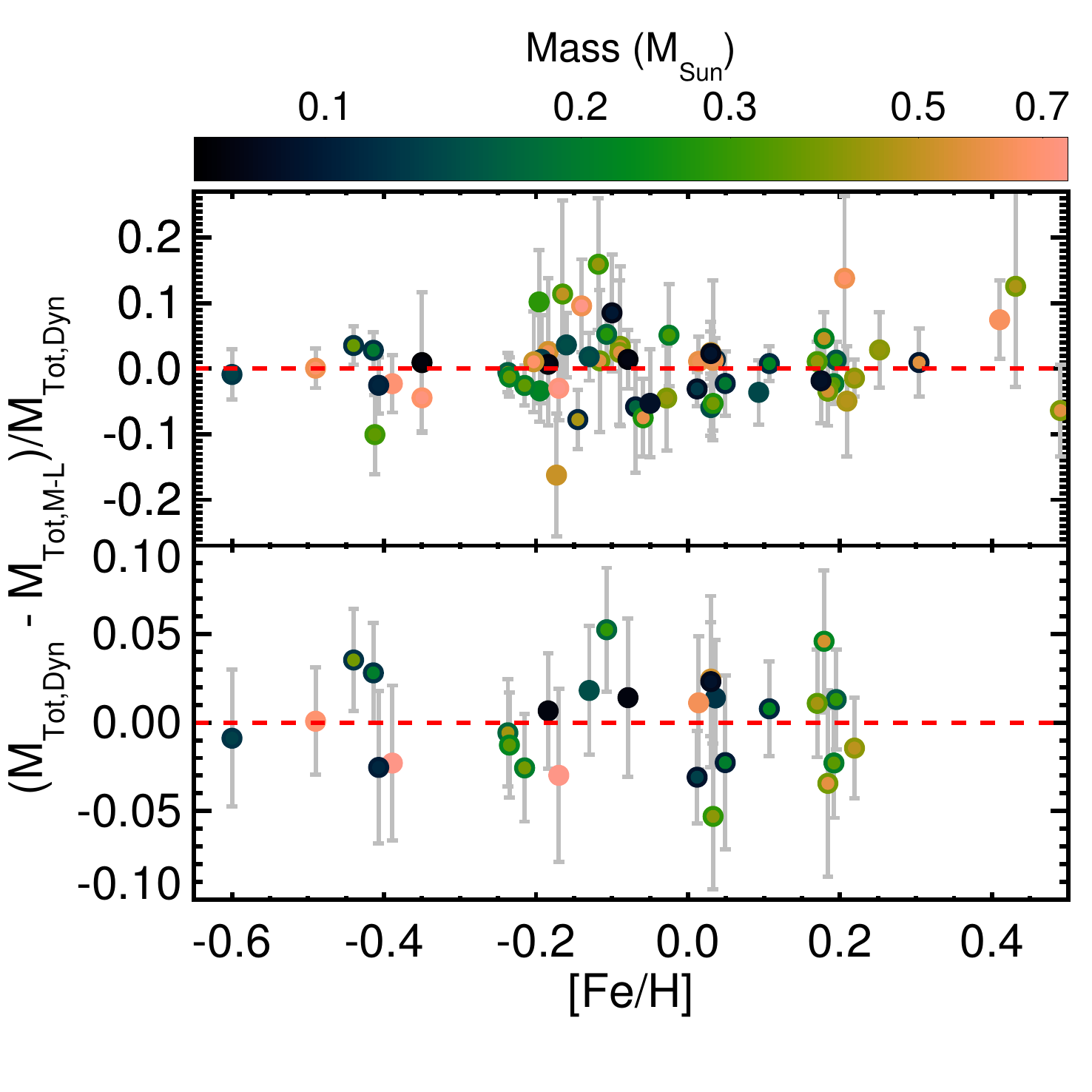}
\caption{Fractional difference between the orbital and predicted total mass as a function of the system metallicity. The top panel contains all systems, while the bottom shows just those with $<5\%$ uncertainties on \mdyn. Note the top and bottom panels have different Y-axis ranges. Points are color-coded by the masses of components, with the inner dot corresponding to the primary star's estimated mass, and the outer circle the companion's estimated mass. Error bars include errors on \mdyn\ {\it and} \mpred\ (including $\sigma_e$).}
\label{fig:metal_resid}
\end{center}
\end{figure}

The residuals from our best-fit indicate a weak (or no) effect on the derived $M_*$ owing to changes in [Fe/H], as we show in Figure~\ref{fig:metal_resid}. A Spearman's rank test yielded no significant correlation between the residuals and [Fe/H]. We tried resampling the measurements using their uncertainties, and $<1\%$ of samples showed a significant correlation. We also repeated this test, but restricted to just the best-characterized systems ($<5\%$ precision on mass) and still found no significant trend with [Fe/H]. 

Our sample is limited in its [Fe/H] range; 67\% of the targets are $-0.2<$[Fe/H]$<+0.2$ and only one target has [Fe/H]$<-0.5$. It is possible that our best-fit relation masked any [Fe/H] term by shifting the fit to match the typical metallicity of stars at a given \mks. We explored [Fe/H] effects in a more robust way by fitting for a term of the form
\begin{eqnarray}\label{eqn:mmk3}
M_{tot,pre} = (1+f[\rm{Fe/H]})\times \nonumber \\ 
\left( 10^{\sum_{i=0}^{n} a_i(M_{K_S,2}-zp)^i} + 10^{\sum_{i=0}^{n} a_i(M_{K_S,1}-zp)^i} \right)
\end{eqnarray}
This is identical to Equation~\ref{eqn:mmk2} multiplied by $(1+f\rm{Fe/H]})$. This assumes that a linear change in [Fe/H] corresponds to a fractional change in $M_*$ (e.g., f=0.1 would correspond to a 10\% change in derived $M_*$ per dex change in [Fe/H] at a fixed \mks). This is generally consistent with the models over the metallicity range considered here (although it is predicted to become increasing nonlienar for [FeH]$<-0.5$). Equation~\ref{eqn:mmk3} also assumes a single $f$ over the whole mass range considered. While this is consistent with the predictions of MIST models, DSEP models show a tightening with decreasing mass (increasing \mks). However, our sample is too small and errors on [Fe/H] are too large to justify adding a term that depends on both [Fe/H] and \mks. We leave higher-order tests for a future investigation with a broader range of metallicities. 

For the metallicity analysis, we excluded the two L dwarfs from the sample because their metallicities are less reliable (extrapolated from an M dwarf calibration). As with our fit to Equation~\ref{eqn:mmk2}, we tested a range of values for $n$ (number of $a_i$ coefficients). Both targets also have masses below the limits of the model grids. Our MCMC fitting method was otherwise identical to that outlined in Section~\ref{sec:relation}. 

\begin{figure*}[p]
\begin{center}
\includegraphics[width=0.97\textwidth]{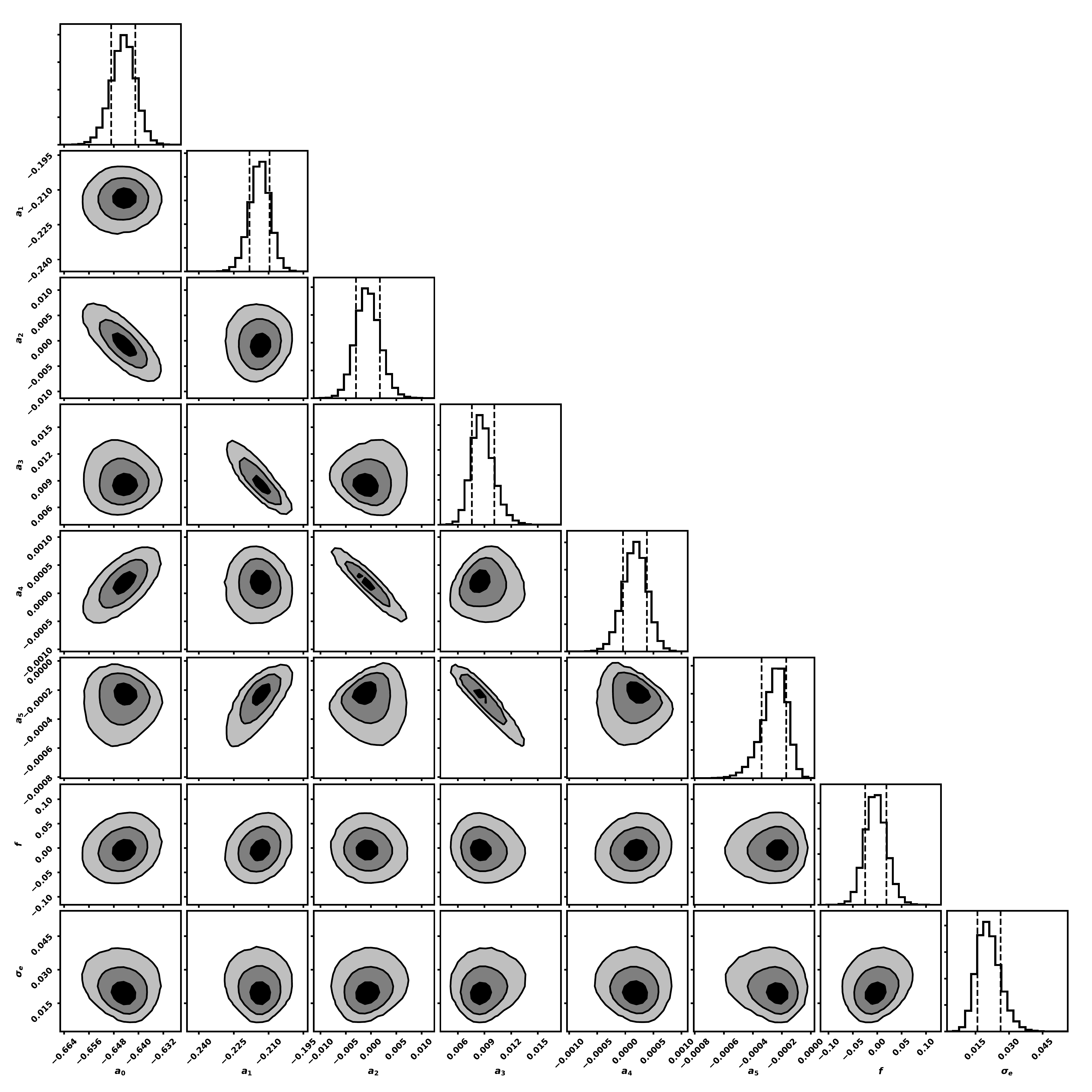}
\caption{Same as Figure~\ref{fig:fitpost}, but for the fit following Equation~\ref{eqn:mmk3}, i.e., including the [Fe/H] term, $f$. }
\label{fig:fitpost_feh}
\end{center}
\end{figure*}

We show the output coefficient posteriors including $f$ in Figure~\ref{fig:fitpost_feh}. We list the corresponding best-fit coefficients in Table~\ref{tab:coeff} along with the median values of $\sigma_e$ and $f$. As with our fits neglecting any [Fe/H] terms, we found significantly better agreement with the lowest-mass objects in the sample using $n=\order$, although $n=4$ and $n=6$ are listed in Table~\ref{tab:coeff} for reference. 

In agreement with our previous analyses, our derived $f$ value is consistent with zero (a 0.0$\pm$2.2\% change in mass per dex change in [Fe/H]). This suggests our relation will work reasonably well even on more extreme metallicity samples. However, it is also possible that [Fe/H] is less important than abundances of elements that specifically impact the strength of molecular features in M dwarf spectra. Higher C/O, for example, suppresses available Oxygen for TiO formation, weakening a major source of opacity in the optical \citep[e.g., C, O, Ti,][]{2012ApJ...747L..27F,2015ApJ...804...40G,Veyette2016a}. This also might explain some of the extra scatter in the relation ($\sigma_e$) if there is sufficient variance of these abundances in the given sample. Testing this will require a means to determine more detailed abundances of M dwarfs \citep[e.g.,][]{Veyette2017}, and/or to add in subdwarf binaries or other systems with more extreme abundances to provide increased leverage on any metallicity effects. 

To compare to the models, we fit the MIST and DSEP grid points in the same manner as the empirical dataset following Equation~\ref{eqn:mmk3}. Our binary sample is not uniformly spaced in [Fe/H] and \mks, so to ensure a fair comparison, we resample the model grid to match the binary sample. For every target, we generated a model-predicted mass at fixed age (5\,Gyr), and alpha abundance (Solar) by linearly interpolating over \mks\ and [Fe/H] (using the assigned values for that target). We used the resulting (model-based) masses with the input \mks\ and [Fe/H] values to fit for a model $f$ value that can be compared to our empirical determination.

\begin{figure}[htp]
\begin{center}
\includegraphics[width=0.47\textwidth]{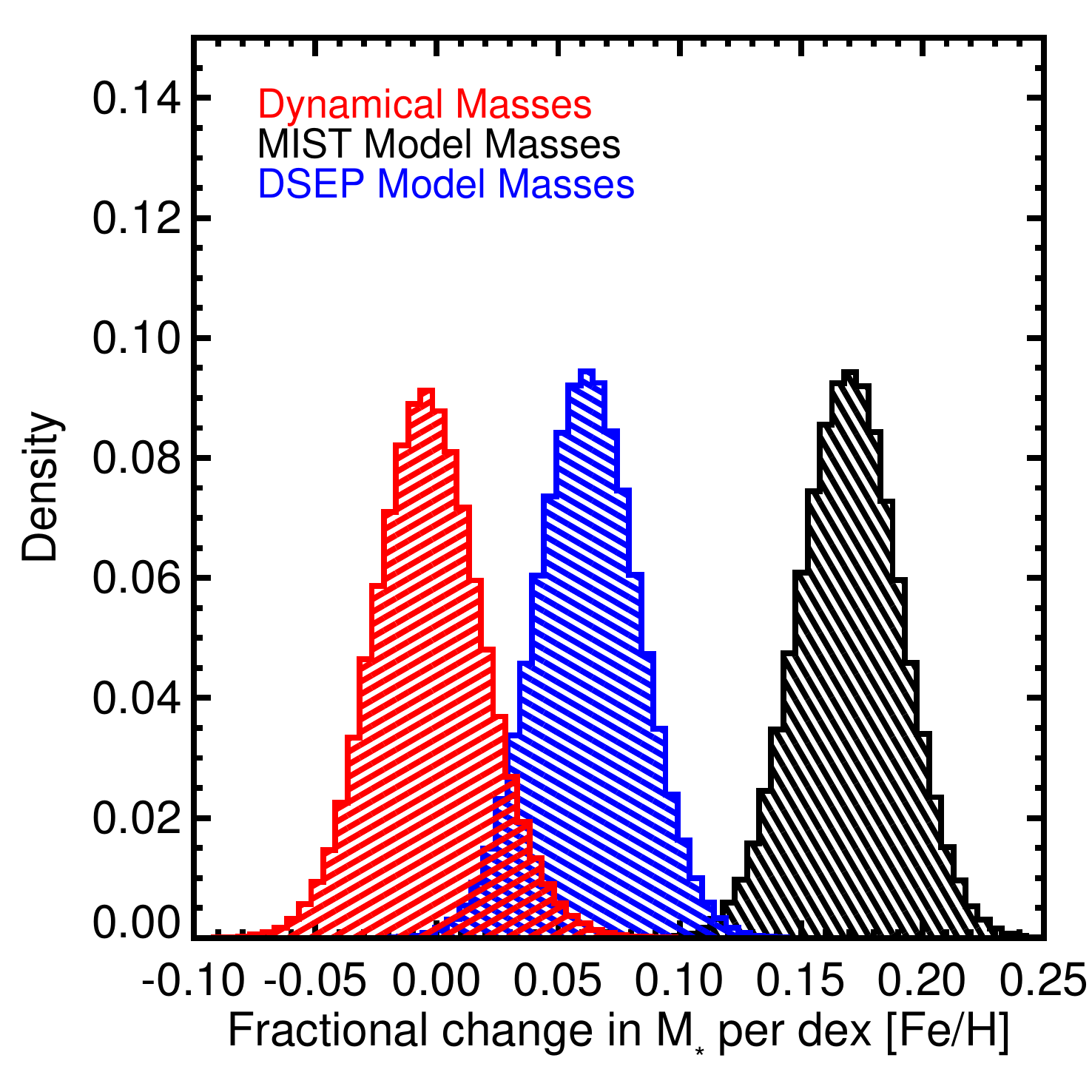}
\caption{Comparison of the posterior on $f$ (fractional change in $M_*$ per dex in metallicity for a fixed \mks; Equation~\ref{eqn:mmk3}) predicted by the MIST models (black) and DSEP (blue) compared to that using our dynamical masses (red). The MIST models significantly overpredict the role of [Fe/H] on the relation, although our results are consistent with predictions from DSEP (at 2$\sigma$). There are an identical number of points in each posterior and the bin sizes are the same. }
\label{fig:f}
\end{center}
\end{figure}

In Figure~\ref{fig:f}, we show the posterior on $f$ from the model grids compared to that from the dynamical masses. MIST models predict a larger [Fe/H] effect then suggested by our binary sample, while DSEP predictions are quite consistent with our own. The difference between the two posteriors ($f_{\rm{model}}-f_{\rm{dynamical}}$) is inconsistent with zero at $5.1\sigma$ for MIST,\footnote{Version 2 of MIST isochrones shows better agreement, but was not available at the time this paper was published} while for DSEP the difference is $2\sigma$. 

The discrepancy between MIST model masses and dynamical masses cannot be explained by $\sigma_e$. Scatter from $\sigma_e$ only amounts to a $\simeq$2\% variation in $M_*$ for a given \mks. The MIST models predict a metallicity effect of $\simeq17\%$ per dex; since our sample covers about 1 dex in [Fe/H], this translates to an expected $\simeq$17\% variation in mass over the full sample, or 8\% if we just consider the majority of the targets. It is possible that $\sigma_e$ is being driven in part by erroneous assigned [Fe/H] (or underestimated errors on [Fe/H]) which would systematically decrease our derived $f$ value, but the effect is too small to reconcile with the MIST models. 

 The discrepancy between MIST and empirical estimates of the impact of [Fe/H] could be owing to missing opacity/molecular lines in the atmospheric models. Recent comparison suggest atmospheric models reproduce optical and NIR spectra of M dwarfs to $\simeq$5\% \citep[e.g.,][]{Lepine2013,Mann2013c}, with the exception of a few molecular features like CaOH, AlH, and NaH \citep{Rajpurohit:2013}. However, these tests have not been performed on the atmospheric models used for MIST isochrones. Missing opacity at optical wavelengths would strengthen the effect of [Fe/H] by underestimating the number of saturated features; if a line is saturated adding [Fe/H] cannot make it stronger, which serves to reduce the impact of [Fe/H]. The effect at NIR wavelengths would be weaker, since there are fewer molecular bands, but underestimated opacity in the optical shifts continuum levels in the NIR (and how those levels change with [Fe/H]). A problem with the input opacities is also consistent with the trend of growing discrepancy at the lowest masses, where molecular bands become increasingly important, and might explain the difference between MIST and DSEP model predictions. 
 
\subsection{Comparison to previous relations}\label{sec:other}

\subsubsection{\citet{Hen1993}}

\citet{Hen1993} provided one of the first \mmkc\ relations, providing the basis for updates from \citet{Delfosse2000} and \citet{Benedict2016}. Although the least precise (scatter of 15-20\% in mass) it covers a large range in mass ($0.08M_\odot \lesssim M_* \lesssim 1M_\odot$). Most of the dynamical mass measurements used for the \citet{Hen1993} relation have since been significantly improved, including many of the astrometric binaries in our sample, but a comparison could reveal any potential changes in results that relied on \citet{Hen1993} with our more precise relation.

\begin{figure}[htb]
\begin{center}
\includegraphics[width=0.47\textwidth]{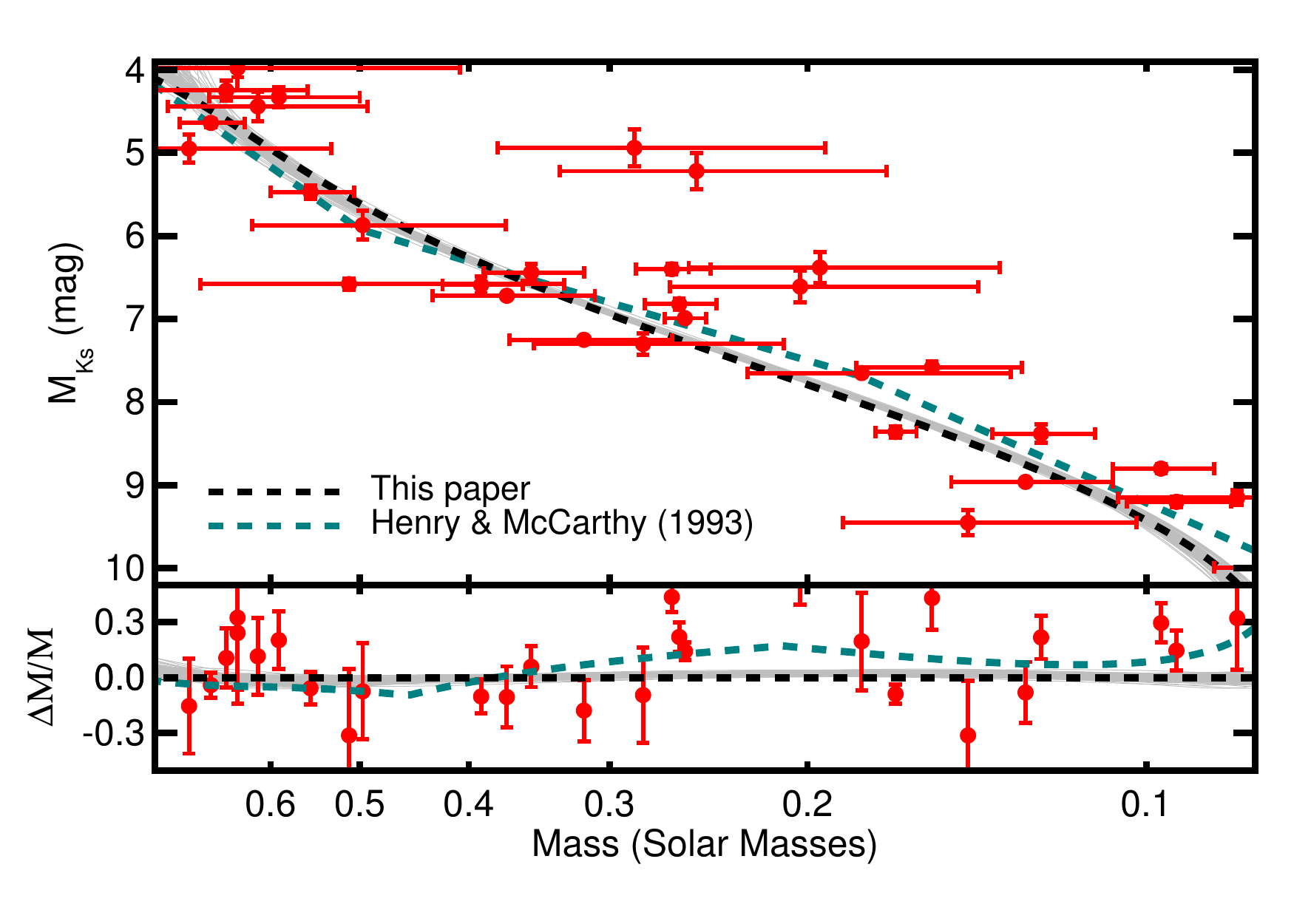}
\caption{Absolute $K_S$-band magnitude as a function of mass for astrometric binaries analyzed by \citet{Hen1993} (red circles). The relation from \citet{Hen1993} is shown as a teal dashed line (converted from $M_K$ to $M_{K_S}$), while the best-fit relation from this paper is shown as a blacked dashed line (with error in grey as in Figure~\ref{fig:relation}). The bottom panel shows the residual of the \citet{Hen1993} points compared to our relation. Note that some extreme outlier masses are not shown in the residual plot. }
\label{fig:henry}
\end{center}
\end{figure}

We show the comparison in Figure~\ref{fig:henry}. The \citet{Hen1993} relation is split into three sections by mass, as can be seen in the sharp change at $\simeq0.2M_\odot$ and $0.5M_\odot$. The scatter in masses from \citet{Hen1993} is large, however, the two relations track each other to within 10-20\% over the entire overlapping mass range, consistent with the 15-20\% uncertainties given by \citet{Hen1993}.

\subsubsection{\citet{Delfosse2000}}

\citet{Delfosse2000} provided one of the most commonly used \mmkc\ relations, covering $0.1M_\odot<M_*<0.6M_\odot$. Like our work, the calibration was built primarily on astrometric binaries. \citet{Delfosse2000} used a mix of individual (targets with radial velocities and/or absolute astrometry) and total masses, with the latter case converted to individual masses using models available at the time. Nearly all the targets in \citet{Delfosse2000} were included in our sample, with the exception of triple stars and eclipsing binaries, both of which were not included in our calibration sample (but were used for tests in Section~\ref{sec:test}). Because of the sample overlap, consistency is expected. However, as with our comparison to \citet{Hen1993}, a comparison can be useful to see how past use of \citet{Delfosse2000} may change with our more precise results.

\begin{figure}[htb]
\begin{center}
\includegraphics[width=0.47\textwidth]{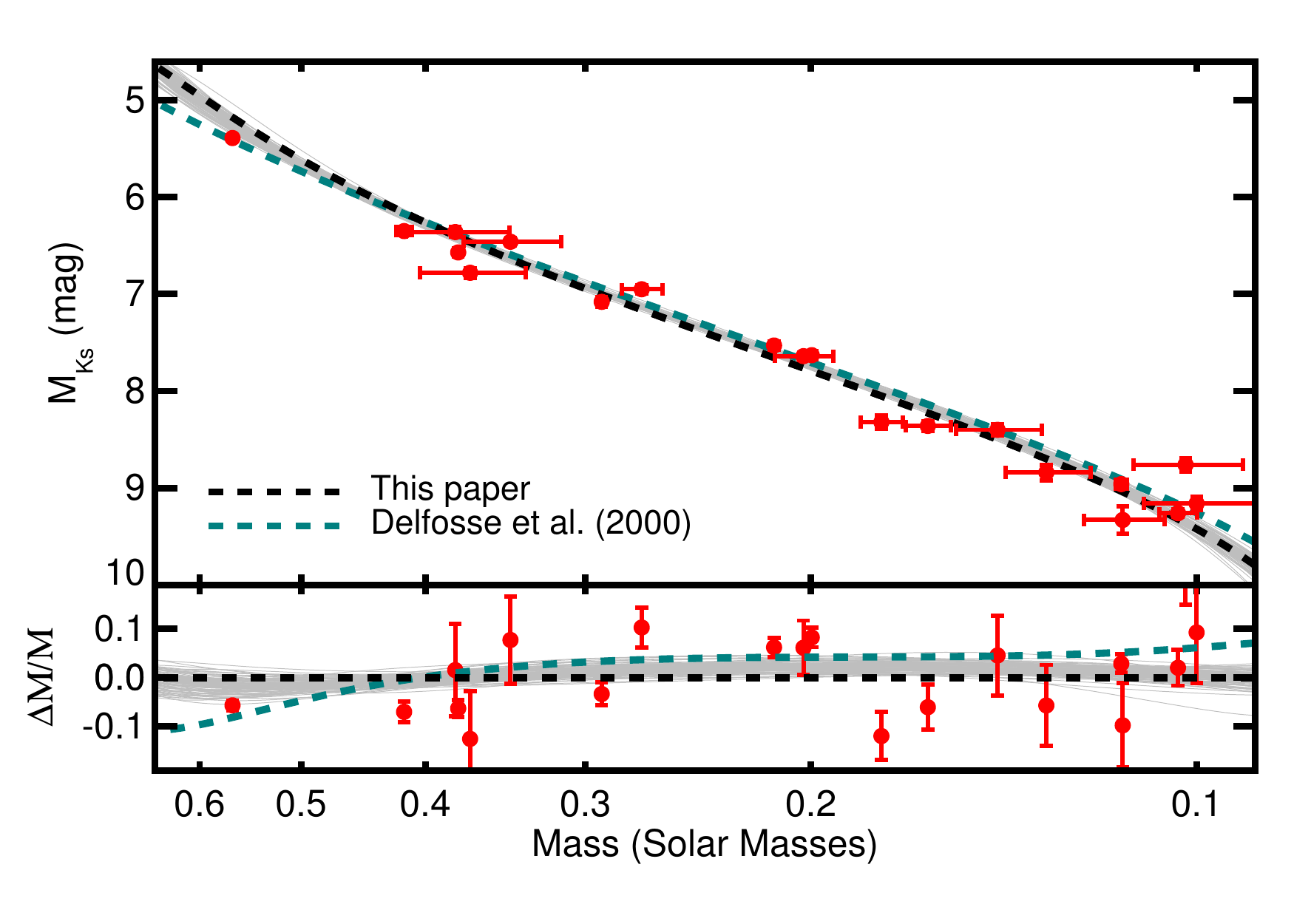}
\caption{Absolute $K_S$-band magnitude as a function of mass for astrometric binaries analyzed by \citet{Delfosse2000} (red circles). The resulting $M_*$-\mks\ relation from \citet{Delfosse2000} is shown as a teal dashed line, while the best-fit relation from this paper is shown as a blacked dashed line (with error in grey as in Figure~\ref{fig:relation}). The bottom panel shows the residual of the \citet{Delfosse2000} points compared to our relation.}
\label{fig:delfosse}
\end{center}
\end{figure}

We show the comparison in Figure~\ref{fig:delfosse}, including the points used in the \citet{Delfosse2000} calibration as well as the two fit lines. Given errors often quoted for the \citet{Delfosse2000} relation (5-10\%), the two fits are in remarkable ($<5\%$) agreement over most of the mass range ($0.15M_\odot \lesssim M_* \lesssim 0.5M_\odot$). Only at the high-mass end do the two relations diverge by as much as 10\%, but \citet{Delfosse2000} had few calibrators in this regime. While the two relations are in excellent agreement, the relation presented here is a factor of 3-5$\times$ more precise over the whole mass regime. 

\subsubsection{\citet{Mann2015b}}

\citet{Mann2015b} built a catalog of 183 M dwarfs with precise \teff\ and $R_*$, calibrated against radii measurements from long-baseline optical interferometry \citep{Boyajian2012} and precision bolometric fluxes \citep[e.g.,][]{Mann2015a, 2015MNRAS.447..846B}. Masses were computed for these stars by interpolating the parameters onto an updated version of the DSEP models as described in \citet{Feiden2013,Feiden2014} and \citet{Muirhead2014}. Although these masses were model-dependent, they accurately reproduced the mass-radius relation from low-mass eclipsing binaries. This suggested that the model-based masses were accurate to $\simeq$3\% or better, and motivated the development of a \mmk\ relation from the \citet{Mann2015b} sample. A comparison to our relation can be seen in part as a test on the updated DSEP models, in addition to the results given in \citet{Mann2015b}. 

\begin{figure}[htb]
\begin{center}
\includegraphics[width=0.47\textwidth]{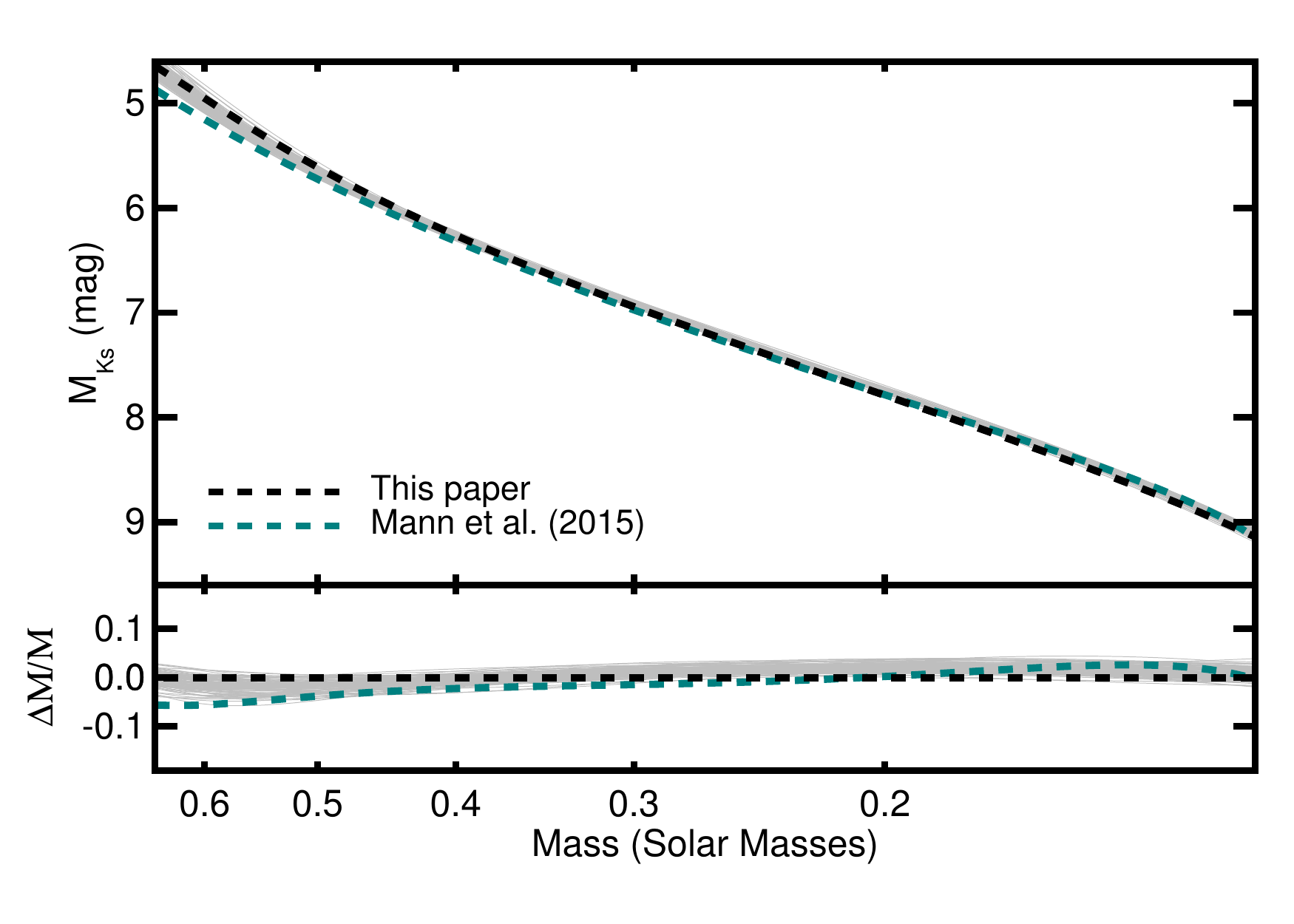}
\caption{Comparison of $M_*$-\mks\ from \citet{Mann2015b}, shown as a teal dashed line, to that from this paper, which is shown in black, with 100 randomly selected realizations of the MCMC (as with Figures \ref{fig:relation} and \ref{fig:delfosse}). Residual is shown in the bottom panel. Individual points from \citet{Mann2015b} on which the calibration is based are not shown (for clarity), but they follow a tight sequence around the teal line. Only the range of masses covered by \citet{Mann2015b} are shown. }
\label{fig:mann}
\end{center}
\end{figure}

We show our fit with uncertainties alongside \citet{Mann2015b}'s in Figure~\ref{fig:mann}. The two fits track each other extremely well, with a maximum divergence of $\simeq$5\%. Given the quoted 2-3\% uncertainties from \citet{Mann2015b} and similar errors in our relation, this difference is not significant. There is a hint of tension at above 0.6$M_\odot$ and around 0.2-0.3$M_\odot$ where the difference is the largest, but the offset never exceeds the quoted uncertainties of the two relations.  

\subsubsection{\citet{Benedict2016}}
Like our work, the \citet{Benedict2016} relation was also based primarily on masses derived from M dwarf astrometric binaries. The \citet{Benedict2016} sample uses absolute astrometry from {\it HST} fine guidance sensors and radial velocities for a subset of systems. In addition to the precision provided by {\it HST}, this combination yields individual (component) masses, and, in many cases, independent constraints on parallaxes. So although our sample is larger and contains most of the targets in \citet{Benedict2016}, their analysis has the advantage of using individual, instead of total masses.

\begin{figure}[htb]
\begin{center}
\includegraphics[width=0.47\textwidth]{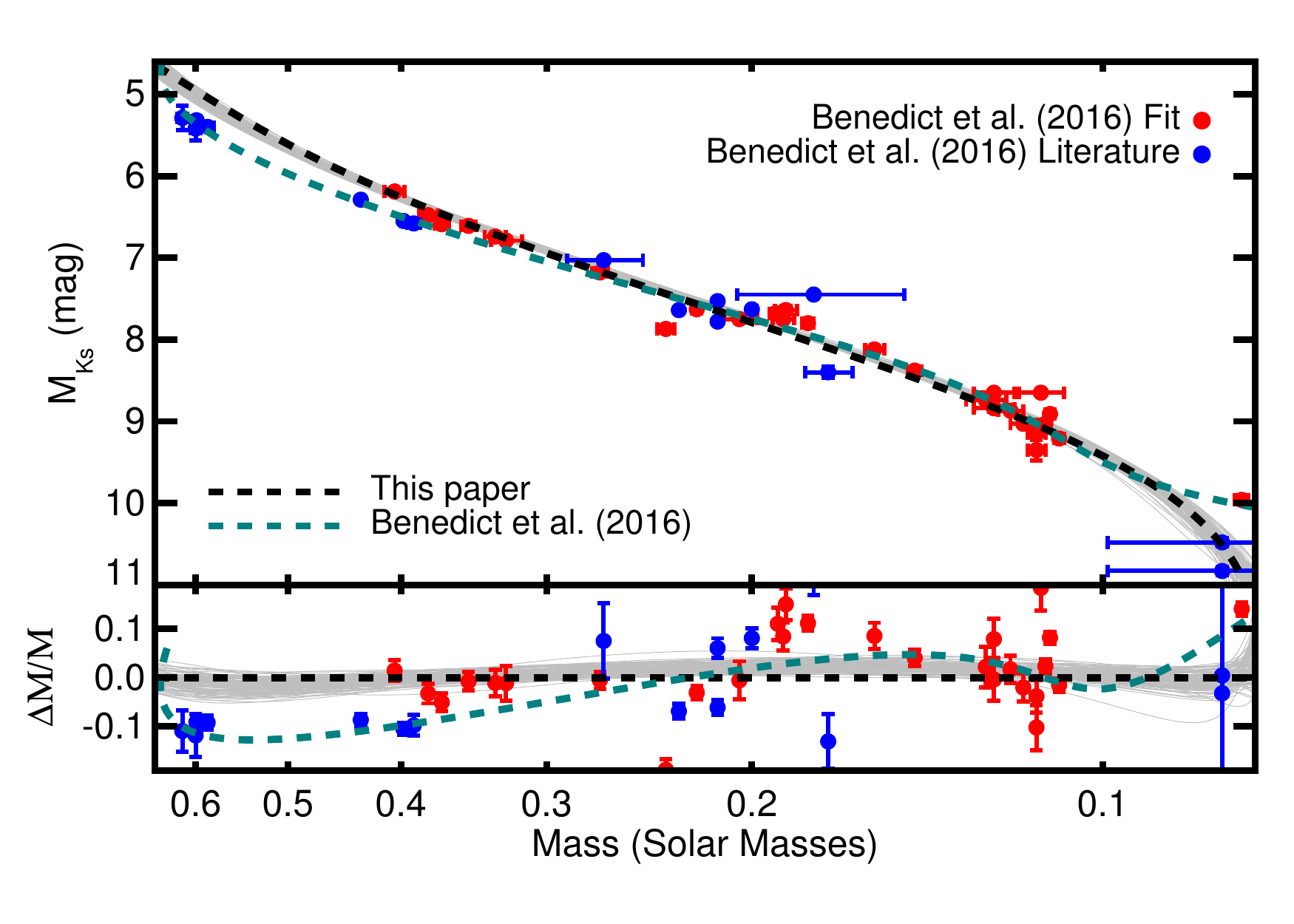}
\caption{Absolute $K_S$-band magnitude as a function of $M_*$ for astrometric binaries analyzed by \citet{Benedict2016} (red circles) and those used in the \citet{Benedict2016} relation, but pulled from the literature (blue circles). The resulting $M_*$-\mks\ relation from \citet{Benedict2016} is shown as a teal dashed line, while the best-fit relation from this paper is as a black dashed line (with random samplings in grey as in earlier figures). The bottom panel shows the residual of the \citet{Benedict2016} points compared to our relation, with the \citet{Benedict2016} relation in teal for reference. Errors in the residuals only reflect errors in $M_*$ and \mks, and do not account for errors in our $M_*$-\mks\ relation. }
\label{fig:benedict}
\end{center}
\end{figure}

We compare our \mmk\ relation to that of \citet{Benedict2016} in Figure~\ref{fig:benedict}. The two relations are in excellent agreement for $0.09M_\odot\lesssim M_* \lesssim 0.25M_\odot$. Below this regime, the \citet{Benedict2016} fit is effectively anchored by one star, GJ1245C, because the two other stars in this low-mass regime (GJ 2005B and C) have relatively large errors. GJ1245AC is in our sample, but we used a parallax from \citet{GaiaDr2} on GJ 1245B for this system, which places it $10\sigma$ (2.5\%) more distant than the parallax adopted by \citet{Benedict2016}. Our orbital parameters for this system are in excellent agreement with \citet{Benedict2016} if we adopt their distance, but the \citet{GaiaDr2} parallax makes the final parameters more consistent with our \mmk\ relation (although still 2$\sigma$ discrepant). If the \citet{Benedict2016} parallax is correct, this reduces the total mass to 0.189$\pm$0.001$M_\odot$, while the predicted mass is 0.207$M_\odot$ for the adjusted $M_K$ values (8.90 and 10.02 for the primary and companion, respectively). To reconcile the dynamical and predicted mass using the \citet{Benedict2016} parallax, we would need to explain why GJ 1245AC is $\simeq$0.3~mag more luminous than predicted by other similar-mass objects. Some of the complications for GJ1245 could be due to youth and/or activity, since the system is known to have a high flare rate \citep{Laurie2015}.

 Above 0.3$M_\odot$, \citet{Benedict2016} predicts masses as much as 10\% higher than our own for a fixed \mks. Our fit agrees reasonably well with the astrometric binaries analyzed by \citet{Benedict2016} in this mass regime. The divergence is driven instead by literature mass determinations that \citet{Benedict2016} included in their \mmk\ fit. Inspection of these literature points makes the origin of the discrepancy more clear: many are eclipsing binaries and have $\Delta K$-band magnitudes of mixed quality and/or lack parallaxes needed for a precise \mks. GU Boo, for example, has absolute magnitudes estimated from an optical eclipse depth combined with bolometric corrections \citep{Lopez2005}, which are drawn from models that perform poorly on M dwarfs \citep{1998A&AS..130...65L,Hauschildt1999}. Similarly, for GJ 2069 AC (CU Cnc) \citet{Benedict2016} adopted \mks\ from \citet{Ribas2003} that disagrees with the 2MASS $K_S$ and {\it Gaia} DR2 parallax (for either AC or B) using any $\Delta K_S$. 
 
\begin{figure}[htb]
\begin{center}
\includegraphics[width=0.47\textwidth]{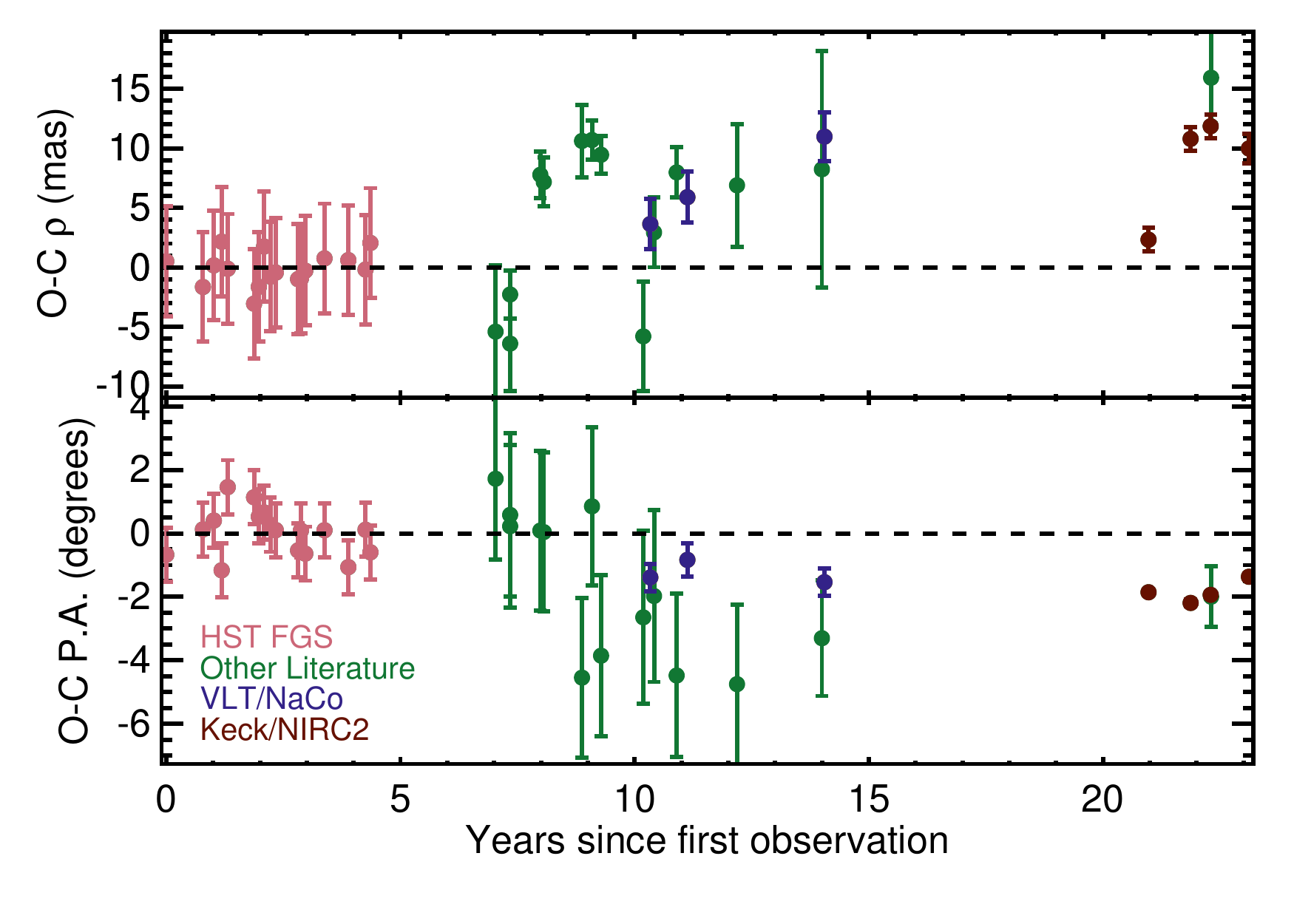}
\caption{Difference between astrometry used in our fit and the orbital parameters from \citet{Benedict2016}. The top shows the residuals in separation, while the bottom shows the residuals in position angle. Points are color-coded according to their source, including {\it HST} FGS data used in the \citet{Benedict2016} analysis. }
\label{fig:gj1005}
\end{center}
\end{figure}

In addition to GJ 1245AC, there are two targets in the \citet{Benedict2016} astrometric sample that are significantly discrepant from our own relation. These are GJ 1005AB and Gl 791.2AB, both of which have masses discrepant from predictions of the \citet{Benedict2016} relation. Our assigned total masses for both systems were much more consistent with both our \mmk\ relation as well the relation from \citet{Benedict2016}.  A comparison of our orbital fits to that of \citet{Benedict2016} revealed the source of the discrepancy; as we show in Figure~\ref{fig:gj1005} for GJ 1005AB, while the \citet{Benedict2016} orbit reproduces the astrometry from {\it HST}, it is highly discrepancy from more recent astrometry (which was not included in the \citet{Benedict2016} fits). Comparing the \citet{Benedict2016} orbit to all astrometry used in our analysis yielded a $\chi^2$ of 1978 (69 degrees of freedom). Our fit showed more tension with the {\it HST} astrometry, but accurately reproduced all measurements within uncertainties, yielding a final $\chi^2$ of 87 ($\chi^2_\nu=1.3$). A similar effect can be seen in Gl 791.2AB. Because the discrepancy between our orbits and those in \citet{Benedict2016} for these two systems can be seen in multiple sources of astrometry (including literature measurements), we consider our orbits and masses to be more accurate. 
 
\section{Conclusions and Discussion}\label{sec:discussion} 

\subsection{Summary}
The mass-luminosity relation has proven to be a critical tool for estimating masses of cool stars for decades, and has broad applications ranging from characterizing extrasolar planets to measuring the initial mass function in massive galaxies. We endeavored to improve on existing \mmk\ relations and evaluate the role of [Fe/H] on the relation by expanding the sample of calibrators, using new techniques to measure the metallicity of binary M dwarfs, and exploring the role of astrophysical scatter on the final precision.

As part of this effort, we combined adaptive optics data from Keck, CFHT, Gemini, and VLT with astrometric measurements from then literature to map the orbits of 55 binaries, which we join with seven systems with orbits from \citet{Dupuy2017}. While the more recent astrometry from Keck/NIRC2 was the most precise, the literature data provided $>$30 yr of data, which was essential to fit systems with $>$10\,year orbital periods. We include all our astrometry as well as those from the literature here, so that future work in this area can continue to grow this baseline and further improve the orbits of these systems.

Using parallaxes from the literature or derived from MEarth astrometry, we converted the orbital information into dynamical mass (\mdyn) measurements for each binary. Six binaries in our final sample of 62 systems had total mass determinations better than 1\%, 13 to better than 2\%, and 28 to better than 5\%. 

We used our dynamical masses and resolved \mks\ magnitudes to fit for an empirical relation between $M_*$ and \mks. Our methodology uses the observed quantity, which is the {\it total} mass of a given binary. This was done by making individual mass predictions from our resolved \mks\ estimates and the \mmk\ relation, and summing component masses to turn them into predictions for the total mass (\mpred) of each binary. The \mpred\ values could then be compared directly to \mdyn\ estimates within the likelihood function. While this halves the number points constraining the fit compared to using individual masses, using \mtot\ is far more robust than using model-based mass ratios, reduces the need for observationally expensive radial velocity and/or absolute astrometry (difficult without {\it HST} for many systems), and helps increase the range of binaries amenable to characterization (e.g., wider systems with small radial velocity amplitudes). 

The resulting \mmk\ relation covers almost an order of magnitude in $M_*$, from $\simeq0.70M\odot$ down to the hydrogen-burning limit ($\simeq0.075M_\odot$) and includes stars spanning the range of [Fe/H] expected for the solar neighborhood ($-0.6<$[Fe/H]$<+0.4$). Accounting for both intrinsic scatter and errors in the fit coefficients, masses from our \mmk\ relation are precise to $\simeq$2\% over most of the M dwarf sequence, rising to $\simeq3\%$ near the edges. 

The primary limit to our precision is scatter in \mdyn$-$\mpred\ above what is expected given measurement uncertainties alone. We characterized this missing error with a free parameter ($\sigma_e$), which we found to be $\simeq$2\% for all fits. It is likely that $\sigma_e$ arose from intrinsic variation in the \mmk\ relation due to a missing astrophysical parameter, such as age/activity/rotation \citep[e.g.,][]{Kraus2011,Feiden2016,Somers2017} or detailed abundances \citep[e.g.,][]{2017A&A...604A..97L,Veyette2017}, although we cannot rule out underestimated errors in the input orbital parameters, parallaxes, or $K_S$ magnitudes. 

We ran a series of tests to confirm that our use of \mtot\ and our assumed functional form did not bias our results or estimated uncertainties. To this end, we generated a set of synthetic binaries with component masses assigned using an assumed \mmk\ relation. We ran our MCMC framework on these synthetic binaries, providing no information about the individual masses (only \mtot). The resulting fits matched the input \mmk\ relation to within uncertainties. Our recovery of the input \mmk\ relation worked even when we assumed a functional form different from our input relation (i.e., a piecewise function with sharp breaks). 

As an additional test of our \mmk\ relation, we compared predicted masses from the \mmk\ relation to stars with {\it individual} dynamical masses from the literature. Our \mmk\ provides individual mass estimates in excellent agreement with those from the literature, with a $\chi^2_\nu$ of 1.06 and an RMS consistent with measurement uncertainties. There is a small ($\lesssim$2\%) systematic offset between the literature eclipsing binaries and predictions from our own relation. With the exception of very young stars, tight eclipsing binaries will generally be more magnetically active and have faster rotation periods than their single-star counterparts. The fact that our \mmk\ relation encompasses these stars within our quoted uncertainties suggests that the relation is effective for the majority of stars, which will have much lower activity levels. 

Using empirically calibrated spectroscopic abundances, we explored the role of [Fe/H] on the \mmk\ relation. Our results indicate the effect of metallicity on the \mmk\ relation is consistent with zero. MIST models significantly overestimate the importance of [Fe/H] in the \mmk\ relation (at $5\sigma$), however predictions from updated DSEP models are consistent with our own. 

We compared our relation to recent similar relations in the literature. Given quoted uncertainties, the \citet{Hen1993}, \citet{Delfosse2000} and \citet{Mann2015b} relations agreed with our own over the full sequence. Our results were consistent with the sample of astrometric binaries analyzed in \citet{Benedict2016}, but our relation diverges from \citet{Benedict2016} above $\simeq0.35M_\odot$. We attribute this difference to literature points included in the \citet{Benedict2016} fit from earlier analyses with uncertain distances and $\Delta K_S$ magnitudes. Our \mmk\ relation represents a significant improvement in precision over all these earlier determinations. 

\subsection{Suggestions when using our \mmk\ relation}\label{sec:caveats}

To help users interested in using \mks\ to compute realistic $M_*$ and $\sigma_{M_*}$ of single stars with parallaxes and $K_S$ magnitudes, we provide a simple code\footnote{\href{https://github.com/awmann/M_-M_K-}{https://github.com/awmann/M\_-M\_K-}} to sample the fit posterior. {\bf Before using that code or the provided MCMC posteriors, take note of the following suggestions:}
\begin{itemize}

\item Our estimate of $\sigma_e$ is only valid for stars comparable to the calibration sample. Targets that are unusual (in terms of activity, metallicity, etc) compared to those in the solar neighborhood may have their mass uncertainties underestimated.

\item The fit behaves more poorly near the edges of the calibration sample. The scatter in the MCMC posterior accounts for this, but restrict use to $4.0<M_{K_S}<11.0$ ($0.075M_\odot<M_*<0.75M_\odot$), and a safer range would be $4.5<M_{K_S}<10.5$ ($0.08M_\odot<M_*<0.70M_\odot$). 

\item Our relation is only valid for main-sequence stars, and the roles of youth and activity were not accounted for in our analysis. Based on the Lyon models \citep{BHAC15}, we advise restricting use to $>100$\,Myr above $0.4M_\odot$, $>300$\,Myr to $0.2M_\odot$, $>500$\,Myr to $0.1M_\odot$, and $>1$\,Gyr below 0.1$M_\odot$. A safer cut would be to only use this on stars $>1$ Gyr, similar to the calibration sample. The comparison to masses from eclipsing binaries suggests possible issues for highly active stars at the $\simeq$2\% level. While these are within our uncertainties, we suggest avoiding highly active stars until this can be directly tested with a more active sample of binaries.

\item The sample metallicity spans $-0.60<\rm{[Fe/H]}<+0.45$, but 84\% of the binaries have $-0.40<\rm{[Fe/H]}<+0.30$. We provide a fit that attempts to take into account changes due to [Fe/H], and the weak impact suggests the [Fe/H]-free relation is safe to use for most stars in the solar neighborhood. However, given the paucity of extreme metallicity systems in our calibration sample, we advise caution when targeting more metal-poor populations ([Fe/H]$\ll-0.6$).

\item The relation is only tested above the hydrogen-burning limit. Since the boundary likely depends on metallicity \citep{2001RvMP...73..719B}, it is also not possible to use a simple $M_{K_S}$ cut. Objects just below the hydrogen-burning limit age slowly \citep{BHAC15}, so the relation given here may give reasonable results, but we urge caution when interpreting resulting $M_*$ values for $M_{K_S}>10.5$. 

\item Since $\sigma_e$ is likely due to astrophysical variation in the relation, we suggest always including this as an irreducible and potentially systematic source of error. 

\end{itemize}

\subsection{Future directions}

We intentionally selected targets that had $\Delta K$ measurements, as \mks\ was known to give the tightest relation with $M_*$ for M dwarfs. Unfortunately, only about \nicefrac{1}{3} of the sample have measurements in an optical band. This limits the utility of the sample, as {\it Gaia} $G$, $BP$, and $RP$ are now widely available for early and mid-M dwarfs, and are generally measured with better precision than 2MASS $K_S$. The growing capabilities of speckle cameras \citep[e.g.,][]{2009AJ....137.5057H} offer the opportunity to add optical contrasts. These can be converted to {\it Gaia} bandpasses, given reasonable assumptions about the component spectra, and used to derive an $M_{G}-M_*$ (or $M_G-M_*$-[Fe/H]) relation that can be applied to millions of K and M dwarfs. Complementary optical data also provide colors for individual components, from which we can measure component \teff\ and luminosity \citep[e.g.,][]{2017ApJ...845...72K}.

We would like to explore changes in the impact of [Fe/H] as a function of \mks or $M_*$, especially given predictions from the DSEP models (Figure~\ref{fig:mk_metal}b). Metallicity effects may also become important only at extremely low metallicities, as was seen for the \mks-$R_*$ relation \citep{Kesseli2018b}. Our sample was heavily biased towards the narrow [Fe/H] distribution of nearby stars. This limited both our ability to explore more complex impacts of [Fe/H] and tighten constraints on $f$. We identified four additional [Fe/H]$<-0.5$ binaries not included in our analysis, including two subdwarf systems. However, these systems have short baselines of astrometry in the literature when compared to their expected orbital periods, and complete orbits will take several more years. The availability of {\it Gaia} parallaxes will also help improve the precision of the known metal-poor systems and aid in the identification of new ones. Lastly, as new methods arrive to measure detailed abundances of M dwarfs \citep{Veyette2016a,Veyette2017} we can explore effects beyond just [Fe/H]. 

Mass ratios were available for some systems \citep[e.g.,][]{Soderhjelm1999,Mlg2007b,Dupuy2017}. However, these determinations were heterogeneous (e.g., some use models, some radial velocities, and some absolute astrometry), and some mass ratios reported in the literature are derived from orbits that disagree with our own determinations \citep[e.g.,][]{Koh2012}. A more robust method would be to include radial velocity or absolute astrometry as part of our analysis. Fortunately, later {\it Gaia} data releases will include full absolute astrometry. When combined with a measure of the flux ratio in the {\it Gaia} $G$ bandpass and our existing astrometry, we will be able to fit for both individual masses and parallaxes simultaneously. The resulting dataset will effectively double our sample size, and may help reveal the origin of $\sigma_e$.

Ages of our binary sample are not known, preventing any study into the effects of age on the \mmk\ relation. However, our larger sample of binaries with orbit measurements still in progress contains known members of binaries in nearby young moving groups, known pre-main-sequence stars, and members of the Hyades cluster. These systems span ages from 10-650\,Myr, offering the chance to both test pre-main-sequence models of M dwarfs \citep{2015ApJ...813L..11M,2016ApJ...817..164R,2016ApJ...818..156C,2016AJ....152..175N} and explore the role of activity on M dwarf parameters \citep[e.g.,][]{Spada2013,Kesseli2018}. The current sample can be included in such work when combined with age indicators like kinematics \citep{2018MNRAS.tmp..966W}, ultraviolet flux \citep{Ansdell2015}, and rotation periods expected from the Transiting Exoplanet Survey Satellite \citep[{\it TESS},][]{2014SPIE.9143E..20R}.

\acknowledgements
The authors thank Meg Schwamb for her help with analysis of the NIRI data. 

AWM was supported through Hubble Fellowship grant 51364 awarded by the Space Telescope Science Institute, which is operated by the Association of Universities for Research in Astronomy, Inc., for NASA, under contract NAS 5-26555. T.J.D.\ acknowledges research support from Gemini Observatory. This work was supported by a NASA Keck PI Data Award (award numbers 1554237, 1544189, 1535910, and 1521162), administered by the NASA Exoplanet Science Institute. 

Data presented herein were obtained at the W. M. Keck Observatory from telescope time allocated to the National Aeronautics and Space Administration through the agency's scientific partnership with the California Institute of Technology and the University of California. The Observatory was made possible by the generous financial support of the W. M. Keck Foundation.

The authors wish to recognize and acknowledge the very significant cultural role and reverence that the summit of Maunakea has always had within the indigenous Hawaiian community. We are most fortunate to have the opportunity to conduct observations from this mountain.

The authors acknowledge the Texas Advanced Computing Center (TACC) at The University of Texas at Austin for providing grid resources that have contributed to the research results reported within this paper. URL: http://www.tacc.utexas.edu.

We would like to thank the University of North Carolina at Chapel Hill and the Research Computing group for providing computational resources and support that have contributed to these research results.

Pyfits is a product of the Space Telescope Science Institute, which is operated by AURA for NASA.

Based on observations obtained at the Gemini Observatory (acquired through the Gemini Observatory Archive), which is operated by the Association of Universities for Research in Astronomy, Inc., under a cooperative agreement with the NSF on behalf of the Gemini partnership: the National Science Foundation (United States), the National Research Council (Canada), CONICYT (Chile), Ministerio de Ciencia, Tecnolog\'{i}a e Innovaci\'{o}n Productiva (Argentina), and Minist\'{e}rio da Ci\^{e}ncia, Tecnologia e Inova\c{c}\~{a}o (Brazil). Observations taken from programs GN-2008B-Q-57, GN-2009B-Q-10, GN-2010B-Q-9, and GN-2011A-Q-26.

Based on observations collected at the European Organisation for Astronomical Research in the Southern Hemisphere under ESO programmes 071.C-0388(A), 072.C-0570(A), 073.C-0155(A), 075.C-0521(A), 075.C-0733(A), 077.C-0783(A), 078.C-0441(A), 079.C-0216(A), 080.C-0424(A), 081.C-0430(A), 082.C-0518(A), 082.C-0518(B), 085.C-0867(B), 086.C-0515(A), 086.C-0515(B), 090.C-0448(A), 091.D-0804(A), 098.C-0597(A), 382.C-0324(A), and 382.D-0754(A).

 This research has made use of the Keck Observatory Archive (KOA), which is operated by the W. M. Keck Observatory and the NASA Exoplanet Science Institute (NExScI), under contract with the National Aeronautics and Space Administration. 

Based on observations obtained at the Canada-France-Hawaii Telescope (CFHT) which is operated by the National Research Council of Canada, the Institut National des Sciences de l'Univers of the Centre National de la Recherche Scientifique of France, and the University of Hawaii. 

This work presents results from the European Space Agency (ESA) space mission Gaia. Gaia data are being processed by the Gaia Data Processing and Analysis Consortium (DPAC). Funding for the DPAC is provided by national institutions, in particular the institutions participating in the Gaia MultiLateral Agreement (MLA). The Gaia mission website is \href{https://www.cosmos.esa.int/gaia}{https://www.cosmos.esa.int/gaia}. The Gaia archive website is \href{https://archives.esac.esa.int/gaia}{https://archives.esac.esa.int/gaia}.

\software{{\tt emcee} \citep{Foreman-Mackey2013}, {\tt corner.py} \citep{corner}, {\tt MPFIT} \citep{Markwart2009}, scipy \citep{jones2001scipy}, pyfits, astropy \citep{2013A&A...558A..33A}, {\tt SpeXTool} \citep{Cushing2004}, {\tt xtellcor} \citep{Vacca2003}, {\tt StarFinder} \citep{2000A&AS..147..335D}.}

\facilities{Keck:II (NIRC2), IRTF (SpeX), CFHT (PUEO, KIR), VLT:Antu (NaCo); Gemini:North (NIRI)}

\bibliography{$HOME/Dropbox/fullbiblio}

\appendix 
\section{Converting Observed $\Delta K_X$ to 2MASS $\Delta K_S$ for M dwarfs}\label{sec:a1}
To place all $K$-band magnitudes on the 2MASS system, we derived a relation between $\Delta K_X$ and $\Delta K_S$ as a function of $\Delta K_X$, where $X$ denotes the particular filter ($K_s$, $K$, $K'$, $Br\gamma$, and $K_{\rm{cont}}$) used for the AO observations. For photometry, we only used observations taken with a filter somewhere in the $K$-band (all wavelengths are used for astrometry).

To derive a conversion between contrasts, we used the 183 absolutely-flux calibrated spectra of nearby single stars from \citet{Mann2015b}, which cover a similar range of \teff\ and $M_*$ as the sample considered here. These spectra are mostly empirical; models are only used to fill in gaps in the spectrum or regions of high telluric contamination, none of which land in the regions covered by the filters considered here.

First we randomly sampled two stars from the sample and scaled the absolute level of each spectrum by the star's distance. We convolved each of the two stars with the relevant filter profiles for NIRC2\footnote{\href{https://www2.keck.hawaii.edu/inst/nirc2/filters.html}{https://www2.keck.hawaii.edu/inst/nirc2/filters.html}}, KIR\footnote{\href{http://www.cfht.hawaii.edu/Instruments/Filters/kir.html}{http://www.cfht.hawaii.edu/Instruments/Filters/kir.html}}, NIRI\footnote{\href{http://www.gemini.edu/sciops/instruments/niri/imaging/filters}{http://www.gemini.edu/sciops/instruments/niri/imaging/filters}}, or NaCo\footnote{\href{http://www.eso.org/sci/facilities/paranal/instruments/naco/inst/filters.html}{http://www.eso.org/sci/facilities/paranal/instruments/naco/inst/filters.html}}, and integrate over all wavelengths to compute the total flux in a given band. The $\Delta K_X$ value for the given pair was then computed as $2.5log_{10}(F_1/F_2)$. We computed the equivalent $\Delta K_S$ for each pair of stars using the 2MASS filter profile from \citet{2003AJ....126.1090C}.  

\begin{figure*}[htp]
\begin{center}
\includegraphics[width=0.47\textwidth]{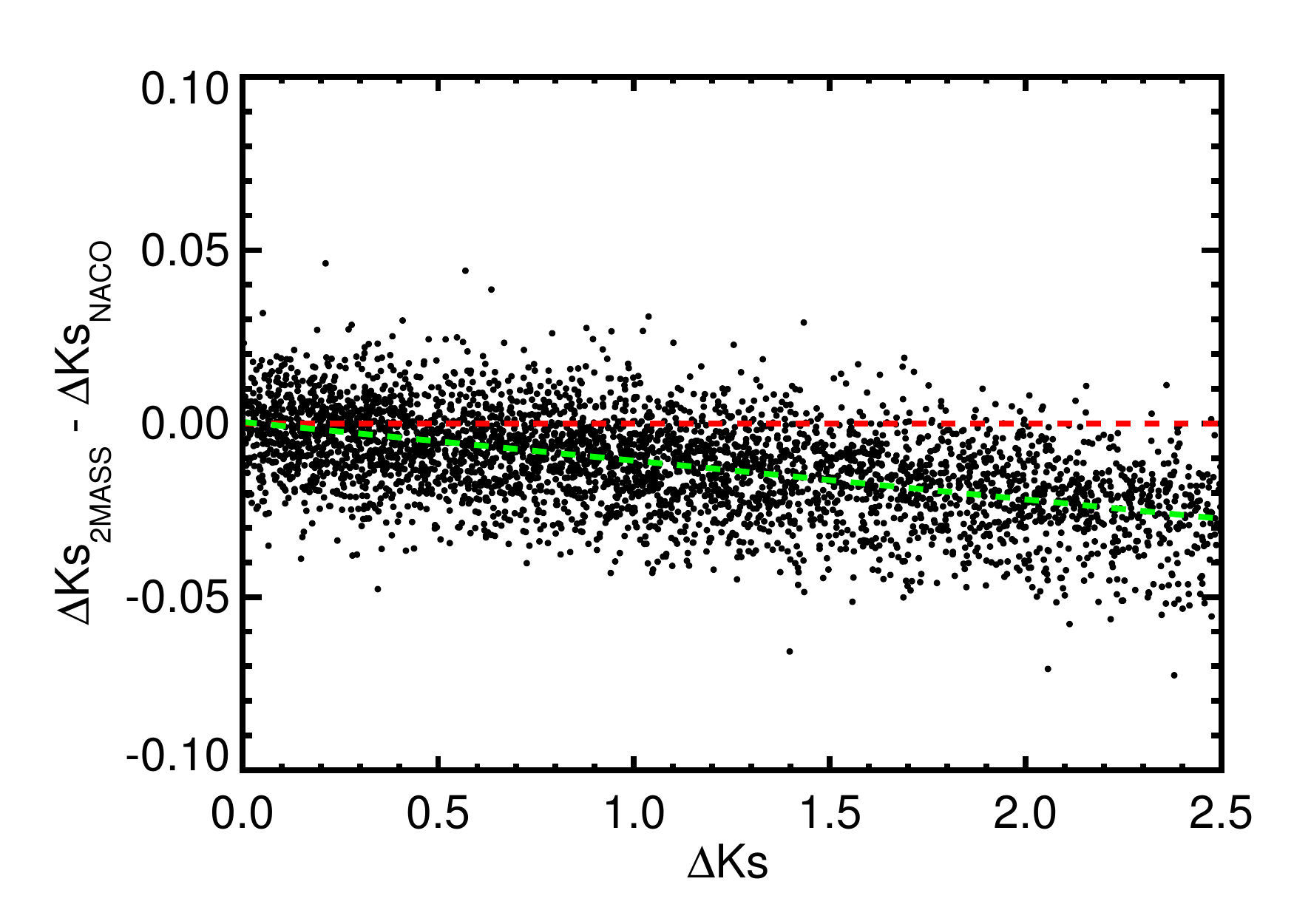}
\includegraphics[width=0.47\textwidth]{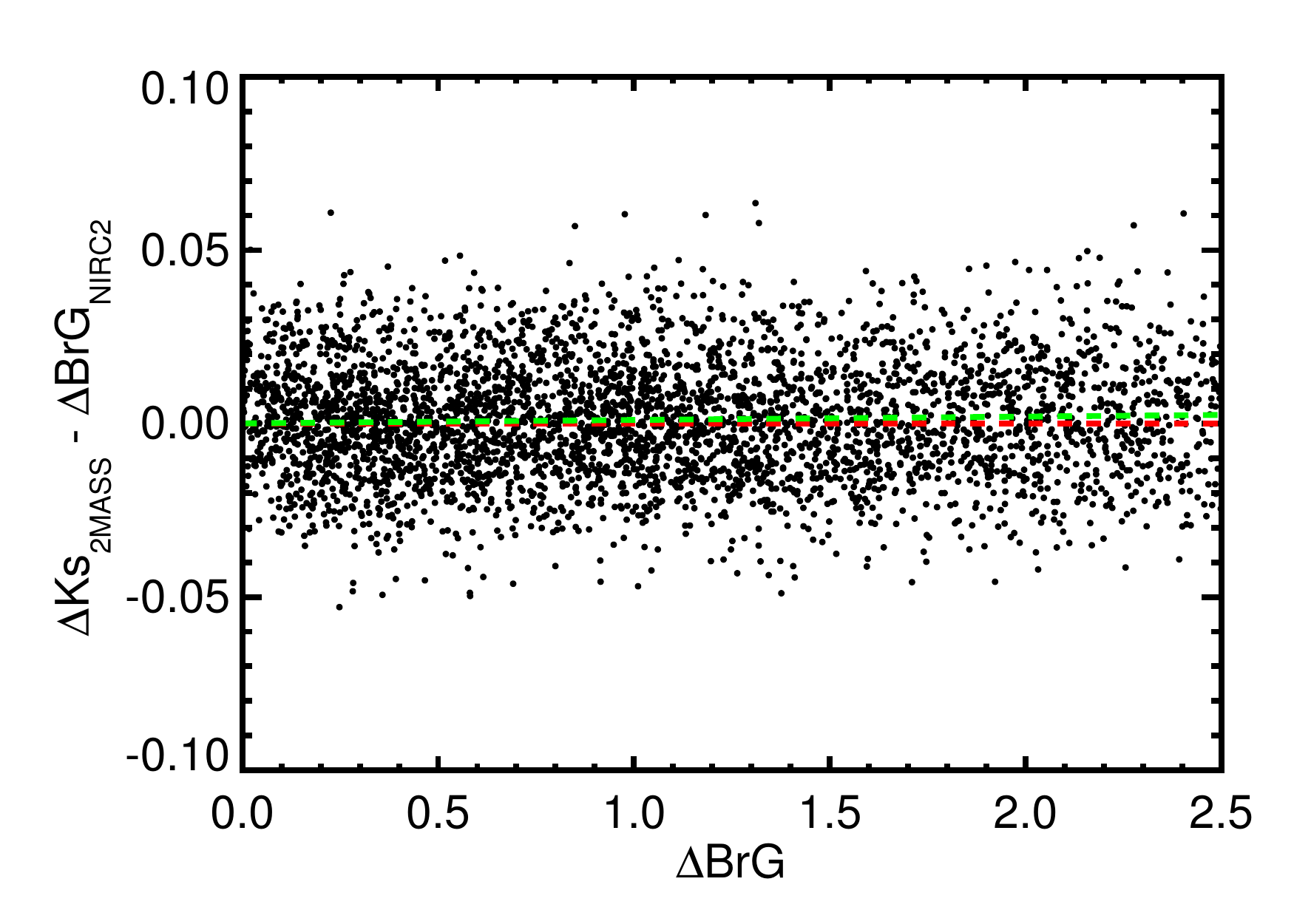}
\includegraphics[width=0.47\textwidth]{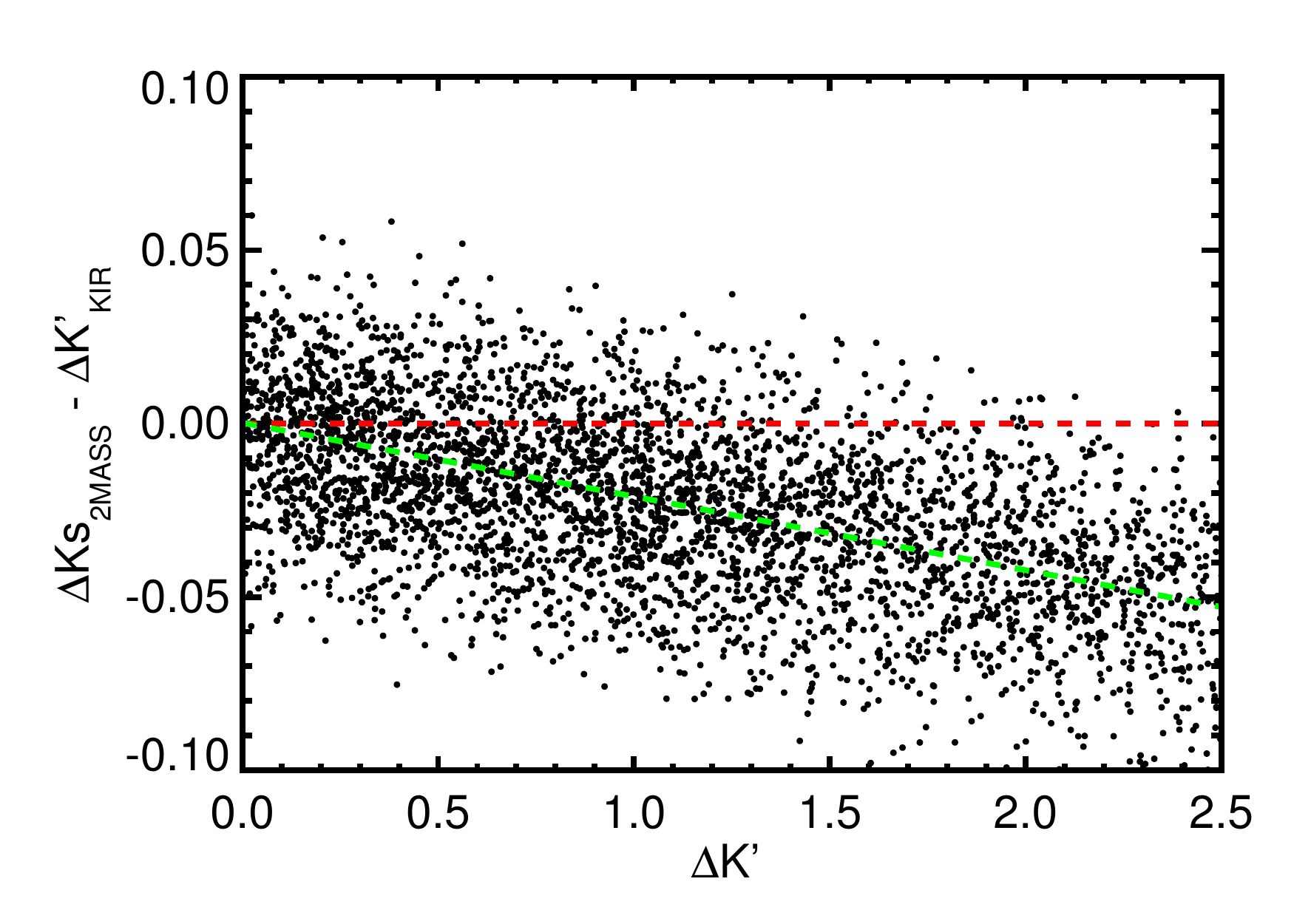}
\includegraphics[width=0.47\textwidth]{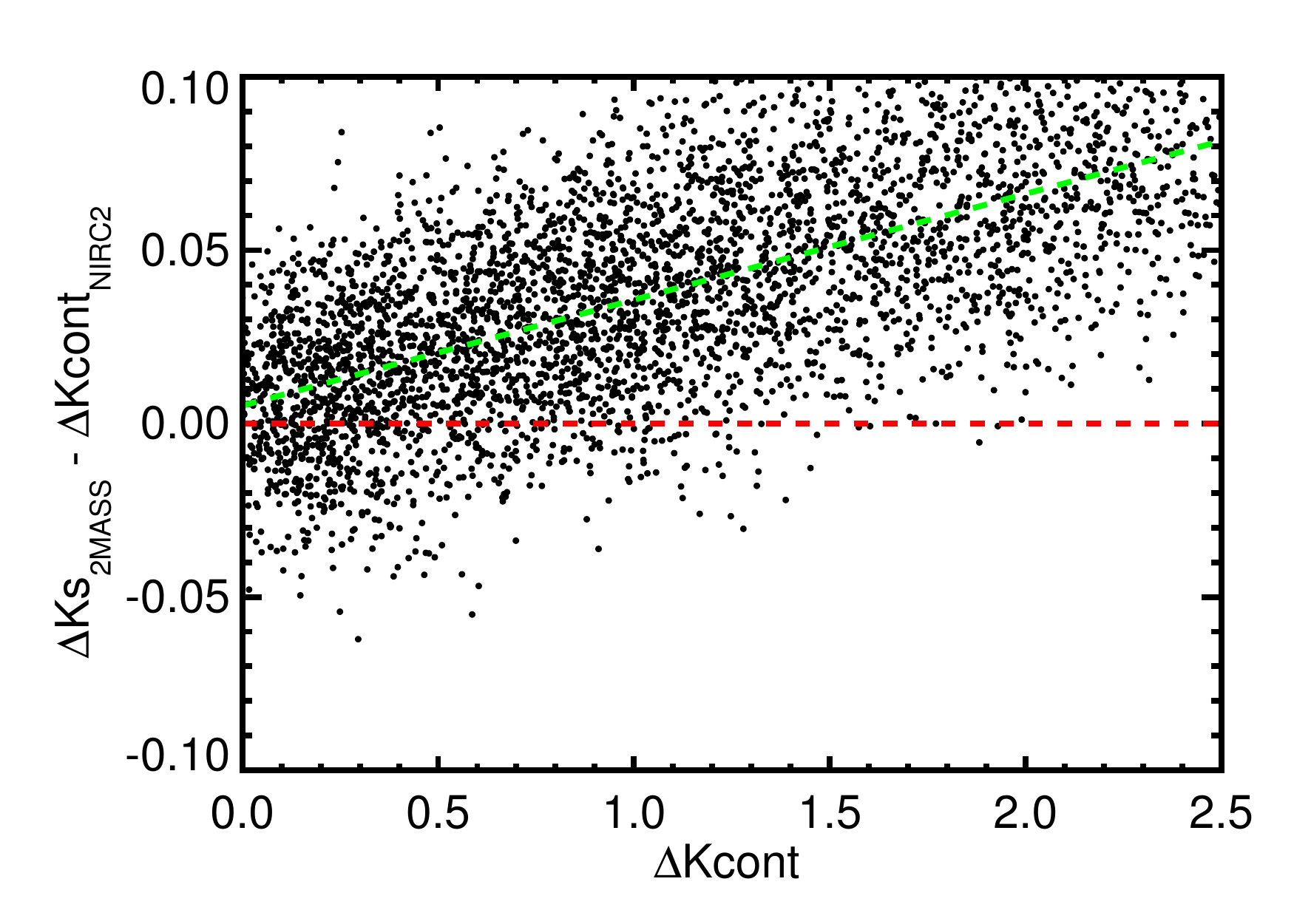}
\caption{Difference between 2MASS $\Delta K_S$ and four example $\Delta m$ values measured from our AO imaging, built from a grid of absolutely-calibrated spectra and the relevant filter profile. No corrections are applied for the NIRC2 $Br\gamma$ and $K_S$ (K-short) filter, as the trend is not significant compared to the calibration precision of the underlying spectra. }
\label{fig:mags}
\end{center}
\end{figure*}

We repeated this process with 5000 unique combinations of the 183 stars for 12 different filter/instrument combinations. For each filter and instrument combination we computed a best-fit line to $\Delta K_S -\Delta K_X$ as a function of $\Delta K_X$. We show four examples in Figure~\ref{fig:mags}. For the majority of the filters considered, the trend is insignificant compared to errors in the underlying spectra and absolute calibration (1-2\%). We did not apply a correction in these cases. 

Most of the scatter seen in Figure~\ref{fig:mags} is due to random errors in the distance of the template star or Poisson noise in the spectra. $K_{\rm{cont}}$, for example, shows a larger apparent scatter in Figure~\ref{fig:mags}, primarily because the narrow band is more sensitive to random Poisson errors in the calibrated spectra, but the final calibration is relatively precise. The uncertainties on applied corrections were 0.01-0.02~mag for all filters, which was driven primarily by potential systematic errors in the underlying spectra. 

We did not see a significant difference in any derived correction based on the metallicity of the component stars. This was expected given how [Fe/H] changes $K$-band flux levels (Figure~\ref{fig:metal}). We also found no significant effect as a function of the mass of the primary.

\clearpage

\end{document}